\documentclass{aa}  
\usepackage{graphicx}
\usepackage{txfonts}
\usepackage{hyperref}
\usepackage{ulem}
\usepackage{comment}
\usepackage{subcaption}
\begin{document} 

\title{The mass function dependence on the dynamical state of dark matter haloes}

\author{
R. Seppi\inst{1}\thanks{E-mail: rseppi@mpe.mpg.de} \and
J. Comparat\inst{1} \and
K. Nandra\inst{1} \and
E. Bulbul\inst{1} \and
F. Prada\inst{2} \and
A. Klypin\inst{3,4}  \and
A. Merloni\inst{1}  \and
P. Predehl\inst{1} \and
J. Ider Chitham\inst{1}
}
\institute{
Max-Planck-Institut f\"{u}r extraterrestrische Physik (MPE), Giessenbachstrasse 1, D-85748 Garching bei M\"unchen, Germany\\
\and
Instituto de Astrofísica de Andalucía (CSIC), Glorieta de la Astronomía, E-18080 Granada, Spain \\
\and 
Astronomy Department, New Mexico State University, Las Cruces, NM, USA \\
\and
Department of Astronomy, University of Virginia, Charlottesville, VA, USA
}

\date{Accepted XXX. Received YYY; in original form ZZZ}


\abstract{
Galaxy clusters are luminous tracers of the most massive dark matter haloes in the Universe.
To use them as a cosmological probe, a detailed description of the properties of dark matter haloes is required. 
}
{
%
We characterize how the dynamical state of haloes impacts the dark matter halo mass function at the high-mass end (i.e., for haloes hosting clusters of galaxies). }
{
%
We used the dark matter-only MultiDark suite of simulations and the high-mass objects $M > 2.7\times 10^{13} M_\odot/h$ therein. 
We measured the mean relations of concentration, offset, and spin as a function of dark matter halo mass and redshift. 
We investigated the distributions around the mean relations. 
We measured the dark matter halo mass function as a function of offset, spin, and redshift. 
%
We formulated a generalized mass function framework that accounts for the dynamical state of the dark matter haloes. 
}
{%
We confirm the recent discovery of the concentration upturn at high masses and provide a model that predicts the concentration for different values of mass and redshift with one single equation.
We model the distributions around the mean values of concentration, offset, and spin with modified Schechter functions.
We find that the concentration of low-mass haloes shows a faster redshift evolution compared to   high-mass haloes, especially in the high-concentration regime.  
We find that the offset parameter is systematically smaller at low redshift, in agreement with the relaxation of structures at recent times. The peak of its distribution shifts by a factor of $\sim 1.5$ from $z=1.4$ to $z=0$.
%
The individual models are combined into a comprehensive mass function model, which predicts the mass function as a function of spin and offset. 
Our model recovers the fiducial mass function with $\sim3\%$ accuracy at redshift 0 and accounts for redshift evolution up to $z\sim 1.5$. 
}
{
This new approach  accounts for the dynamical state of the halo when measuring the halo mass function. It offers a connection with dynamical selection effects in galaxy cluster observations. 
This is key toward precision cosmology using cluster counts as a probe.
}

\maketitle

%
%
%
%
%
\section{Introduction}

Galaxy clusters are the most massive virialized, gravitationally bound structures in the Universe. 
They grow hierarchically, starting from matter perturbations in the initial density field. This makes them  good tracers of the underlying cosmic web and its densest regions. 
Clusters in cosmology are used to construct the halo mass function, which indicates the mass density of haloes in a specific volume, in a small mass interval   between M and M + dM \citep{weinberg_2013_review}. 
An early theoretical description of the mass function was given by \citet[PS][]{PressSchechter1974} based on the assumption that Gaussian density perturbations overcoming a fixed density contrast collapse into haloes. This formally accounts for only half of the total halo mass in the Universe. 
An alternative approach, employing the excursion set theory, solved these shortcomings by considering the probability of crossing a given barrier with random walks \citep{Bond1991}. 
This provides a good prediction for high-mass haloes, but it predicts too many low-mass objects.
The introduction of ellipsoidal collapse corrected these differences between simulations and theory \citep{ShethTormen1999, Sheth2002}.\\

The mass function has been extensively studied in more recent works in an  attempt to find a universal model that is  independent from cosmology \citep{Jenkins2001, Tinker2008, Bhattacharya2011, Despali2016, Bocquet2016, Comparat2017, Bocquet2020arXiv200312116B}.  
A robust way to build such a mass function model is using N-body simulations \citep{Kravtsov1997, Springel2005}, for example MultiDark \citep{Prada2012, Klypin2016}. 
The generalization of such models as a function of cosmological parameters is best handled by emulating the mass function based on large sets of simulations \citep[e.g.,][]{McClintock2019ApJ...872...53M,Nishimichi2019ApJ...884...29N,Bocquet2020arXiv200312116B}.
It is important to precisely predict the halo mass function to fulfill the potential of current and future X-ray, SZ, or optical cluster surveys such as eROSITA \citep{Merloni2012, Predehl2020arXiv201003477P}, Planck \citep{Zubeldia2019MNRAS}, SPT-3G \citep{Benson2014SPIE.9153E..1PB}, CMB S4 \citep{Abazajian2019CMBS4}, SPIDERS/eBOSS \citep{Dawson2016,Finoguenov2020}, DESI \citep{DESI2016}, 4MOST \citep{deJong2011_4MOST,Finoguenov2019Msngr}, Euclid \citep{Laureijs2011}, LSST \citep{2009lsst}, and WFIRST \citep{Spergel2015_WFIRST}. 
These future surveys will provide tighter constraints on mass-observable scaling relations. This means that systematic uncertainties due to the accuracy of the halo mass function model and its evolution with redshift will contribute to the total error budget significantly more than in the previous cluster counts experiments \citep{Salvati2020}. 
Therefore, a detailed prediction of the theoretical dark matter halo statistic is required. In this work we calibrate a mass function model that  includes the dynamical properties of dark matter haloes.
Relaxed dark matter haloes can be selected according to multiple diagnostics \citep{Neto2007, Maccio2008, Prada2012, Klypin2016}: (i) the virial parameter $2K/|W|-1$, where $K$ and $W$ are respectively the kinetic and potential energy within the virial radius; 
(ii) the spin parameter $\lambda = J\sqrt{E}/GM^{5/2}$ \citep{Peebles1969angular_momentum}, which  
traces the dynamical state of the halo;  
(iii) the fraction of substructures, which is higher in unrelaxed haloes; 
(iv) the offset parameter {\tt $X_{\rm off} = |R_{\rm peak} - R_{\rm cm}|/R_{\rm vir}$} \citep{Behroozi2013, Klypin2016}, which  
is the difference between the position of the peak of the density profile and the center of mass, normalized by virial radius. 
If the halo is perfectly relaxed, the peak of the profile will correspond to the center of mass, and  {\tt $X_{\rm off}$} will be small.
On the other hand, higher {\tt $X_{\rm off}$} values will indicate an unrelaxed halo (e.g., merger, accretion). 
A combination of these quantities can be used.

\begin{figure*}
    \centering
    \includegraphics[width=\columnwidth]{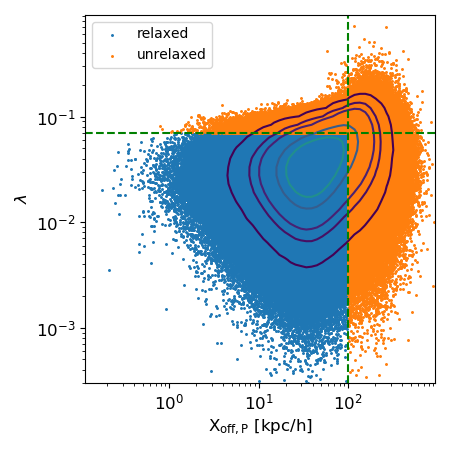}
    \includegraphics[width=\columnwidth]{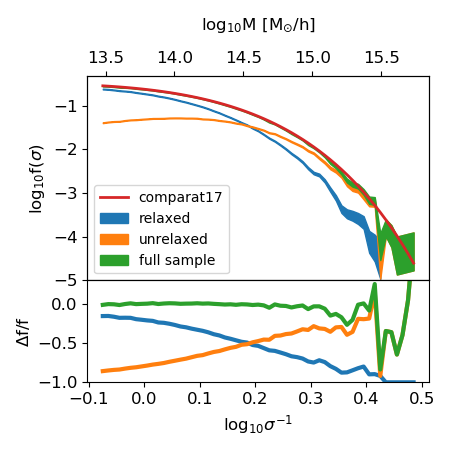}
        \caption{Contribution of relaxed and unrelaxed haloes to the total mass function. \textbf{
        Left panel}: Distribution of redshift zero dark matter haloes in the $X_{\rm off,P}$ and $\lambda$ plane. Cuts in $X_{\rm off,P}$ and $\lambda$ are applied to divide relaxed (blue) and disturbed (orange) structures. The contours contain 1$\%$, 5$\%$, 10$\%$, 30$\%$, and  50$\%$ of the data. \textbf{Right panel}: Halo mass functions vs. mass ($\sigma$) at redshift 0, as defined in Sect. \ref{sec:formalism} and using Equations \ref{eq:f_counts} and \ref{eq:fsig}. This mass function is built with different subsets of haloes from HMD at $z=0$ (see Sect. \ref{sec:N-body:data}): the red line indicates the model from \citet{Comparat2017}, the shaded areas represent the relaxed (blue), unrelaxed (orange), and the full sample (green) of haloes. The areas cover $1\sigma$ uncertainties. The lower panel shows the residuals fraction of each component compared to the red model in the upper panel.}
    \label{fig:relaxed-unrelaxed-MF}
\end{figure*}

To illustrate the respective contribution to the mass function of relaxed and disturbed haloes, we divided the sample of haloes in HugeMultiDark (hereafter HMD, see Sect. \ref{sec:N-body:data}) at $z=0$ using $X_{\rm off}$ and $\lambda$. 
We consider the offset parameter in physical scale $X_{\rm off,P}=|R_{\rm peak} - R_{\rm cm}|$ (i.e., not normalizing by the virial radius). We consider haloes with $X_{\rm off,P} < 100$ kpc/$h$ and $\lambda < 0.007$ as relaxed. In Fig. \ref{fig:relaxed-unrelaxed-MF} we show the halo mass function of the complete halo population: the relaxed and disturbed halo population. 
The left panel shows the distribution of redshift zero haloes in the $X_{\rm off,P}$ and $\lambda$ plane. 
The right panel shows the three multiplicity functions (see formalism in Sect. \ref{sec:formalism}) sampled by all haloes (green), relaxed haloes (blue), and disturbed haloes (orange). 
The bottom panel shows their relative contribution as a function of halo mass. 
It is clear how at $10^{13.5} M_{\odot}/h$ the contribution from relaxed structures dominates by a factor of about 0.8 dex. 
The multiplicity functions of the two samples cross each other at $10^{14.5} M_{\odot}/h$, then the unrelaxed structures take over at the high-mass end. 
This approach offers a possible connection to selection effects in observations, such as the cool core bias in X-rays \citep{Eckert2011, Kafer2019, Kaefer2020A&Awvdet}. It possibly offers a solution to mitigate biases in a cosmological interpretation of clusters abundance. This might improve cosmological constraints using X-ray selected clusters \citep{IderChitham2020MNRAS.499.4768I}.

In this article, we investigate the variations of the dark matter halo mass function as a function of the dynamical state of the constituting haloes. 
To trace the dynamical state we use {\tt $X_{\rm off}$} and $\lambda$.
This paper is structured as follows.
We define the formalism in Sect. \ref{sec:formalism}. 
We present the N-body data used in Sect. \ref{sec:N-body:data}. 
We present the average relations between concentration, $X_{\rm off}$, spin, and mass, as well as their distribution around mean values, in Sect. \ref{sec:relations}.
We define the generalized mass function framework in Sect. \ref{sec:general_mf_framework}.
We present the generalized model of the halo mass function as a function of {\tt $X_{\rm off}$} and $\lambda$ in Sect. \ref{sec:model}.
We present the results of the fit and the best-fit parameters in Sect. \ref{sec:results}.
We summarize our findings in Sect. \ref{sec:discussion}, and discuss their implications for cosmological studies. 
%
%
%
%
%
%
\section{Formalism and definitions}
\label{sec:formalism}



The growth of the density perturbations in the matter field is described by the evolution of the overdensity field and its variance as a function of scale.
The variance of the smoothed density field is defined as
\begin{equation}
    \sigma^2(M,z) = \int \frac{d^3k}{(2\pi)^3}\hat{W}^2(k,R)P(k,z),
    \label{eq:sigma_definition}
\end{equation}
where $k$ is the wavenumber,  
$\hat{W}$ the Fourier transform of a top-hat filtering function, 
and $P(k,z)=D_{+}(z)^2 P(k,0)$ the linear matter power spectrum. Its redshift evolution is encoded in the growth factor $D_+(z)$. 
The quantity
\begin{equation}
    \nu(M,z) = \delta_{cr}(z)/\sigma(M,z)
    \label{eq:nu}
\end{equation}
is the peak height, where $\delta_{cr}$ 
is the critical overdensity required for a structure to 
collapse 
in a dark matter halo \citep{PressSchechter1974, Sheth2002}. 
We report explicit values of mass, $\sigma$, and $\nu$ in Table \ref{tab:mass_sigma_nu} for z=0 and z=0.5.

\noindent We write the mass function in its differential form as in Eq. \ref{eq:MF}  \citep[see][for a recent review]{Allen2011}, 
\begin{equation}
    \frac{\text{d}n}{\text{dln}M} = \frac{\rho_m}{M} \Big|\frac{\text{dln}\sigma}{\text{dln}M}\Big|f(\sigma),
    \label{eq:MF}
\end{equation}
where $f(\sigma)$ is the multiplicity function. 
A comprehensive list of models of the multiplicity function is available in Table 1 of \citet{Murray2013}. 

\noindent In this paper as mass variable we  use   $\sigma$ (Equation \ref{eq:sigma_definition}),  peak height (Equation \ref{eq:nu}),  and mass interchangeably.

\begin{table}
    \centering
    \caption{Correspondence between mass, peak height, and variance  of the linear density field at z=0 and z=0.5.}    
    \begin{tabular}{c c c c c c c}
    \hline \hline
    Mass & \multicolumn{2}{c}{z=0} & \multicolumn{2}{c}{z=0.5} &  \\
    $\log_{10}M_{\odot}/h$ & $\nu=\delta_c/\sigma$ & $\log_{10}(1/\sigma)$ & $\nu=\delta_c/\sigma$ & $\log_{10}(1/\sigma)$ \\
    \hline
10 & 0.446 & -0.577 & 0.579 & -0.465 \\
10.5 & 0.506 & -0.523 & 0.657 & -0.41 \\
11 & 0.58 & -0.463 & 0.752 & -0.35 \\
11.5 & 0.673 & -0.399 & 0.872 & -0.286 \\
12 & 0.79 & -0.329 & 1.02 & -0.216 \\
12.5 & 0.942 & -0.253 & 1.22 & -0.14 \\
13 & 1.14 & -0.17 & 1.48 & -0.0568 \\
13.5 & 1.41 & -0.0779 & 1.83 & 0.035 \\
14 & 1.78 & 0.0236 & 2.31 & 0.137 \\
14.5 & 2.31 & 0.137 & 3 & 0.25 \\
15 & 3.09 & 0.263 & 4.01 & 0.376 \\
15.5 & 4.29 & 0.405 & 5.56 & 0.518 \\

    \hline
    \end{tabular}
    \footnotesize{These quantities are described by Eq. \ref{eq:sigma_definition} and \ref{eq:nu}.}
    \label{tab:mass_sigma_nu}
\end{table}

%
%
%
%
%
\section{Simulations}
\label{sec:N-body:data}

We describe the set of gravity-only simulations and the halo finding post-process in this section.

\subsection{MultiDark}
We use the MultiDark simulations \citep{Prada2012, Riebe2013, Klypin2016}. 
They are computed in a flat $\Lambda \rm CDM$ Planck \citep{Planck_2014} cosmology ($H_0=67.77$ km s$^{-1}$ Mpc$^{-1}$, $\Omega_{m0}=0.307115$, $\Omega_{b0}=0.048206$, $\sigma_8=0.8228$) with the \textsc{gadget-2} code \citep{Springel2005}. 
It is one of the largest sets of high-resolution ($\sim4000^3$ particles) N-body simulations. 

We used three MultiDark simulations: \textsc{HMD}, \textsc{BigMD}, \textsc{MDPL2} (see details in Table \ref{tab:MD}). 
Alternative simulations that could be used for this project include \textsc{Millennium-XXL, DarkSkies, Q Continuum, v$^2$GC simulation}, described in \citet{Angulo2012,Skillman2014,Heitmann2015}, and \citet{Ishiyama2015} respectively.

\begin{table}
        \centering
        \caption{N-body simulations used in this analysis. 
        L: length of the box in Gpc$/h$. 
        M$_p$: mass of the particle in M$_\odot/h$. 
        M$_{\rm min}$: minimum halo mass considered $M_{\rm vir}>M_{\rm min}$ in M$_\odot/h$. 
        Number of haloes in the snapshots at z=0.}
        \label{tab:MD}
        \begin{tabular}{cc cc cc cc} 
        \hline
    Name &  L & M$_{\rm p}$ & M$_{\rm min}$ & N haloes      \\
    \hline
\textsc{HMD}  & 4.0    & $7.9\times10^{10}$ & $2\times10^{13}$ &13 330 574\\ 
\textsc{BigMD}  & 2.5    & $2.4\times10^{10}$ & $5\times10^{12}$ &27 575 832 \\ 
\textsc{MDPL2}  & 1.0    & $1.51\times10^{9}$ & $4\times10^{11}$ &17 036 888\\ 
\hline
        \end{tabular}
\end{table}

The complete list of simulation outputs (snapshots) utilized are given in  Table \ref{tab:snapshots:1}, where the expansion parameter $a$ and the corresponding redshifts are reported for each snapshot.

\subsection{Halo finding}
Finding haloes in dark matter simulations is not an easy task \citep[see][for a review]{Knebe2013MNRAS.435.1618K, Behroozi2015MNRAS.454.3020B}.
In this study haloes are identified by the Robust Overdensity Calculation using K-Space Topologically Adaptive Refinement (\textsc{rockstar}) and  \textsc{consistentTrees} algorithms \citep{Behroozi2013}. 
They are  based 
on adaptive hierarchical refinement of friends-of-friends (FOF) groups. They work with six phase-space dimensions (halo positions and velocities), and one time dimension. This allows us to track 
relative motions and merging history between substructures in different snapshots.
\textsc{rockstar} computes the halo mass of identified objects by removing the unbound particles inside the virial radius. Virial mass and virial radius are related by

\begin{equation}
    M_{\rm vir}(z) = \frac{4}{3}\pi \Delta_{\rm vir}(z)\Omega_{\rm M}(z)\rho_{\rm b}(z)R_{\rm vir}^3,
    \label{eq:Mvir}
\end{equation}
where $R_{\rm vir}$ encompasses a mean halo density equal to the background matter density multiplied by $\Delta_{\rm vir}$, $\Omega_{\rm M}$ is the matter density parameter, and $\rho_{\rm b}$ is the matter density of the Universe. The overdensity over the matter background $\Delta_{\rm vir}$ is defined according to \citet{BryanNorman1998} as
\begin{align}
    \Delta_{\rm vir} &= (18\pi^2 + 82x- 39x^2)/\Omega(z), \label{eq:delta_vir} \\ \notag
    \Omega(z) &= \Omega_{\rm M,0}(1+z)^3/E(z)^2, 
\end{align}
where $x=\Omega(z)-1$, $\Omega_{M,0}$ is the matter density parameter at the present day, and E(z) is the Hubble parameter as a function of redshift in units of $H_0$:
\begin{equation*}
    E(z) = \frac{H(z)}{H_0} = \sqrt{\Omega_{\rm 0,M} (1+z)^3 + \Omega_\Lambda}.
\end{equation*}
Given the reference cosmology adopted in this work, the virial overdensity is equal to 332.5 at z=0 and asymptotically tends to 178 at high redshift.\\
The recovery of main properties such as position, mass, and circular velocity is consistent between different finders. However, derived properties such as spin show a $\sim$20$\%$ scatter \citep{Knebe2013MNRAS.435.1618K}. This holds especially for low-mass haloes with fewer than 30-40 particles, where the identification of substructures is not straightforward \citep{Knebe2011}. The low-mass limits in Table \ref{tab:MD} are set at more than $\sim$200 particles per halo in all boxes. This ensures accurate halo properties \citep{Behroozi2013, Knebe2013MNRAS.435.1618K}. \\
In this work we use the virial overdensity (Equation \ref{eq:delta_vir}). The virial mass function has been shown to be the one that comes closest to universality \citep{Despali2016}.

%
%
%
%
%
\section{Concentration, offset, spin: Empirical relations with peak height and redshift}
\label{sec:relations}

In the footsteps of \citet{Klypin2016} and \citet{Rodriguez-Puebla2016}, 
we analyze the average relations linking concentration, $\lambda$, and $X_{\rm off}$ to the peak height, respectively in sections \ref{subsection:conc:mass}, \ref{subsection:spin-mass}, and \ref{subsection:xoff-mass}. 
We also analyze the distributions of these quantities around the mean relations. 
The mean relations are fitted by models that simultaneously account for the mass and redshift dependence of the relations (Eqs. \ref{eq:conc_peak}, \ref{eq:lambda_sigma_rel}, and \ref{eq:xoff_sigma_rel}).
The probability density functions (PDFs) of concentration, spin, and $X_{\rm off}$ are fitted by modified Schechter models, respectively in equations \ref{eq:modified_schechter_conc}, \ref{eq:modified_schechter}, and  \ref{eq:modified_schechter_xoff}. The 
PDFs at different redshifts are modeled independently.



\subsection{Concentration--mass--redshift relation}
\label{subsection:conc:mass}

We study the relation between concentration and mass.
Numerical simulations \citep{Navarro_Frenk_White_1996} show that to a good approximation ($\sim10-20\%$), the density of dark matter haloes is described by the profile in Eq. \ref{eq:NFW},

\begin{equation}
    \rho(r) = \frac{\rho_s}{(r/R_s)(1+r/R_s)^2},
    \label{eq:NFW}
\end{equation}
where $R_s$ is the scale radius. 
The characteristic density of the halo $\rho_s$ is   equal to
\begin{equation}
        \rho_s 
               = \rho_{\rm crit} \frac{\Delta}{3}\frac{c^3}{\ln(1+c)-c/(1 + c)}, 
\end{equation}
where $c$ is the concentration, $\Delta$ the overdensity, and $\rho_{\rm crit}$ is the critical density of the Universe. 
The concentration is a dimensionless quantity defined by 
\begin{equation}
c_{\Delta} = R_{\Delta}/R_s,
\end{equation}
where $\Delta$ can refer to any threshold. In this work we consider the virial overdensity over the matter background $\Delta_{\rm vir}$ (see Equation \ref{eq:delta_vir}). Therefore, we study the virial concentration $c_{\rm vir} = R_{\rm vir}/R_{\rm s}$.
The concentration--mass relation  is an important part of models describing galaxy clusters \citep[e.g., reviews from][]{Allen2011,umetsu2020clustergalaxy} or gravitational lensing \citep[e.g., reviews from][]{Bartelmann_2010,Kilbinger_2015}.

This relation has been extensively studied in simulations. 
The concentration anti-correlates with mass with negative redshift trend \citep[e.g.,][]{Diemer_Joyce2019_conc,Ragagnin2019conc_magneticum}. 
Its detailed trend depends on the measurement method,  in simulations \citep{Meneghetti2013conc, Lang2015concvoroni, Poveda-Ruiz2016ApJ...832..169P} and in observations \citep{Foex2014A&A...572A..19Fconc, Phriksee2020MNRAS.491.1643P, Du2015conc_mass_obs, Shan2017ApJ840104S}. This introduces possibles biases in the measure of concentration \citep{Sereno2015conc_selection, Cibirka2017conc_codex, vanUitert2016A&A...586A..43V}.
\citet{Diemer2015conc_universal} found that the relation shows the smaller deviation from universality when adopting the definition $c_{200c}$. 
However, it is not completely universal, meaning that concentration is described not only by mass or $\nu$, but also by assembly history. 
Leaving aside biases in definition and measurement, different models have been proposed to describe it: 
power laws \citep[{e.g.,}][]{Duffy_2008,Dutton_Maccio_2014}; a 
combination of power laws  to describe the high-mass upturn \citep[{e.g.,}][]{Klypin2016, Diemer_Joyce2019_conc}; 
semi-analytic models based on the Press-Schechter theory \citep{Correa2015Mconc_model}.
In this section we extend the models from \citet{Klypin2016}. We adjust a global model that includes a redshift dependence for high-mass haloes at relatively low redshift ($z < 1.5$), which is particularly interesting for haloes hosting galaxy clusters.

\subsubsection{Model}

We model the concentration $c$ as a function of the rms of the overdensity field $\sigma(M,z)$ (not as the function of halo mass $M$). 
Our parameterization of the concentration--$\sigma$ relation is a generalization of the  \citet{Klypin2016} relation and reads
\begin{align}
    &c(\sigma,z) = \frac{b_0}{(1+z)^{0.2}}  \Big[1 + 7.37 \Big(\frac{\sigma}{a_0(1+z)^{1/2}}\Big)^{3/4} \Big] ...\nonumber \\
    &\Big[1 + 0.14 \Big(\frac{\sigma}{a_0(1+z)^{1/2}}\Big)^{-2} \Big].
    \label{eq:conc_peak}
\end{align}

\begin{figure}
    \centering
    \includegraphics[width=\columnwidth]{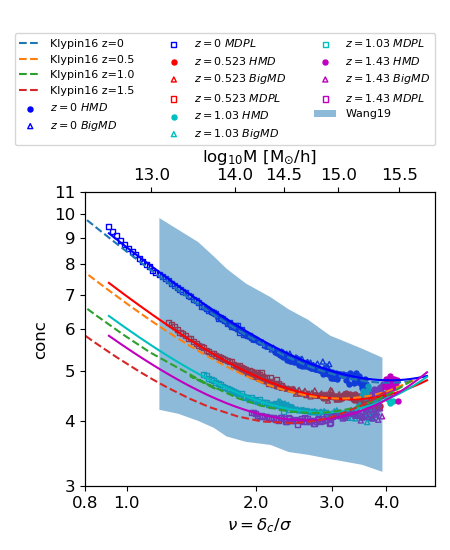}
        \caption{Concentration--$\sigma$ relation (Equation \ref{eq:conc_peak}). Circular dots, triangles, and squares represent \textsc{HMD, BigMD, MDPL2,} respectively. They are color-coded by redshift. Straight lines indicate our best-fit model, dotted lines show the model from \citet{Klypin2016}, and  the shaded blue area indicates the distribution from \citet{Wang_Springel_White2019} at z=0. The upper x-axis converts peaks values into mass at z=0.}
    \label{fig:conc_mass_relation}
\end{figure}

The best-fit values are in Table \ref{tab:conc_sigma_pars}. We find best-fit values of $a_0 = 0.754091 \pm 0.000004, b_0 = 0.574413\pm 0.000002$, in agreement with \citet{Klypin2016}. 
This model is fitted using haloes with $M > 10^{12.5} M_\odot/h$ ($\nu \sim 0.95$ at z=0). The low-mass end is not sampled due to the particle mass resolution. We do not consider a high-redshift regime where there is a statistical limitation of high-mass haloes. We chose the binning limits following \citet{Klypin2016}.
The strength of our model resides in the ability to predict the concentration--$\sigma$ relation for a variety of masses and redshifts with a single equation.
We recover the hockey stick shape of the relation (see Fig. \ref{fig:conc_mass_relation}). This is consistent with \citet{Prada2012}, \citet{Klypin2016}, and \citet{Diemer_Joyce2019_conc} (the last also includes a cosmology dependence). 
This upturn feature is absent in \citet{Duffy_2008} and \citet{ Wang_Springel_White2019}, 
due to the smaller volumes analyzed (400 Mpc/$h$ and 500 Mpc/$h$, respectively). They are not large enough to obtain a significant number of high-mass objects and probe the upturn. 

\begin{figure*}
    \centering
    \includegraphics[width=1.8\columnwidth]{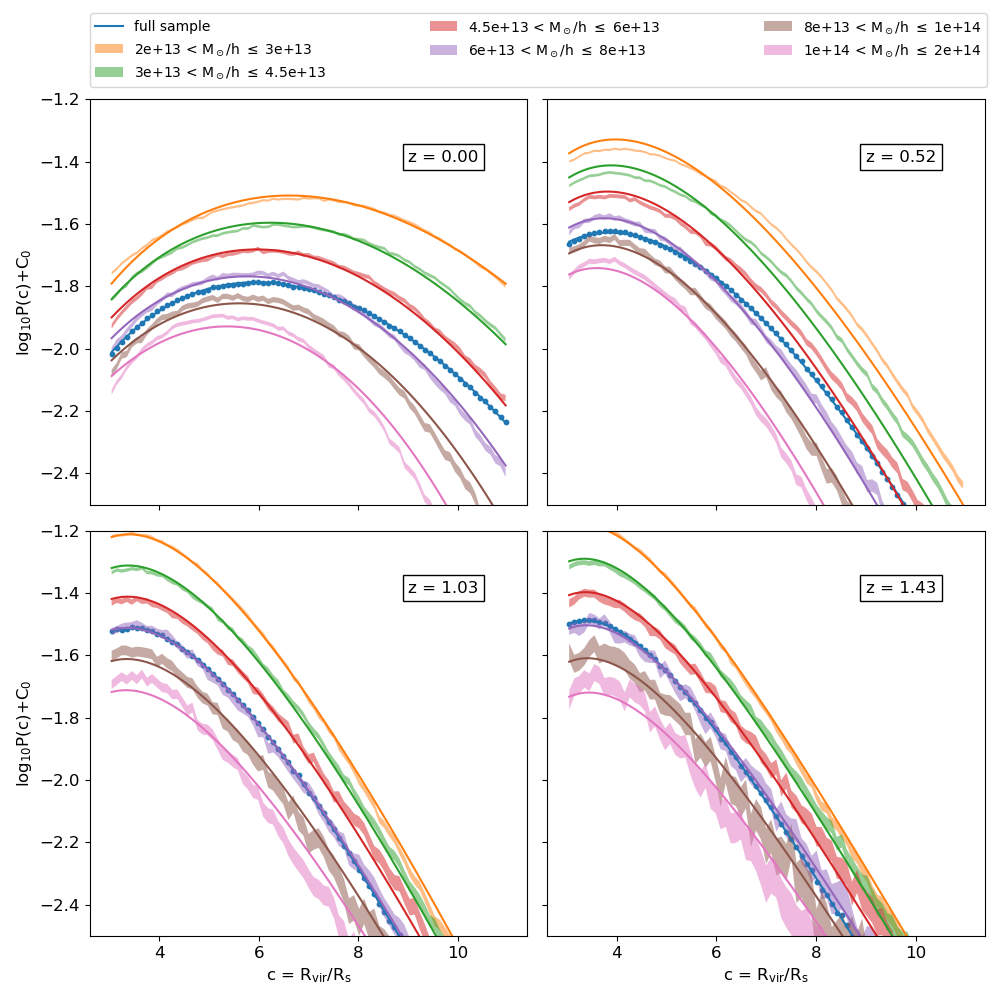}
        \caption{Probability density function of the concentration (Equation \ref{eq:modified_schechter_conc}) at the different redshifts values given in each panel. Each set is divided into mass slices, and is color-coded accordingly. The shaded areas represent the data with $1\sigma$ error, while the straight lines indicate the best-fit model. The blue points and   line represent the total sample not sliced in mass. For clarity, each line and its fit is shifted by 0.1 dex on the y-axis. This means that the constant $C_0$ assumes values of (+0.3, +0.2, +0.1, 0.0, -0.1, -0.2). The purple line is not shifted, and is thus  the one with the proper normalization. }
    \label{fig:conc_pdf}
\end{figure*}

We describe the distribution of concentration around its mean value and we model its probability density function.
We use snapshots from \textsc{HMD}, at four redshifts. 
Each snapshot is divided into six slices of mass. 
The PDFs obtained at each redshift are fitted simultaneously.
It  includes a $\sigma$ dependence in our model (see Equation \ref{eq:modified_schechter_conc}): 
\begin{equation}
    P(c, \sigma) = A \Big(\frac{c}{x_0\sigma^{e_0}} \Big)^{\alpha\sigma^{e_1}} \text{exp} \Big[-\Big(\frac{c}{x_0\sigma^{e_0}} \Big)^{\beta\sigma^{e_2}} \Big].
    \label{eq:modified_schechter_conc}    
\end{equation}
The best-fit parameters are included in Table \ref{tab:conc_sigma_pars}. Uncertainties are negligible for most parameters. 
The distribution of concentration around its mean value is well described by a modified Schechter function with mass-dependent terms (Equation \ref{eq:modified_schechter_conc}).
We use the best-fit model to predict the distribution of concentration in fixed mass slices as a function of redshift. The result is
shown in Fig. \ref{fig:conc_pdf}. The full distribution of concentration (not sliced in different mass intervals) is well described by a modified Schechter as well.



\subsubsection{Discussion}
The average value of concentration increases at late times. It goes from 4 at z=1.43 to 5.8 at z=0 for haloes corresponding to peak height $\nu=2$.
We confirm the recent discovery of the concentration upturn at high masses \citep{Klypin2016,Diemer_Joyce2019_conc}. \citet{Klypin2016} suggest that the high-mass upturn is caused by the tendency of regions with smaller root mean square variance in the overdensity field to be more spherical. 
These regions are the ones that evolve into high-mass dark matter haloes. 
Because of this aspect, gravitational accretion of matter toward the center is more efficient and results in higher concentration. In addition, very massive haloes correspond to the highest peaks of the density field. This means that the probability of finding them in a phase characterized by strong accretion and therefore higher concentration is higher, as shown by \citet{Ludlow2012MNRAS.427.1322L}.

We find that the concentration of haloes with different masses shows a different evolution with redshift (see Fig. \ref{fig:conc_pdf}).
At z=0 the concentration of high-mass haloes gives a smaller contribution to the total  PDF in the high-concentration tail. 
At higher redshift we see high-concentration haloes with both high and low mass. The opposite holds for low-mass haloes, which contribute more to the total PDF at higher z in the low-concentration regime. 
At high redshift the distribution of concentration for different mass bins has the same shape. 
Conversely, at low redshift this statement does not hold. 
The low-mass distribution is much broader than the high-mass one.
There is also a general redshift trend: with decreasing redshift the number of high-concentration, low-mass haloes increases. 
More recently in time the distribution is flatter; in fact, the slope of the power-law is smaller, going from $\sim 4.5$ at z=1.43 to $\sim 1.4$ in the present day. Moreover, the mass trend of the exponential decay at z=0 is negative with sigma ($e_2 = -0.959$), which translates into a faster decrease at high mass.  This means that the distribution of concentration of low-mass haloes evolves more than that of high-mass objects. For example, at concentration $c=8$ the PDF of high-mass haloes ($1\times 10^{14} < M < 2\times10^{14} M_\odot/h$) changes by 0.32 dex between $z=1.43$ and $z=0$, while for low-mass objects ($2\times 10^{13} < M < 3\times10^{13} M_\odot/h$) it varies by 0.49 dex. There is a net difference of 0.17 dex between the two, meaning that the high end of concentration evolves $\sim$1.5 times faster for lower mass haloes compared to high-mass ones. 
This is in agreement with the fact that these haloes evolve in different environments. This causes a different redshift evolution of the distribution of concentration of haloes with different masses. Structures with the same mass but different formation histories present different properties. 
This aspect is related to the notion of assembly bias \citep{Gao2005, Croton2007, Angulo2008MNRAS.387..921A}. 
There is a correlation between halo concentration and the age of the universe when the progenitor reached a fixed fraction of its current mass \citep{Zhao2009conc_mah}. Moreover, a faster mass accretion causes a slower increase of concentration. Very massive objects reside in highly populated environments, which slows down the evolution of concentration \citep{Zhao2003MNRAS.339...12Z}. This can be used to predict concentration as a function of assembly history \citep{Giocoli2012concentration, Ludlow2013MNRAS.432.1103L, Ludlow2014MNRAS.441..378L}. The latter is also related to the evolution of cosmological parameters. This allows modeling of the concentration mass relation for cold and warm dark matter haloes \citep{Ludlow2016conc_cold_warm}. 
In general, high-mass haloes cluster more than low-mass ones. Low-mass objects, some of which are found in isolated environments, will experience different histories. 
Isolated low-mass haloes will be highly concentrated, while the concentration of low-mass haloes in dense environments will stay low. 
\citet{Contreras2019} found that low-concentration haloes cluster more than haloes with high concentration at high redshift. 
This is in agreement with our results.

\subsection{Spin--mass--redshift relation}
\label{subsection:spin-mass}

In this subsection we study the relation between the spin parameter and mass.
An accurate theoretical description of the distribution of the halo spin is key to understanding the  possible implications of the systemic rotation on cluster sample definitions.
This rotation is induced by a combination of the initial spin of the halo, the infalling material, and the merging activity.\\ 
Measurements of cluster rotation were obtained at low redshift, using member galaxies to infer the rotation movement \citep{Hwang_2007_spin_galaxy, Tovmassian2015rotat, Manolopouloucluster_rot, Bilton2019MNRAS.490.5017Bspin}. The spin in galaxy clusters has also been studied in X-rays \citep{Bianconi2013spin_xray, Eckert2019A&A...621A..40E}. 
High-resolution data are needed to explore this topic,  in the optical \citep{Song2018ApJ...869..124Sspin} and  X-ray \citep{Hitomi2018Perseus} bands. Moreover, cluster cores can be analyzed by distortion of the 6.7 keV line \citep{Sunyaev2003AstL...29..783Sspin}, fluctuations in the CMB due to rotation \citep{Rephaeli1995SZ_review, Cooray_2002_sz_spin}, and rotational kSZ effect \citep{Baxter2019JCAP...06..001Bspin_data}. 
Inferring motion with the SZ effect was also demonstrated for relaxed clusters with a significant spin on simulated (hydrodynamically) clusters \citep{Baldi2018spin, Baldi2019JPhCS1226a2003Bspin}. 
Future SZ and X-rays surveys might enable such measurements on a large number of clusters. 
A statistical description of the halo population as a function of spin is thus of interest, and is developed in this section.

\subsubsection{Model}

We analyze the spin--mass--redshift relation and its PDF. 
The spin is defined as $\lambda= JE^{1/2}/GM^{5/2}$
\citep{Peebles1969angular_momentum}.
We fit the mean relation with mass as a linear relation (Eq. \ref{eq:lambda_sigma_rel}). 

\begin{equation}
    \lambda(\sigma) = a_0 + b_0 \sigma.
    \label{eq:lambda_sigma_rel}
\end{equation}
\begin{figure}
    \centering
    \includegraphics[width=\columnwidth]{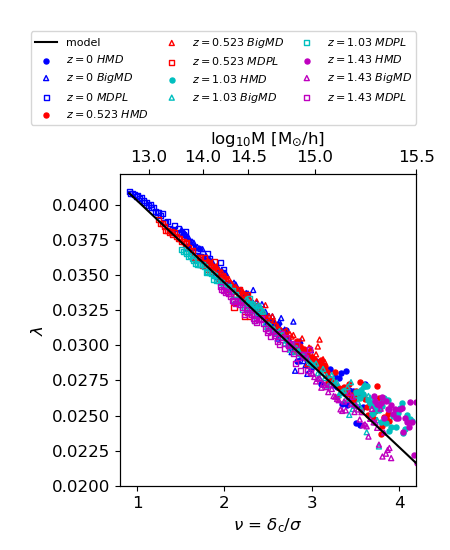}
        \caption{Spin--mass relation ($\lambda$-$\sigma$, Equation \ref{eq:lambda_sigma_rel}).
        The model is a linear relation with no redshift trend. 
        Data points are color-coded by redshift, while the different geometrical shapes refer to different simulations: squares for HMD, triangles for BigMD, and circles for MDPL. The straight line indicates the best-fit model, which considers all simulations and redshift at the same time. The best-fit parameters are given in Table \ref{tab:lambda_sigma_pars}. The upper x-axis converts peaks values into mass at z=0.
        }
    \label{fig:spin_sigma_relation}
\end{figure}

\begin{figure}
    \centering
    \includegraphics[width=\columnwidth]{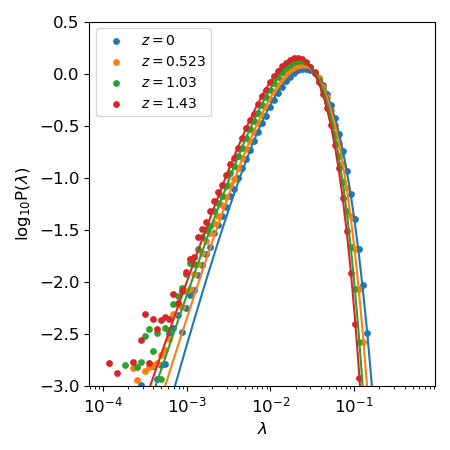}
    \caption{PDF of $\lambda$ (Equation \ref{eq:modified_schechter}). Individual points represent spin bins used to compute the distribution; straight lines refer to the best-fit modified Schechter model. They are color-coded by redshift. We do not consider mass dependence in this relation. Each redshift slice is fitted independently. The best-fit parameters are given in Table \ref{tab:lambda_sigma_pars}.}
    \label{fig:pdf_spin}
\end{figure}
There is no noticeable dependence on redshift, so we do not consider a redshift evolution in this model. 
The best-fit parameters are reported in Table \ref{tab:lambda_sigma_pars}. 
The relation is shown in Fig. \ref{fig:spin_sigma_relation}. 
We find that the spin correlates weakly with halo mass, as \citet{Bett_2007_spin} and \citet{Rodriguez-Puebla2016} did.
The PDF of the spin parameter is best fit by a modified Schechter law with no mass dependence: 

\begin{equation}
    P(\lambda) = A \Big(\frac{\lambda}{x_0} \Big)^\alpha \text{exp} \Big[-\Big(\frac{\lambda}{x_0} \Big)^\beta \Big].
    \label{eq:modified_schechter}    
\end{equation}

Distributions for different redshift snapshots are shown in Fig. \ref{fig:pdf_spin} (see Table \ref{tab:lambda_sigma_pars} for best-fit parameters). 
The result is consistent with previous findings, \citep[e.g.,][]{Rodriguez-Puebla2016}. 
The evolution with redshift of the PDF shows that with time haloes build up higher spins: the maximum of the PDF shifts to higher spin values when redshift decreases (Fig. \ref{fig:pdf_spin}). 
We also tried a lognormal distribution as a model, but it was not successful.


\subsubsection{Discussion}
The small correlation with mass and the well-modeled evolution with redshift makes it a rather simple dependence to account for in statistical studies of the halo population. 
From the perspective of the measurement of the halo mass function based on a cluster sample, marginalizing over the spin is possible, with limited complications. 
It also shows the spin cannot be considered  a candidate for assembly bias. 

\subsection{Offset--mass--redshift relation}
\label{subsection:xoff-mass}

The  offset parameter $X_{\rm off}$ traces the relaxation state of the halo \citep{Thomas2001, Neto2007, Henson2017MNRAS.465.3361H}. 
\citet{Hollowood2019ApJS..244...22H} analyzed the miscentering in SDSS galaxies followed up with Chandra X-ray observations. 
Provided there is  a link between $X_{\rm off}$ and the brightest cluster galaxy (BCG) to X-ray displacement, an estimation of the bi-variate mass and $X_{\rm off}$ function from observations of clusters is possible. 
To interpret it requires a detailed description of the link between  $X_{\rm off}$ and mass, which is detailed in this section.
We find that low-mass haloes have smaller offset than high-mass ones at each redshift. Moreover, the offset parameter is reduced by a factor of $\sim 1.5$ from $z\sim 1.5$ to $z=0$. This is in agreement with the fact that structures relax in time. 

\subsubsection{Model}

\begin{figure}
    \centering
    \includegraphics[width=\columnwidth]{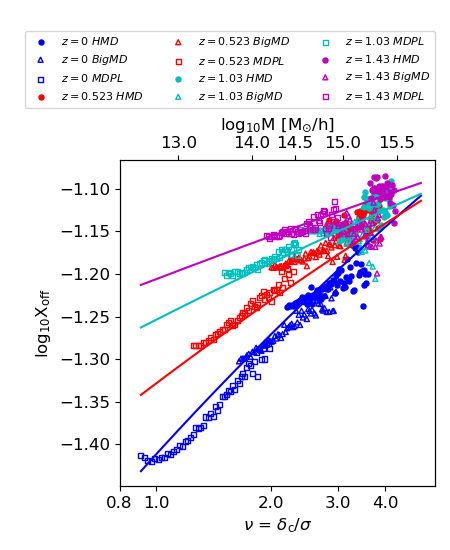}
        \caption{Offset--mass relation ($X_{\rm off}$-$\sigma$, Equation \ref{eq:xoff_sigma_rel}). 
        Circular dots, triangles, and squares represent \textsc{HMD, BigMD, and MDPL2} respectively. They are color-coded by redshift. Straight lines show the best-fit model. The best-fit parameters are given in Table \ref{tab:xoff_sigma_pars}. The upper x-axis converts peaks values into mass at z=0.}
    \label{fig:xoff_sigma_relation}
\end{figure}

We model the $X_{\rm off}-\sigma$ relation with a redshift dependent power law, see Equation \ref{eq:xoff_sigma_rel},
\begin{equation}
    \log_{10}X_{\rm off} = \frac{a_0}{E(z)^{0.136}}\sigma^{b_0E(z)^{-1.11}},
    \label{eq:xoff_sigma_rel}
\end{equation}
where $E(z)$ is the dimensionless Hubble parameter.
The best-fit parameters are given in Table \ref{tab:xoff_sigma_pars}. We find $a_0 = -1.30418 \pm 0.00001$, $b_0 = 0.15084 \pm 0.00001$. 
We find a significant redshift evolution of the normalization and the slope of the relation. 
The data obtained from MultiDark simulations and the best-fit models are shown in Fig. \ref{fig:xoff_sigma_relation}. The best-fit parameters of the $X_{\rm off}-\sigma$ relation and the distribution of $X_{\rm off}$ are given in Table \ref{tab:xoff_sigma_pars}. 
As for concentration, the strength of this equation relies upon its ability to predict an average $X_{\rm off}$ value given the mass and the redshift of a dark matter halo. 

\begin{figure}
    \centering
    \includegraphics[width=\columnwidth]{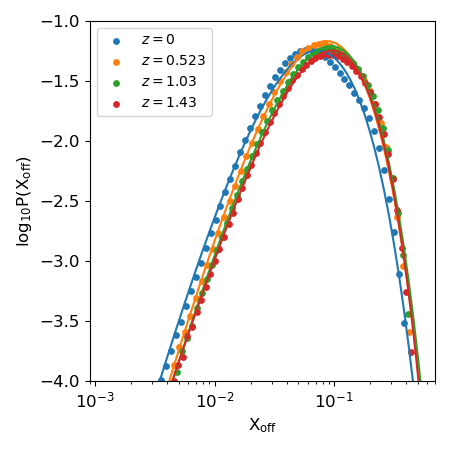}
        \caption{Probability density function of $X_{\rm off}$. Each panel shows the distribution at a specific redshift. Scatter points indicate the data, while straight lines represent the modified Schechter model. The samples are color-coded by redshift. Each redshift slice is fitted independently by Equation \ref{eq:modified_schechter_xoff}. The best-fit parameters are given in Table \ref{tab:xoff_sigma_pars}.
        }
\label{fig:pdf_xoff}
\end{figure}

We describe the distribution of $X_{\rm off}$ around its mean value with a modified Schechter function (Equation \ref{eq:modified_schechter_xoff}).
The PDF of $X_{\rm off}$ does not show mass dependency. It is included in the normalization to the virial radius $R_{\rm vir} \alpha M_{\rm vir}^{1/3} $. 
Therefore, we do not consider any $\sigma$ dependence:

\begin{equation}
    P(X_{\rm off}) = A \Big(\frac{X_{\rm off}}{x_0} \Big)^{\alpha} \text{exp} \Big[-\Big(\frac{X_{\rm off}}{x_0} \Big)^{\beta} \Big].
    \label{eq:modified_schechter_xoff}    
\end{equation}

We fit all haloes in each redshift snapshot together. Figure \ref{fig:pdf_xoff} shows distributions at different redshifts, fitted independently from one another. The parameters are given in Table \ref{tab:xoff_sigma_pars}. 

\subsubsection{Discussion}

We find a non-zero slope for the relation between $X_{\rm off}$ and mass. On average, high-mass haloes have a larger 
offset parameter. This is described by the negative $a_0$ parameter. This is in agreement with the hierarchical picture of structure formation. High-mass haloes formed recently and have not had time to dynamically relax. A further contribution is given by the environment surrounding these structures. High-mass haloes form in the knots of the large-scale structure where more matter is available for inflows and mergers, making these objects more disturbed.
In a picture where $X_{\rm off}$ possibly traces the cool core--non-cool core classification of galaxy clusters in X-rays \citep{Eckert2011}, massive structures have a higher fraction of non-cool cores. 
This influences the high-mass tail of the mass function (see Fig. \ref{fig:relaxed-unrelaxed-MF}). Therefore, given an X-ray flux limit, the mass function of a relaxed galaxy clusters sample will be complete to lower masses than an unrelaxed one.
This effect also evolves with redshift. 
At high z, haloes show a larger offset. In this context it means that it is more difficult to detect structures at earlier times. 
More recently, structures have had time to relax and therefore show smaller values of the offset parameter. 
This is linked to the development of cool cores at low redshift \citep{Ettori2008cool_core_evo}. 
The PDF shows a power-law growth from low offset values (slope $\alpha=3.71$ at z=0) and exponential cutoff at high $X_{\rm off}$. It is described by a modified Schechter function (Equation \ref{eq:modified_schechter_xoff}). Its maximum shifts by a factor of $\sim1.5$, from $X_{\rm off} \sim 0.09$ to $X_{\rm off} \sim 0.06$, between redshift 1.4 and 0, confirming that haloes have more time to relax.
The shape of the distribution does not show a significant redshift trend. The width of the probability density functions, measured at $\log_{10}P(X_{\rm off})=-2.5$, at $z=0$ and $z=1.4$ agree with $2.8\%$ accuracy. At z=0 these values span from $X_{\rm off}\sim 0.01$ to $X_{\rm off}\sim 0.3$.
This precludes the offset from being a possible assembly bias candidate.
\section{Generalized mass function}
\label{sec:general_mf_framework}

Generalizations of the mass function have been made in a number of directions: cosmological parameters, angular momentum, and friction \citep{Achitouv_2012,Achitouv_2014,Popolo2017} via detailed modeling of the collapse barrier. 
Nevertheless, it is technically demanding to connect these parameters to observations. 
In this section we generalize the mass function formalism to include additional variables 
($X_{\rm off}$,$\lambda$) in its formulation. 
These two quantities describe the properties of dark matter haloes and will hopefully become connectable to observational properties.

\subsection{Definitions}
We compute the mass function as follows:
\begin{equation}
    \frac{dn}{dlnM} = \frac{\Delta N_M}{V\Delta\ln M}.
    \label{eq:f_counts}
\end{equation}
Here $\Delta N_M$ is the number of haloes in each mass bin $\Delta\ln M$ and $V$ is the total volume of the simulation. It corresponds to Eq. \ref{eq:MF}.
The $M(\sigma)$ relation and its first derivative are computed with \textsc{colossus} \citep{Diemer2018}. 
By convention, the relation between a certain mass and its corresponding scale is normalized with the matter density at z=0.
By combining the previous equations with Eq. \ref{eq:MF}, we estimate a multiplicity function:
\begin{equation}
    f(\sigma) = \frac{dn}{dlnM}\frac{M}{\rho_m}\left(\frac{dln\sigma^{-1}}{dlnM}\right)^{-1}.
    \label{eq:fsig}
\end{equation}

\subsection{Generalization}
We include the offset parameter and spin by generalizing the approach described in the previous section.
The number density of haloes as a function of mass and dynamical state is described by Equation \ref{eq:counts_h}:
\begin{equation}
    \frac{dn}{\text{dln}M\ \text{dlog}X_{\rm off}\ \text{dlog}\lambda} = \frac{\Delta N_{M,X_{\rm off},\lambda}}{V\ s_M\ s_{X_{\rm off}}\ s_\lambda}.
    \label{eq:counts_h}
\end{equation}
Here 
$\Delta N_{M,X_{\rm off},\lambda}$ is the number of haloes in each mass, $X_{\rm off}$ and $\lambda$ bin, $V$ is the total volume of the simulated cube and $s_M$, $s_{X_{\rm off}}$, and $s_\lambda$ are the natural (base 10) logarithm of mass ($X_{\rm off}, \lambda$) binning. 
Equivalently to equation \ref{eq:fsig}, we calculate
\begin{equation}
    h(\sigma,X_{\rm off},\lambda) = \frac{dn}{\text{dln}M\ \text{dlog}X_{\rm off}\ \text{dlog}\lambda}\frac{M}{\rho_m}\left(\frac{dln\sigma^{-1}}{dlnM}\right)^{-1}.  
    \label{eq:h_sigma_xoff_lambda}
\end{equation}
We consider a single integration of $h(\sigma,X_{\rm off},\lambda)$, which results in the set of Eqs. \ref{eq:integral_of_h}. 
The notation $g_X$ designates the marginalization of $h(\sigma,X_{\rm off},\lambda)$ over the variable $X$: 
\begin{align}
    g_\lambda(\sigma,X_{\rm off}) & = \int h(\sigma,X_{\rm off},\lambda)d\lambda , \nonumber \\
    g_{X_{\rm off}}(\sigma,\lambda) & = \int h(\sigma,X_{\rm off},\lambda)dX_{\rm off} , \nonumber \\
    g_\sigma(X_{\rm off},\lambda) & = \int h(\sigma,X_{\rm off},\lambda)d\sigma . \nonumber \\    
    \label{eq:integral_of_h}
\end{align}
Integrating again, we obtain
\begin{align}
    f_{X_{\rm off},\lambda}(\sigma) & = \int g_{\lambda}(\sigma,X_{\rm off})dX_{\rm off} = \int g_{X_{\rm off}}(\sigma,\lambda)d\lambda , \nonumber \\
    f_{\sigma,\lambda}(X_{\rm off}) & = \int g_{\lambda}(\sigma,X_{\rm off})d\sigma = \int g_{\sigma}(X_{\rm off},\lambda)d\lambda , \nonumber \\
    f_{\sigma,X_{\rm off}}(\lambda) & = \int g_{X_{\rm off}}(\sigma,\lambda)d\sigma = \int g_{\sigma}(X_{\rm off},\lambda)dX_{\rm off} . \nonumber \\    
    \label{eq:integral_of_g}
\end{align}
The functions $f, g, h$ are thus linked by derivatives as 
\begin{align}
    g(X,Y) & = \frac{\partial f(X)}{\partial Y} , \nonumber \\
    h(X,Y,Z) & = \frac{\partial^2 f(X)}{\partial Y \partial Z} = \frac{\partial g(X,Y)}{\partial Z} ,
    \label{eq:derivative_of_f}
\end{align}
where X, Y, Z are permutations of the variables $\sigma, X_{\rm off}, \lambda$. 
With this method we recover the multiplicity function f($\sigma$), which in this notation is f$_{X_{\rm off}, \lambda}$($\sigma$).
This allows us to study the behavior of the dark matter halo mass function according to different variables, making sure that in the end our analysis provides an accurate multiplicity function. 

\subsection{Mass--offset--spin function $h(\sigma, X_{\rm off}, \lambda)$}

Here we present a model for the generalized mass function $h(\sigma, X_{\rm off}, \lambda)$ introduced in the previous subsection. 
The relaxation state of a dark matter halo is related to the values of {\tt $X_{\rm off}$} and $\lambda$ parameters.
We consider \textsc{HMD}, \textsc{BigMD}, and \textsc{MDPL2} to build a 3D histogram of halo counts in bins of $\sigma, X_{\rm off},\lambda$, according to equations \ref{eq:counts_h} and \ref{eq:h_sigma_xoff_lambda}.

Mass functions are expressed as multiplicity functions $f(\sigma)$. It allows the inclusion of part of the redshift evolution in $\sigma$. 
We focus on the high-mass end of the mass function, using haloes with $M > 2.7\times 10^{13} M_\odot/h$.

We measure the halo number density in bins of $\log_{10}\sigma^{-1}$ instead of mass. 
We consider linear spaced bins from $-0.09$ to $0.6$, with 0.01 width, corresponding to values close to $2.8\times10^{13}$ and $1.3\times10^{16}$ $M_\odot/h$. We consider 50 bins spanning logarithmically from $10^{-3.8}$ to $10^{-0.2}$ for $X_{\rm off}$ and 50 bins spanning logarithmically from $10^{-4.5}$ to $10^{-0.1}$ for $\lambda$.
This is almost three orders of magnitude higher than the mass resolution in HMD ($7.9 \times 10^{10} M_\odot/h$). 
Therefore, our results 
will not be impacted by the mass resolution of the simulations. 
The total sample consists of 8,051,654 haloes for HMD, 2,103,896 for BigMD and 142,527 for MDPL. 
First, we estimate directly $f(\sigma)$ for each simulation, according to Equations \ref{eq:f_counts} and \ref{eq:fsig}. 
Then, we estimate $h$ by computing a 3D histogram in the same mass bins and different $X_{\rm off}$ and $\lambda$ bins. 

Uncertainties on histogram values are computed considering a Poisson number count term and a cosmic variance term:

\begin{align}
    &\delta h(\sigma,X_{\rm off},\lambda) = h(\sigma,X_{\rm off},\lambda)\sqrt{\frac{1}{N_{M,Xoff,\lambda}} + C^2} , \nonumber \\
    &\delta \log_{10} h(\sigma,X_{\rm off},\lambda) = \frac{1}{\ln{10}}\frac{\delta h(\sigma,X_{\rm off},\lambda)}{h(\sigma,X_{\rm off},\lambda)}.
    \label{eq:error}
\end{align}
Here $C$ is a term accounting for cosmic variance, which is set differently according to the type of simulation. Values for the cosmic variance are given in Table \ref{tab:cosmic_variance}. These values are estimated by \citet{Comparat2017}, using a jackknife method for variance at low masses. 
We fitted bins containing more than 50 haloes, which according to equation \ref{eq:error} means an uncertainty of around $15\%$; in this way our measurement is not dominated by  Poisson uncertainty.

\begin{table}
        \centering
        \caption{Cosmic variance in different MD simulations.}
        \label{tab:cosmic_variance}
        \begin{tabular}{cc cc} 
        \hline
        simulation & cosmic variance \\
    HMD &  0.02\\
    BigMD &  0.03\\
    MDPL & 0.04 \\
    \hline
        \end{tabular}
\end{table}






%
%
%
%
%

\section{Model}
\label{sec:model} 

We create a single model for the $h(\sigma, X_{\rm off}, \lambda)$ function. We describe in detail its features at z=0 and its redshift evolution.

\subsection{Redshift zero}
We consider that both $\lambda$ and $X_{\rm off}$ probability density functions are described by a modified Schechter function (Eqs. \ref{eq:modified_schechter} and \ref{eq:modified_schechter_xoff}), and combine these two functions with the multiplicity function along the mass axis. 
We obtain the following model, 
\begin{align}
    &h(\sigma, X_{\rm off}, \lambda, z, A,a,q,\mu,\alpha,\beta,\gamma,\delta,e) = ...\nonumber \\
    &A\sqrt{\frac{2}{\pi}}\Big(\sqrt{a}\frac{\delta_c}{\sigma}\Big)^q \exp\Big[-\frac{a}{2}\frac{\delta_c^2}{\sigma^2}\Big]\Big(\frac{X_{\rm off}}{\mu'} \Big)^{\alpha} ... \nonumber \\
    & \exp\Big[- \Big( \frac{X_{\rm off}}{\mu'}\Big)^{0.05\alpha}\Big] \Big(\frac{\lambda}{\mu}\Big)^{\gamma}\exp\Big[- \Big( \frac{X_{\rm off}}{\mu'\sigma^{e}}\Big)^{\beta} \Big(\frac{\lambda}{\mu} \Big)^{\delta}\Big], 
    \label{eq:model}
\end{align}
with $\mu' = 10^{1.83\log_{10}\mu}$, to disentangle the degeneracy between the two knees of the modified Schechter functions.


This model recalls the \citet{Bhattacharya2011} formulation along the mass axis 
and considers the combination of a power law and an exponential cutoff (i.e., a modified Schechter function) along the $X_{\rm off}$ and $\lambda$ axes.
The last exponential contains crossed terms between $X_{\rm off}$ and $\lambda$, which takes into account their correlation. Both $X_{\rm off}$ and $\lambda$ modified Schechter functions do not have mass dependency (as suggested by Equations \ref{eq:modified_schechter} and \ref{eq:modified_schechter_xoff}). 
In the \citet{Bhattacharya2011} formulation, there is a double power law; here we consider a single $\sigma$ power law. 
Additional $\sigma$ dependencies are described by crossed mass-dependent terms in the exponential cutoff, which relates $X_{\rm off}$ and $\lambda$ modified Schechter functions. 
The position of the knee of the $X_{\rm off}$ function (i.e., $\mu'$) is the same in its two exponential cutoffs, but in the second one we introduce the scaling with mass through the parameter $e$. 
We correlate directly the knees of the modified Schechter functions for $\lambda$ and $X_{\rm off}$ ($\mu$, $\mu'= 10^{1.83\log_{10}\mu}$), and the slopes of the power law and exponential cutoff of $X_{\rm off}$ ($\alpha$, $0.05\alpha$), to write the model in the most compact way possible.



\subsection{Evolution with redshift}
\label{subsection:z_evo_model}
Since structures accrete matter with time and grow, the mass function depends on redshift \citep{Springel2005b}. Part of this redshift evolution is in the mass--$\sigma$ relation through the matter power spectrum (Equation \ref{eq:sigma_definition}). However, this does not make the mass function completely universal at different times. \citet{Tinker2008} showed that a spherical overdensity mass function evolves up to $30\%$ from z=0 to z=2.5. \citet{Despali2016} highlighted how only virial overdensity nears the universality for the mass function. \citet{Crocce2010} considered a FOF mass function and found a $10\%$ evolution up to z=2. In this work we use \textsc{rockstar} to identify haloes, its base is a FOF algorithm as well. \\
Departures from universality are partially explained by the cosmology dependence of the mass function on the power spectrum and growth rate \citep{Ondaro-Mallea2021arXiv210208958O}. We find that the distribution functions of $X_{\rm off}$ and $\lambda$ show a redshift trend as well (see Sect. \ref{sec:relations}, and Equations \ref{eq:modified_schechter} and \ref{eq:modified_schechter_xoff}). We provide a detailed description of the evolution of $h(\sigma,X_{\rm off},\lambda)$ with redshift. Further investigation, using simulations in different cosmologies, is needed to assess a possible relation between parameters in Equation \ref{eq:zevolution} and cosmological parameters.\\
Given our model at z=0, we use the latter as the benchmark model to fit the halo mass$-X_{\rm off}-\lambda$ function at a higher redshift. For this goal we concatenate again samples from \textsc{HMD}, \textsc{BigMD}, and \textsc{MDPL2} at redshifts 0.045, 0.117, 0.221, 0.425, 0.523, 0.702, 0.779, 1.032, and 1.425. We note that \textsc{BigMD} is not tabulated at exactly the same  redshift snapshots as the other two simulations. 
Nonetheless, we use snapshots that are as close as possible, resulting in a $1.3\%$ difference for the worst-case scenario at z=1.425. 
Further details about the snapshots are available in Appendix \ref{appendix:tables}. 
For all these snapshots we consider the same $X_{\rm off}$ and $\lambda$ binning as we did for z=0. 
However, we shift  the $\sigma$ binning slightly  upward compared to the z=0 case, which allows us to reach masses of 
$7\times10^{12}\ M_\odot/h$ at z=0.702 and $10^{12}\ M_\odot/h$ at z=1.425.\\
We include a redshift evolution for all the parameters $A$, $a$, $q$, $\mu$, $\alpha$, $\beta$, $\gamma$, $\delta$, and $e$. 
We note that we did not consider an evolving critical density contrast with redshift, fixing it at z=0. 
So $\delta_c$ in Equation \ref{eq:model} is fixed at the value of 1.68647. 
Considering its evolution, even if tiny, introduces the need for additional evolution of the parameters, as pointed out by \citet{Bhattacharya2011}. 
We model the redshift evolution for these parameter using exponents $k_0$, $k_1$, $k_2$, $k_3$, $k_4$, $k_5$, $k_6$, $k_7$, and $k_8$ as follows in Equation \ref{eq:zevolution}:

\begin{align}
    \log_{10}A(z) &= \log_{10}A_0(1+z)^{k_0}, \nonumber \\ 
    a(z) &= a_0(1+z)^{k_1}, \nonumber \\  
    q(z) &= q_0(1+z)^{k_2}, \nonumber \\
    \log_{10}\mu(z) &= \log_{10}\mu_0(1+z)^{k_3}, \nonumber \\  
    \alpha(z) &= \alpha_0(1+z)^{k_4}, \nonumber \\  
    \beta(z) &= \beta_0(1+z)^{k_5}, \nonumber \\  
    \gamma(z) &= \gamma_0(1+z)^{k_6}, \nonumber \\
    \delta(z) &= \delta_0(1+z)^{k_7}, \nonumber \\
    e(z) &= e_{0}(1+z)^{k_8}. \nonumber \\  
\label{eq:zevolution}          
\end{align}

\section{Results}
\label{sec:results}
We present the result of the fits to the data (Sect. \ref{sec:N-body:data}) and the parameters of the model (Sect. \ref{sec:model}). We fit directly $\log_{10}h(\sigma,X_{\rm off},\lambda)$, which allows better modeling of the high-mass end.
We consider a Gaussian likelihood 

\begin{equation}
    \log \mathcal{L} = -0.5 \sum \Big(\frac{D-M}{E}\Big)^2,
    \label{eq:likelihood}
\end{equation}
where $D$ is $\log_{10}h(\sigma,X_{\rm off},\lambda)$ computed from Equation \ref{eq:counts_h}, $M$ is the base 10 logarithm of the model (Equation \ref{eq:model}), and $E$ is the uncertainty of $\log_{10}h(\sigma,X_{\rm off},\lambda)$ (see log error in Equation \ref{eq:error}). 

\subsection{Redshift zero}
The best-fit parameters at z=0 are obtained maximizing the likelihood in Eq. \ref{eq:likelihood}. 
We derive posterior probability distributions and the Bayesian evidence with the nested sampling Monte Carlo algorithm MLFriends \citep{Buchner2014, Buchner2019}, using the
UltraNest \footnote{\url{https://johannesbuchner.github.io/UltraNest/}} software. 
The results are shown in Fig. \ref{fig:corner_z_0}. We used flat priors. 
The description of the parameters, priors, and posteriors is summarized in Table \ref{tab:parameters}.
We obtain  $\log_{10}A = -22.004\pm 0.006$, $a =  0.885\pm 0.004$,  $q=2.284 \pm 0.016$, $\log_{10}\mu =  -3.326 \pm 0.001$, $\alpha =  5.623\pm 0.002$, $\beta = -0.391\pm 0.001$, $\gamma = 3.024\pm 0.003$, $\delta = 1.209\pm 0.001$, $e = -1.105\pm 0.005$. The parameter $a$ is in agreement with \citet{Bhattacharya2011}. Our $q$ parameter shows higher values, but this is expected because an additional mass trend is described by the knee of the exponential cutoff with mixed $X_{\rm off}, \lambda$ terms. The parameters $\alpha$ and $\gamma$ describe the power-law increment from small $X_{\rm off}$ and  $\lambda$ values, respectively. The second  is similar to the values computed for the spin distribution in Sect. \ref{sec:relations}, while the first  is bigger than almost a factor of two. This is expected because an additional offset trend is described by the negative $\beta$ parameter. Moreover, together with $\beta$, the parameter $e$  accounts for the relation between offset and spin, including mass dependency as well. The negative $e$ allows the shifting of the peak  along the $X_{\rm off}$ axis to higher values with mass, according to the findings in Sect. \ref{sec:relations}.

\begin{figure*}
    \centering
    \includegraphics[width=0.8\columnwidth]{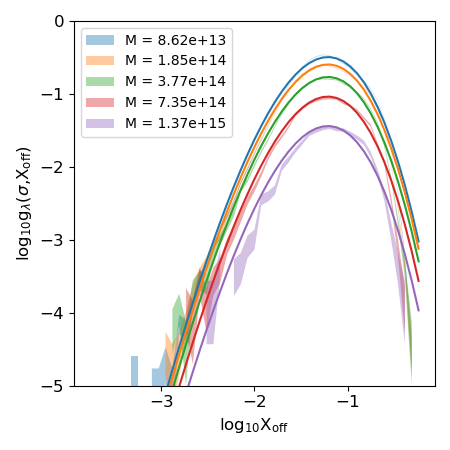}
    \includegraphics[width=0.8\columnwidth]{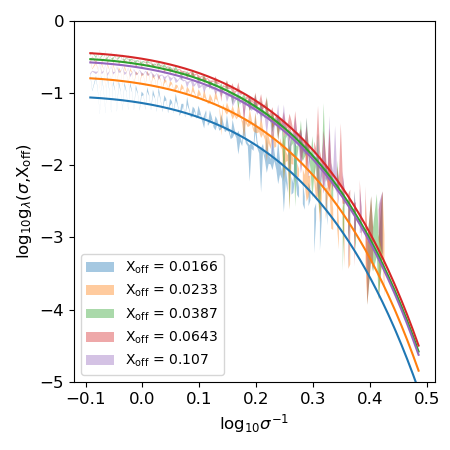}
    \includegraphics[width=0.8\columnwidth]{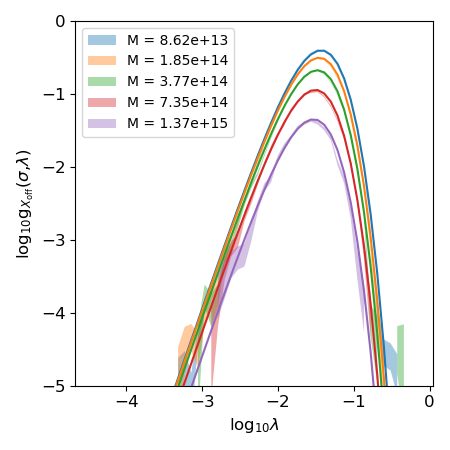}
    \includegraphics[width=0.8\columnwidth]{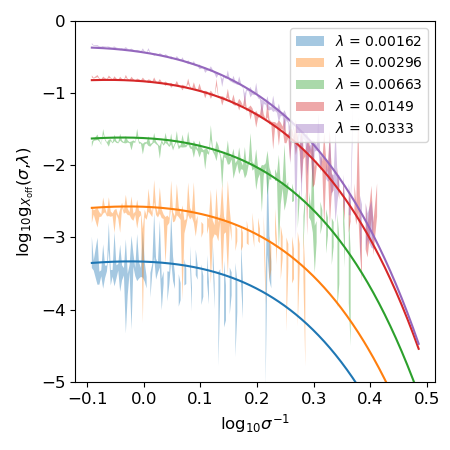}
    \includegraphics[width=0.8\columnwidth]{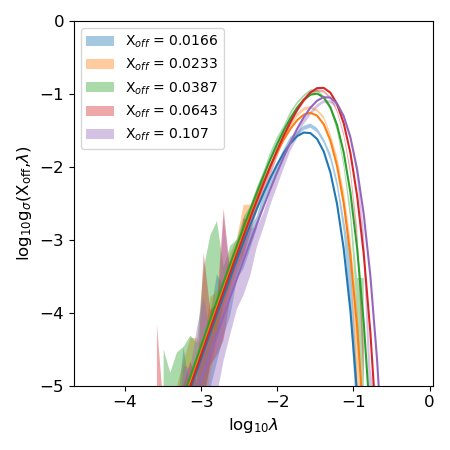}
    \includegraphics[width=0.8\columnwidth]{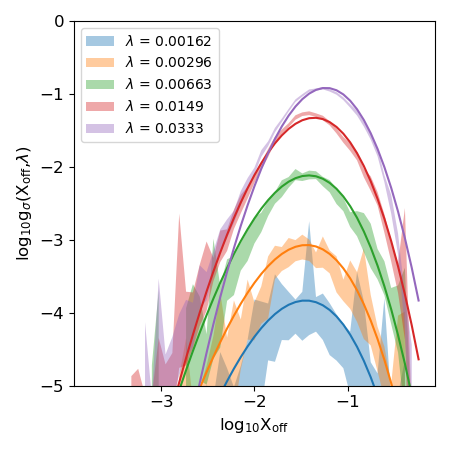}
        \caption{Single integration of the 3D model. In each panel straight lines indicate the best-fit model, while shaded areas represent the data with 1$\sigma$ uncertainties. Top left: $g_{\lambda}(\sigma,X_{\rm off})$ as a function of $X_{\rm off}$ in different mass slices. Top right: $g_{\lambda}(\sigma,X_{\rm off})$ as a function of $\sigma$ in different $X_{\rm off}$ slices. Middle left: $g_{X_{\rm off}}(\sigma,\lambda)$ as a function of $\lambda$ in different mass slices. Middle right: $g_{X_{\rm off}}(\sigma,\lambda)$ as a function of $\sigma$ in different $\lambda$ slices. Bottom left: $g_{\sigma}(X_{\rm off},\lambda)$ as a function of $\lambda$ in different $X_{\rm off}$ slices. Bottom right: $g_{\sigma}(X_{\rm off},\lambda)$ as a function of $X_{\rm off}$ in different $\lambda$ slices. The integrals are defined in Equations \ref{eq:integral_of_h}.}
        \label{fig:g_data_model}
\end{figure*}

Since this is a 3D model, we show in Fig. \ref{fig:g_data_model} all six combinations of h($\sigma$,$X_{\rm off}$,$\lambda$) integrated in 1D (Equation \ref{eq:integral_of_h}). To perform the integrals we exploit the Simpson's rule method for numerical integration \footnote{\url{https://docs.scipy.org/doc/}}. We show each 2D distribution in five different slices of a single quantity. 
We recover the typical exponential cutoff along the mass axis, as well as the modified Schechter shapes for offset and spin. Our model describes the  $\sigma$ and $\lambda$ evolution very well, $g_{X_{\rm off}}(\sigma,\lambda)$ also agrees with the data  in the tails of the distribution. We note the  small deviations in the distribution of $X_{\rm off}$ when $X_{\rm off}$ values approach the spatial resolution limit. For low-redshift samples, each version of MultiDark has its own resolution limit: 25 kpc/$h$ for \textsc{HMD}, 10 kpc/$h$ for \textsc{BigMD}, 5 kpc/$h$ for \textsc{MDPL2}. 
The mass trend of the low $X_{\rm off}$ slice is slightly underpredicted by the model at low mass. The same holds for the spin trend; there is a 0.1 dex, $3.1\sigma$ tension between the peak values of model and data. This is expected within the resolution limit. An improvement toward efficient computation and future generation of N-body simulations will be needed to probe the kiloparsec scales of Dark Matter haloes in simulation cubes of the gigaparsec scale.  
All the panels in Fig. \ref{fig:g_data_model} involving $\sigma$ show different uncertainties between succeeding mass bins as a result of the concatenation of bins from different MultiDark versions. \textsc{MDPL2} bins contain fewer haloes than \textsc{BigMD}, which contain fewer haloes than \textsc{HMD}. This translates into smaller uncertainties for bins containing a higher number of haloes. Overall, the model provides an excellent representation of the data.    

\begin{figure*}
    \centering
    \includegraphics[width=\columnwidth]{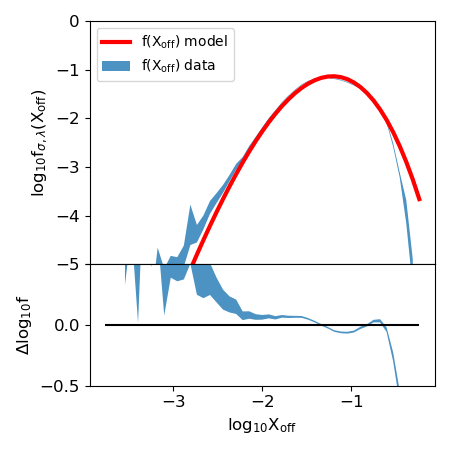}
    \includegraphics[width=\columnwidth]{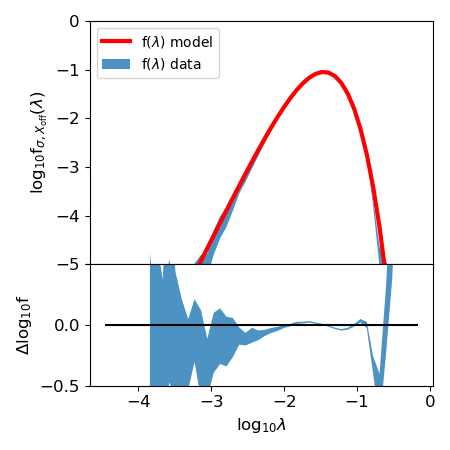}
        \caption{Comparison between data and model of $f_{\sigma,\lambda}(X_{\rm off})$ and $f_{\sigma,X_{\rm off}}(\lambda)$. 
        In the top panels the  straight red lines indicate the integral on the best-fit model, while the shaded blue areas represent the integral on the 3d $h(\sigma,X_{\rm off},\lambda)$ data with 1$\sigma$ uncertainties. Each bottom panel shows the residual trend with $\sigma$ error; the straight black line represents the perfect match between data and model with null residual. Top left: $f(X_{\rm off})$ as a function of $X_{\rm off}$. Bottom left: Residual between $f_{\sigma,\lambda}(X_{\rm off})$ data and model in logarithmic scale. Top right: $f(\lambda)$ as a function of $\lambda$. Bottom right: Residual between $f_{\sigma,X_{\rm off}}(\lambda)$ data, and model in logarithmic scale.}
        \label{fig:f_data_model}
\end{figure*}

We further integrate the model in Equation \ref{eq:model}, obtaining the distributions of $X_{\rm off}$ $f_{\sigma,\lambda}(X_{\rm off})$ and $\lambda$ $f_{\sigma,X_{\rm off}}(\lambda)$ (Equation \ref{eq:integral_of_g}). Figure \ref{fig:f_data_model} shows the result. The distributions around the peaks are well described by the model. This is important because these functions are dominated by objects described by the peak of the PDF. The spin distribution is better behaved than the $X_{\rm off}$ in the tails. This is expected due to spatial resolution limits. Moreover, this is a further confirmation of the fact that $X_{\rm off}, \lambda$ scatter around their mean values with modified Schechter distributions (see Sect. \ref{sec:relations}). 

We obtain the multiplicity function $f_{X_{\rm off},\lambda}(\sigma)$ marginalizing $h(\sigma,X_{\rm off},\lambda)$ on $X_{\rm off}, \lambda$ (i.e., performing the double integral): 

\begin{equation*}
 f_{X_{\rm off},\lambda}(\sigma) = \int \int h(\sigma,X_{\rm off},\lambda) dX_{\rm off}d\lambda.    
\end{equation*}

\begin{figure*}
    \centering
    \includegraphics[width=1.8\columnwidth]{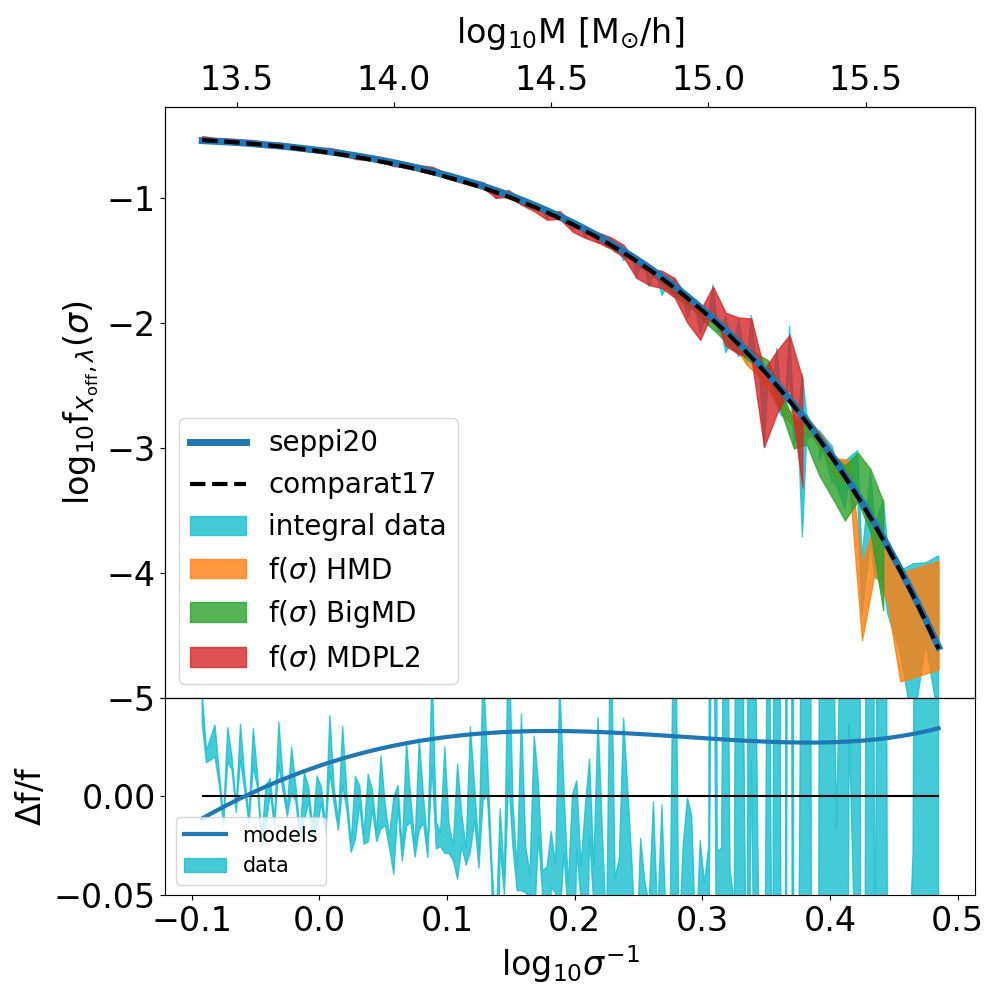}
        \caption{Comparison of multiplicity functions. Top panel: The three shaded regions show the 1$\sigma$ contours of $f(\sigma)$ data directly computed on different simulations (orange for \textsc{HMD}, green for \textsc{BigMD}, red for \textsc{MDPL2}). The light blue shaded region is the $1\sigma$ contour of the 2D integral computed on the concatenated sample containing all three simulations; the dashed pink line indicates the mass function from \citet{Comparat2017};   the blue solid line is the $f(\sigma)$  recovered by integrating our model along $X_{\rm off}$ and $\lambda$. Bottom panel: The blue thick line is the fractional difference between our $f(\sigma)$ and that of \citet{Comparat2017}. The light blue shaded area denotes the $1\sigma$ contours of the residual between the integrated data and our  best-fit model; the black horizontal line indicates the perfect match with null residual.}
        \label{fig:integral}
\end{figure*}

The result is shown in Fig. \ref{fig:integral}.
In the top panel, we show a comparison between our model, the data obtained from simulations, and the \citet{Comparat2017} model, fitted on these same simulations at z=0. The multiplicity functions computed directly on each simulation cube, without taking $X_{\rm off}$ and $\lambda$ into account, are shown by three shaded regions in different colors: red for \textsc{MDPL2}, green for \textsc{BigMD}, and orange for \textsc{HMD}. Bigger simulation boxes extend to higher mass values. 
The light blue shaded region represents the 2D integral computed on the concatenated sample of all three simulations. The solid blue line is the integral of our model and the dashed pink line is the \citet{Comparat2017} model.  
In the lower panel, we show the percentage difference between the multiplicity function $f(\sigma)$ we recover and the \citet{Comparat2017} model, obtained on the same MultiDark simulations. This difference is always under $3.3\%$ in the mass range of interest. It is also compatible with uncertainty on the data. Our model is able to recover the halo mass distribution in the simulations, with the advantage of taking into account parameters that describe the dynamical state as well. 
Once again, we note that our model is adapted to masses higher than $2.7\times 10^{13} M_\odot/h$ at z=0 ($10^{12} M_\odot/h$ at z=1.4). 

\subsection{Evolution with redshift}

To study the redshift evolution we start from the fiducial model at z=0. We add the redshift dependence (Equation \ref{eq:zevolution}) to the best-fit parameters in Equation \ref{eq:model} at z=0. We concatenate samples for ten redshift values, as described in \ref{subsection:z_evo_model}.

We obtain the values of the exponents in equation \ref{eq:zevolution} fitting the z trend of each parameter for all the concatenated snapshots simultaneously. 
We obtain $k_0 = -0.0441\pm 0.0001$, $k_1 = -0.161\pm 0.001$, $k_2 = 0.041\pm 0.002$, $k_3 = -0.1286\pm 0.0002$, $k_4 = 0.1081\pm 0.0002$, $k_5 =  -0.311 \pm 0.001$, $k_6 =  0.0902\pm 0.0004$, $k_7 = -0.0768\pm 0.0004$, and $k_8 = 0.612 \pm 0.002$. The full result is shown by the triangular plot in Fig. \ref{fig:corner_zevo}. Priors and posteriors for each parameter are given in Table \ref{tab:parameters_zevo}. 

\begin{figure*}
    \centering
    \includegraphics[width=2\columnwidth]{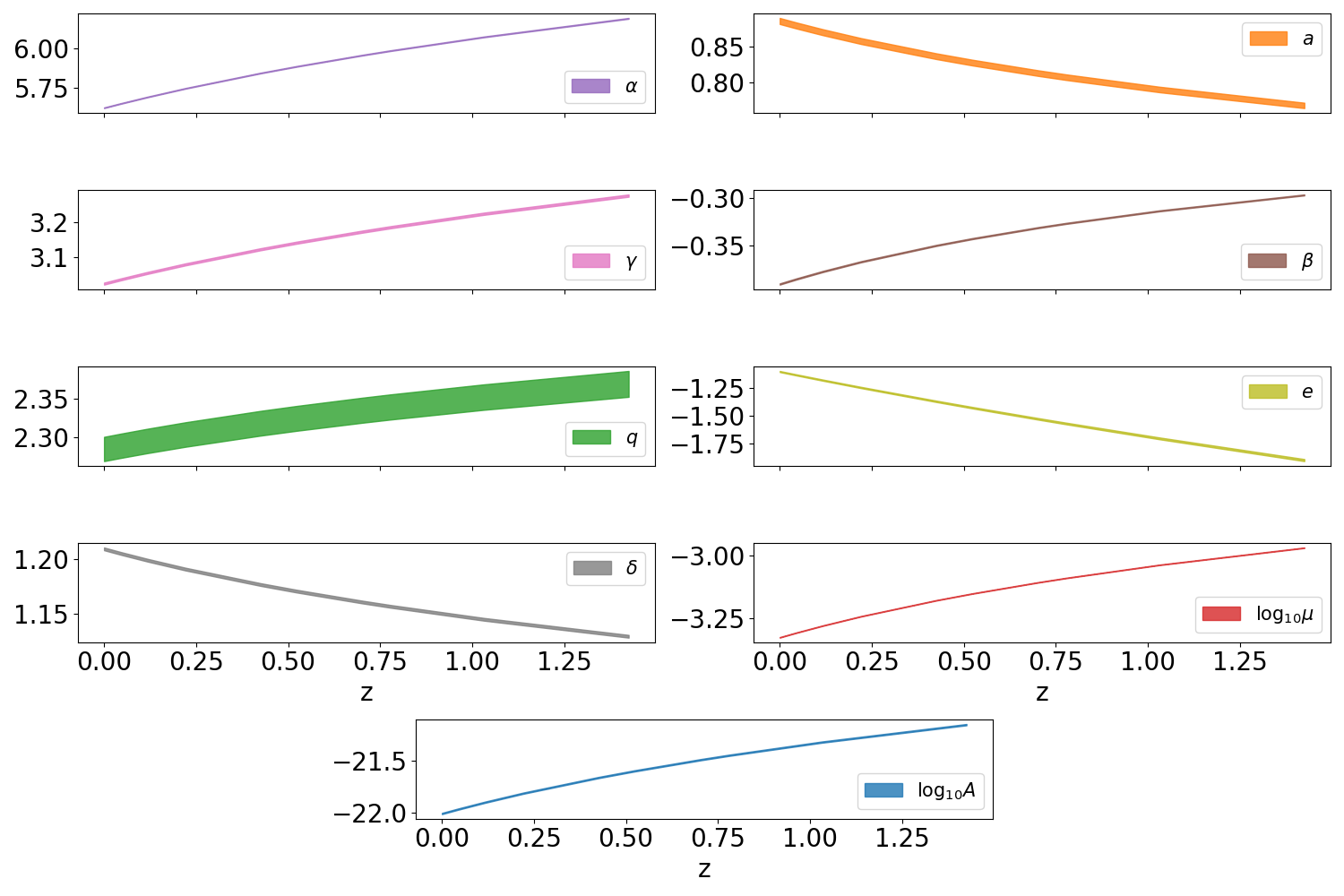}
        \caption{Redshift evolution of the best-fit parameters of our model. Each panel shows a single parameter. The values at z=0 are reported in Table \ref{tab:parameters}. The redshift evolution is described by Equation \ref{eq:zevolution}; the slopes are given in Table \ref{tab:parameters_zevo}. }
        \label{fig:zevo}
\end{figure*}

Redshift dependence is shown in Fig. \ref{fig:zevo}. The shaded areas include the uncertainty on  the best-fit parameter at z=0 and on the z evolution, according to equation \ref{eq:error_parameter}

\begin{equation}
    \delta P = \Big[\Big(\frac{\partial P}{\partial P_0} \delta P_0\Big)^2 + \Big(\frac{\partial P}{\partial k} \delta k\Big)^2\Big]^{1/2},
    \label{eq:error_parameter}
\end{equation}

where $P$ is each parameter in equation \ref{eq:model}, $P_0$ is its value at z=0, and $k$ indicates each parameter describing the evolution in equation \ref{eq:zevolution}.

The parameters A, q, $\mu$, $\alpha$, $\gamma$, and $\beta$ show an increasing redshift trend. On the other hand a, $\beta$, and e decrease with redshift. This means that with increasing redshift the modified Schechter functions need to increase more quickly and decrease more slowly. The knee describing the mass trend (parameter a) decreases with redshift, in agreement with \citet{Bhattacharya2011}. They find no redshift dependence for the slope of mass trend (parameter q). This is not true for this work, where q increases with z. However, this is mitigated by the mass trend of the position of the knee in the crossed $X_{\rm off},\lambda$ exponential cutoff, which decreases at early times. The fact that the position of the $X_{\rm off}$ knee (parameter $\mu$) moves to higher values at high z confirms the results of Sect. \ref{sec:relations}, with the higher average value of $X_{\rm off}$ early in time (Figures \ref{fig:xoff_sigma_relation} and \ref{fig:pdf_xoff}).

\section{Summary and conclusions}
\label{sec:discussion}

In the context of the hierarchical model of structure formation, the evolution of the number density of galaxy clusters is a powerful cosmological probe. 
In order to achieve precision cosmology with the next generation of galaxy clusters samples (such as the eROSITA All Sky Survey; \citealt{Merloni2012}), precise modeling of the theoretical mass function is necessary. We calibrated a model that includes a dynamical description of dark matter haloes. 
Using the formalism described here and the MultiDark simulations, we quantified the impact on the mass function of the lack of unrelaxed structures (see Fig. \ref{fig:relaxed-unrelaxed-MF}). 
We explored relations between quantities that describe different aspects of dark matter haloes, including their dynamical state. We investigated the concentration-mass relation. We confirmed the recent discovery of the concentration upturn at high masses, in agreement with previous results from \citet{Prada2012} and \citet{Klypin2016}, based on a similar set of MultiDark simulations. 
In addition, our model provides a prediction of concentration according to mass and redshift with one single equation (Equation \ref{eq:conc_peak}).  
The probability density function of concentration is a modified Schechter law, with mass dependency (Equation \ref{eq:modified_schechter_conc}). We find that the concentration of low-mass haloes has a faster redshift evolution than high-mass objects, especially in the high-concentration regime. For concentration $c=8$, the PDF for high-mass haloes shifts by 0.32 dex from $z=1.43$ to $z=0$, while the low-mass value changes by 0.49 dex.
We find the spin parameter $\lambda$ to be modeled by a linear relation with mass and a probability density function well described by a modified Schechter function (Equation \ref{eq:modified_schechter}), in agreement with \citet{Rodriguez-Puebla2016}. 
The offset parameter evolves with mass and redshift according to Equation \ref{eq:xoff_sigma_rel}. The negative slope of the relation suggests that low-mass haloes are typically more relaxed compared to high-mass objects. This is true at every redshift. The offset distribution around the mean value is well described by a modified Schechter function (Equation \ref{eq:modified_schechter_xoff}). The peak of the distribution shifts by a factor of 1.5 between $z\sim1.4$ and $z=0$. This is in agreement with haloes relaxing with time and the recent formation of cool cores in galaxy clusters.\\
We define a general mass function framework, where dark matter haloes are not only described as a function of mass but also by the two additional variables $X_{\rm off}, \lambda$. 
This approach   considers the mass, offset parameter, and spin of each halo at the same time in a $\sigma-X_{\rm off}-\lambda$ function (Equation \ref{eq:h_sigma_xoff_lambda}). We model it in Sect. \ref{sec:model} combining terms of a fiducial mass function \citep{Bhattacharya2011} with modified Schechter functions for $X_{\rm off}, \lambda$, as obtained in Sect. \ref{sec:relations}. This new approach  accounts for the dynamical state of dark matter haloes directly in the context of the halo mass function, providing 2D and 1D distributions at the same time.\\
We describe the fitting procedure and results in Sect. \ref{sec:results}. Our result at z=0 recovers the \citet{Comparat2017} mass function, which is fitted on the same set of simulations, with $3.3\%$ accuracy. This means that our model is able to account for the dynamical state of dark matter haloes simultaneously with mass and to describe the multiplicity function with great precision. 
In addition, our model includes the redshift evolution, according to Equation \ref{eq:zevolution}. The result is shown in Fig. \ref{fig:zevo}. \\
The link with observations will be  explored further in future work.

\section*{Acknowledgements}

We thank the anonymous referee for the constructive feedback.

JC thanks Dominique Eckert for insightful discussions about this project.

The CosmoSim database used in this paper is a service by the Leibniz-Institute for Astrophysics Potsdam (AIP).
The MultiDark database was developed in cooperation with the Spanish MultiDark Consolider Project CSD2009-00064.

The authors gratefully acknowledge the Gauss Centre for Supercomputing e.V. (www.gauss-centre.eu) and the Partnership for Advanced Supercomputing in Europe (PRACE, www.prace-ri.eu) for funding the MultiDark simulation project by providing computing time on the GCS Supercomputer SuperMUC at Leibniz Supercomputing Centre (LRZ, \href{lrz.de}{www.lrz.de}).

 We derive posterior probability distributions and the Bayesian evidence with the nested sampling Monte Carlo algorithm MLFriends (\citet{Buchner2014}, \citet{Buchner2019}) using the UltraNest software.
 
To fit relations and distributions in Sect. \ref{sec:relations}, we use a $\chi^2$ minimization algorithm with \textsc{curve\_fit} python package \citep{2020SciPy-NMeth} (\url{https://scipy.org/}). 

In order to perform cosmology calculation, we used the python toolkit \textsc{Colossus} \citep{Diemer2018} (\url{https://bdiemer.bitbucket.io/colossus/}).



\bibliographystyle{aa}
\bibliography{biblio} 

\begin{thebibliography}{114}
\expandafter\ifx\csname natexlab\endcsname\relax\def\natexlab#1{#1}\fi

\bibitem[{{Abazajian} {et~al.}(2019){Abazajian}, {Addison}, {Adshead}, {Ahmed},
  {Allen}, {Alonso}, {Alvarez}, {Anderson}, {Arnold}, {Baccigalupi}, {Bailey},
  {Barkats}, {Barron}, {Barry}, {Bartlett}, {Basu Thakur}, {Battaglia},
  {Baxter}, {Bean}, {Bebek}, {Bender}, {Benson}, {Berger}, {Bhimani},
  {Bischoff}, {Bleem}, {Bocquet}, {Boddy}, {Bonato}, {Bond}, {Borrill},
  {Bouchet}, {Brown}, {Bryan}, {Burkhart}, {Buza}, {Byrum}, {Calabrese},
  {Calafut}, {Caldwell}, {Carlstrom}, {Carron}, {Cecil}, {Challinor}, {Chang},
  {Chinone}, {Cho}, {Cooray}, {Crawford}, {Crites}, {Cukierman}, {Cyr-Racine},
  {de Haan}, {de Zotti}, {Delabrouille}, {Demarteau}, {Devlin}, {Di Valentino},
  {Dobbs}, {Duff}, {Duivenvoorden}, {Dvorkin}, {Edwards}, {Eimer}, {Errard},
  {Essinger-Hileman}, {Fabbian}, {Feng}, {Ferraro}, {Filippini}, {Flauger},
  {Flaugher}, {Fraisse}, {Frolov}, {Galitzki}, {Galli}, {Ganga}, {Gerbino},
  {Gilchriese}, {Gluscevic}, {Green}, {Grin}, {Grohs}, {Gualtieri}, {Guarino},
  {Gudmundsson}, {Habib}, {Haller}, {Halpern}, {Halverson}, {Hanany},
  {Harrington}, {Hasegawa}, {Hasselfield}, {Hazumi}, {Heitmann}, {Henderson},
  {Henning}, {Hill}, {Hlozek}, {Holder}, {Holzapfel}, {Hubmayr},
  {Huffenberger}, {Huffer}, {Hui}, {Irwin}, {Johnson}, {Johnstone}, {Jones},
  {Karkare}, {Katayama}, {Kerby}, {Kernovsky}, {Keskitalo}, {Kisner}, {Knox},
  {Kosowsky}, {Kovac}, {Kovetz}, {Kuhlmann}, {Kuo}, {Kurita}, {Kusaka},
  {Lahteenmaki}, {Lawrence}, {Lee}, {Lewis}, {Li}, {Linder}, {Loverde},
  {Lowitz}, {Madhavacheril}, {Mantz}, {Matsuda}, {Mauskopf}, {McMahon},
  {McQuinn}, {Meerburg}, {Melin}, {Meyers}, {Millea}, {Mohr}, {Moncelsi},
  {Mroczkowski}, {Mukherjee}, {M{\"u}nchmeyer}, {Nagai}, {Nagy}, {Namikawa},
  {Nati}, {Natoli}, {Negrello}, {Newburgh}, {Niemack}, {Nishino}, {Nordby},
  {Novosad}, {O'Connor}, {Obied}, {Padin}, {Pandey}, {Partridge}, {Pierpaoli},
  {Pogosian}, {Pryke}, {Puglisi}, {Racine}, {Raghunathan}, {Rahlin},
  {Rajagopalan}, {Raveri}, {Reichanadter}, {Reichardt}, {Remazeilles}, {Rocha},
  {Roe}, {Roy}, {Ruhl}, {Salatino}, {Saliwanchik}, {Schaan}, {Schillaci},
  {Schmittfull}, {Scott}, {Sehgal}, {Shandera}, {Sheehy}, {Sherwin},
  {Shirokoff}, {Simon}, {Slosar}, {Somerville}, {Spergel}, {Staggs}, {Stark},
  {Stompor}, {Story}, {Stoughton}, {Suzuki}, {Tajima}, {Teply}, {Thompson},
  {Timbie}, {Tomasi}, {Treu}, {Tristram}, {Tucker}, {Umilt{\`a}}, {van
  Engelen}, {Vieira}, {Vieregg}, {Vogelsberger}, {Wang}, {Watson}, {White},
  {Whitehorn}, {Wollack}, {Kimmy Wu}, {Xu}, {Yasini}, {Yeck}, {Yoon}, {Young},
  \& {Zonca}}]{Abazajian2019CMBS4}
{Abazajian}, K., {Addison}, G., {Adshead}, P., {et~al.} 2019, arXiv e-prints,
  arXiv:1907.04473

\bibitem[{Achitouv {et~al.}(2014)Achitouv, Wagner, Weller, \&
  Rasera}]{Achitouv_2014}
Achitouv, I., Wagner, C., Weller, J., \& Rasera, Y. 2014, Journal of Cosmology
  and Astroparticle Physics, 2014, 077–077

\bibitem[{Achitouv \& Corasaniti(2012)}]{Achitouv_2012}
Achitouv, I.~E. \& Corasaniti, P.~S. 2012, Journal of Cosmology and
  Astroparticle Physics, 2012, 002–002

\bibitem[{{Allen} {et~al.}(2011){Allen}, {Evrard}, \& {Mantz}}]{Allen2011}
{Allen}, S.~W., {Evrard}, A.~E., \& {Mantz}, A.~B. 2011, \araa, 49, 409

\bibitem[{{Angulo} {et~al.}(2008){Angulo}, {Baugh}, \&
  {Lacey}}]{Angulo2008MNRAS.387..921A}
{Angulo}, R.~E., {Baugh}, C.~M., \& {Lacey}, C.~G. 2008, \mnras, 387, 921

\bibitem[{{Angulo} {et~al.}(2012){Angulo}, {Springel}, {White}, {Jenkins},
  {Baugh}, \& {Frenk}}]{Angulo2012}
{Angulo}, R.~E., {Springel}, V., {White}, S.~D.~M., {et~al.} 2012, \mnras, 426,
  2046

\bibitem[{{Baldi} {et~al.}(2018){Baldi}, {De Petris}, {Sembolini}, {Yepes},
  {Cui}, \& {Lamagna}}]{Baldi2018spin}
{Baldi}, A.~S., {De Petris}, M., {Sembolini}, F., {et~al.} 2018, \mnras, 479,
  4028

\bibitem[{{Baldi} {et~al.}(2019){Baldi}, {De Petris}, {Sembolini}, {Yepes},
  {Cui}, \& {Lamagna}}]{Baldi2019JPhCS1226a2003Bspin}
{Baldi}, A.~S., {De Petris}, M., {Sembolini}, F., {et~al.} 2019, in Journal of
  Physics Conference Series, Vol. 1226, Journal of Physics Conference Series,
  012003

\bibitem[{Bartelmann(2010)}]{Bartelmann_2010}
Bartelmann, M. 2010, Classical and Quantum Gravity, 27, 233001

\bibitem[{{Baxter} {et~al.}(2019){Baxter}, {Sherwin}, \&
  {Raghunathan}}]{Baxter2019JCAP...06..001Bspin_data}
{Baxter}, E.~J., {Sherwin}, B.~D., \& {Raghunathan}, S. 2019, \jcap, 2019, 001

\bibitem[{{Behroozi} {et~al.}(2015){Behroozi}, {Knebe}, {Pearce}, {Elahi},
  {Han}, {Lux}, {Mao}, {Muldrew}, {Potter}, \&
  {Srisawat}}]{Behroozi2015MNRAS.454.3020B}
{Behroozi}, P., {Knebe}, A., {Pearce}, F.~R., {et~al.} 2015, \mnras, 454, 3020

\bibitem[{{Behroozi} {et~al.}(2013){Behroozi}, {Wechsler}, \&
  {Wu}}]{Behroozi2013}
{Behroozi}, P., {Wechsler}, R., \& {Wu}, H.-Y. 2013, \apj, 762, 109

\bibitem[{{Benson} {et~al.}(2014){Benson}, {Ade}, {Ahmed}, {Allen}, {Arnold},
  {Austermann}, {Bender}, {Bleem}, {Carlstrom}, {Chang}, {Cho}, {Cliche},
  {Crawford}, {Cukierman}, {de Haan}, {Dobbs}, {Dutcher}, {Everett}, {Gilbert},
  {Halverson}, {Hanson}, {Harrington}, {Hattori}, {Henning}, {Hilton},
  {Holder}, {Holzapfel}, {Irwin}, {Keisler}, {Knox}, {Kubik}, {Kuo}, {Lee},
  {Leitch}, {Li}, {McDonald}, {Meyer}, {Montgomery}, {Myers}, {Natoli},
  {Nguyen}, {Novosad}, {Padin}, {Pan}, {Pearson}, {Reichardt}, {Ruhl},
  {Saliwanchik}, {Simard}, {Smecher}, {Sayre}, {Shirokoff}, {Stark}, {Story},
  {Suzuki}, {Thompson}, {Tucker}, {Vanderlinde}, {Vieira}, {Vikhlinin}, {Wang},
  {Yefremenko}, \& {Yoon}}]{Benson2014SPIE.9153E..1PB}
{Benson}, B.~A., {Ade}, P.~A.~R., {Ahmed}, Z., {et~al.} 2014, Society of
  Photo-Optical Instrumentation Engineers (SPIE) Conference Series, Vol. 9153,
  {SPT-3G: a next-generation cosmic microwave background polarization
  experiment on the South Pole telescope}, 91531P

\bibitem[{Bett {et~al.}(2007)Bett, Eke, Frenk, Jenkins, Helly, \&
  Navarro}]{Bett_2007_spin}
Bett, P., Eke, V., Frenk, C.~S., {et~al.} 2007, Monthly Notices of the Royal
  Astronomical Society, 376, 215

\bibitem[{{Bhattacharya} {et~al.}(2011){Bhattacharya}, {Heitmann}, {White},
  {Luki{\'c}}, {Wagner}, \& {Habib}}]{Bhattacharya2011}
{Bhattacharya}, S., {Heitmann}, K., {White}, M., {et~al.} 2011, \apj, 732, 122

\bibitem[{{Bianconi} {et~al.}(2013){Bianconi}, {Ettori}, \&
  {Nipoti}}]{Bianconi2013spin_xray}
{Bianconi}, M., {Ettori}, S., \& {Nipoti}, C. 2013, \mnras, 434, 1565

\bibitem[{{Bilton} {et~al.}(2019){Bilton}, {Hunt}, {Pimbblet}, \&
  {Roediger}}]{Bilton2019MNRAS.490.5017Bspin}
{Bilton}, L.~E., {Hunt}, M., {Pimbblet}, K.~A., \& {Roediger}, E. 2019, \mnras,
  490, 5017

\bibitem[{{Bocquet} {et~al.}(2020){Bocquet}, {Heitmann}, {Habib}, {Lawrence},
  {Uram}, {Frontiere}, {Pope}, \& {Finkel}}]{Bocquet2020arXiv200312116B}
{Bocquet}, S., {Heitmann}, K., {Habib}, S., {et~al.} 2020, arXiv e-prints,
  arXiv:2003.12116

\bibitem[{{Bocquet} {et~al.}(2016){Bocquet}, {Saro}, {Dolag}, \&
  {Mohr}}]{Bocquet2016}
{Bocquet}, S., {Saro}, A., {Dolag}, K., \& {Mohr}, J.~J. 2016, \mnras, 456,
  2361

\bibitem[{{Bond} {et~al.}(1991){Bond}, {Cole}, {Efstathiou}, \&
  {Kaiser}}]{Bond1991}
{Bond}, J.~R., {Cole}, S., {Efstathiou}, G., \& {Kaiser}, N. 1991, \apj, 379,
  440

\bibitem[{{Bryan} \& {Norman}(1998)}]{BryanNorman1998}
{Bryan}, G.~L. \& {Norman}, M.~L. 1998, \apj, 495, 80

\bibitem[{{Buchner}(2014)}]{Buchner2014}
{Buchner}, J. 2014, arXiv e-prints, arXiv:1407.5459

\bibitem[{{Buchner}(2019)}]{Buchner2019}
{Buchner}, J. 2019, \pasp, 131, 108005

\bibitem[{{Cibirka} {et~al.}(2017){Cibirka}, {Cypriano}, {Brimioulle}, {Gruen},
  {Erben}, {van Waerbeke}, {Miller}, {Finoguenov}, {Kirkpatrick}, {Henry},
  {Rykoff}, {Rozo}, {Dupke}, {Kneib}, {Shan}, \&
  {Spinelli}}]{Cibirka2017conc_codex}
{Cibirka}, N., {Cypriano}, E.~S., {Brimioulle}, F., {et~al.} 2017, \mnras, 468,
  1092

\bibitem[{{Comparat} {et~al.}(2017){Comparat}, {Prada}, {Yepes}, \&
  {Klypin}}]{Comparat2017}
{Comparat}, J., {Prada}, F., {Yepes}, G., \& {Klypin}, A. 2017, \mnras, 469,
  4157

\bibitem[{{Contreras} {et~al.}(2019){Contreras}, {Zehavi}, {Padilla}, {Baugh},
  {Jim{\'e}nez}, \& {Lacerna}}]{Contreras2019}
{Contreras}, S., {Zehavi}, I., {Padilla}, N., {et~al.} 2019, \mnras, 484, 1133

\bibitem[{Cooray \& Chen(2002)}]{Cooray_2002_sz_spin}
Cooray, A. \& Chen, X. 2002, The Astrophysical Journal, 573, 43

\bibitem[{{Correa} {et~al.}(2015){Correa}, {Wyithe}, {Schaye}, \&
  {Duffy}}]{Correa2015Mconc_model}
{Correa}, C.~A., {Wyithe}, J. S.~B., {Schaye}, J., \& {Duffy}, A.~R. 2015,
  \mnras, 452, 1217

\bibitem[{{Crocce} {et~al.}(2010){Crocce}, {Fosalba}, {Castand er}, \&
  {Gazta{\~n}aga}}]{Crocce2010}
{Crocce}, M., {Fosalba}, P., {Castand er}, F.~J., \& {Gazta{\~n}aga}, E. 2010,
  \mnras, 403, 1353

\bibitem[{{Croton} {et~al.}(2007){Croton}, {Gao}, \& {White}}]{Croton2007}
{Croton}, D.~J., {Gao}, L., \& {White}, S. D.~M. 2007, \mnras, 374, 1303

\bibitem[{{Dawson} {et~al.}(2016){Dawson}, {Kneib}, {Percival}, {Alam},
  {Albareti}, {Anderson}, {Armengaud}, {Aubourg}, {Bailey}, {Bautista},
  {Berlind}, {Bershady}, {Beutler}, {Bizyaev}, {Blanton}, {Blomqvist},
  {Bolton}, {Bovy}, {Brandt}, {Brinkmann}, {Brownstein}, {Burtin}, {Busca},
  {Cai}, {Chuang}, {Clerc}, {Comparat}, {Cope}, {Croft}, {Cruz-Gonzalez}, {da
  Costa}, {Cousinou}, {Darling}, {de la Macorra}, {de la Torre}, {Delubac}, {du
  Mas des Bourboux}, {Dwelly}, {Ealet}, {Eisenstein}, {Eracleous}, {Escoffier},
  {Fan}, {Finoguenov}, {Font-Ribera}, {Frinchaboy}, {Gaulme}, {Georgakakis},
  {Green}, {Guo}, {Guy}, {Ho}, {Holder}, {Huehnerhoff}, {Hutchinson}, {Jing},
  {Jullo}, {Kamble}, {Kinemuchi}, {Kirkby}, {Kitaura}, {Klaene}, {Laher},
  {Lang}, {Laurent}, {Le Goff}, {Li}, {Liang}, {Lima}, {Lin}, {Lin}, {Lin},
  {Long}, {Lundgren}, {MacDonald}, {Geimba Maia}, {Malanushenko},
  {Malanushenko}, {Mariappan}, {McBride}, {McGreer}, {M{\'e}nard}, {Merloni},
  {Meza}, {Montero-Dorta}, {Muna}, {Myers}, {Nandra}, {Naugle}, {Newman},
  {Noterdaeme}, {Nugent}, {Ogando}, {Olmstead}, {Oravetz}, {Oravetz},
  {Padmanabhan}, {Palanque-Delabrouille}, {Pan}, {Parejko}, {P{\^a}ris},
  {Peacock}, {Petitjean}, {Pieri}, {Pisani}, {Prada}, {Prakash}, {Raichoor},
  {Reid}, {Rich}, {Ridl}, {Rodriguez-Torres}, {Carnero Rosell}, {Ross},
  {Rossi}, {Ruan}, {Salvato}, {Sayres}, {Schneider}, {Schlegel}, {Seljak},
  {Seo}, {Sesar}, {Shandera}, {Shu}, {Slosar}, {Sobreira}, {Streblyanska},
  {Suzuki}, {Taylor}, {Tao}, {Tinker}, {Tojeiro}, {Vargas-Maga{\~n}a}, {Wang},
  {Weaver}, {Weinberg}, {White}, {Wood-Vasey}, {Yeche}, {Zhai}, {Zhao}, {Zhao},
  {Zheng}, {Ben Zhu}, \& {Zou}}]{Dawson2016}
{Dawson}, K.~S., {Kneib}, J.-P., {Percival}, W.~J., {et~al.} 2016, \aj, 151, 44

\bibitem[{{de Jong}(2011)}]{deJong2011_4MOST}
{de Jong}, R. 2011, The Messenger, 145, 14

\bibitem[{{Del Popolo} {et~al.}(2017){Del Popolo}, {Pace}, \& {Le
  Delliou}}]{Popolo2017}
{Del Popolo}, A., {Pace}, F., \& {Le Delliou}, M. 2017, \jcap, 3, 032

\bibitem[{{DESI Collaboration} {et~al.}(2016){DESI Collaboration}, {Aghamousa},
  {Aguilar}, {Ahlen}, {Alam}, {Allen}, {Allende Prieto}, {Annis}, {Bailey},
  {Balland}, \& et~al.}]{DESI2016}
{DESI Collaboration}, {Aghamousa}, A., {Aguilar}, J., {et~al.} 2016, ArXiv
  e-prints [\eprint[arXiv]{1611.00036}]

\bibitem[{{Despali} {et~al.}(2016){Despali}, {Giocoli}, {Angulo}, {Tormen},
  {Sheth}, {Baso}, \& {Moscardini}}]{Despali2016}
{Despali}, G., {Giocoli}, C., {Angulo}, R.~E., {et~al.} 2016, \mnras, 456, 2486

\bibitem[{{Diemer}(2018)}]{Diemer2018}
{Diemer}, B. 2018, \apjs, 239, 35

\bibitem[{{Diemer} \& {Joyce}(2019)}]{Diemer_Joyce2019_conc}
{Diemer}, B. \& {Joyce}, M. 2019, \apj, 871, 168

\bibitem[{{Diemer} \& {Kravtsov}(2015)}]{Diemer2015conc_universal}
{Diemer}, B. \& {Kravtsov}, A.~V. 2015, \apj, 799, 108

\bibitem[{{Du} {et~al.}(2015){Du}, {Fan}, {Shan}, {Zhao}, {Covone}, {Fu}, \&
  {Kneib}}]{Du2015conc_mass_obs}
{Du}, W., {Fan}, Z., {Shan}, H., {et~al.} 2015, \apj, 814, 120

\bibitem[{{Duffy} {et~al.}(2008){Duffy}, {Schaye}, {Kay}, \& {Dalla
  Vecchia}}]{Duffy_2008}
{Duffy}, A.~R., {Schaye}, J., {Kay}, S.~T., \& {Dalla Vecchia}, C. 2008,
  \mnras, 390, L64

\bibitem[{{Dutton} \& {Macci{\`o}}(2014)}]{Dutton_Maccio_2014}
{Dutton}, A.~A. \& {Macci{\`o}}, A.~V. 2014, \mnras, 441, 3359

\bibitem[{{Eckert} {et~al.}(2019){Eckert}, {Ghirardini}, {Ettori}, {Rasia},
  {Biffi}, {Pointecouteau}, {Rossetti}, {Molendi}, {Vazza}, {Gastaldello},
  {Gaspari}, {De Grandi}, {Ghizzardi}, {Bourdin}, {Tchernin}, \&
  {Roncarelli}}]{Eckert2019A&A...621A..40E}
{Eckert}, D., {Ghirardini}, V., {Ettori}, S., {et~al.} 2019, \aap, 621, A40

\bibitem[{{Eckert} {et~al.}(2011){Eckert}, {Molendi}, \&
  {Paltani}}]{Eckert2011}
{Eckert}, D., {Molendi}, S., \& {Paltani}, S. 2011, \aap, 526, A79

\bibitem[{{Ettori} \& {Brighenti}(2008)}]{Ettori2008cool_core_evo}
{Ettori}, S. \& {Brighenti}, F. 2008, \mnras, 387, 631

\bibitem[{{Finoguenov} {et~al.}(2019{\natexlab{a}}){Finoguenov}, {Merloni},
  {Comparat}, {Nandra}, {Salvato}, {Tempel}, {Raichoor}, {Richard}, {Kneib},
  {Pillepich}, {Sahl{\'e}n}, {Popesso}, {Norberg}, {McMahon}, \& {4MOST
  Collaboration}}]{Finoguenov2019Msngr}
{Finoguenov}, A., {Merloni}, A., {Comparat}, J., {et~al.} 2019{\natexlab{a}},
  The Messenger, 175, 39

\bibitem[{{Finoguenov} {et~al.}(2019{\natexlab{b}}){Finoguenov}, {Rykoff},
  {Clerc}, {Costanzi}, {Hagstotz}, {Ider Chitham}, {Kiiveri}, {Kirkpatrick},
  {Capasso}, {Comparat}, {Damsted}, {Dupke}, {Erfanianfar}, {Henry}, {Kaefer},
  {Kneib}, {Lindholm}, {Rozo}, {van Waerbeke}, \& {Weller}}]{Finoguenov2020}
{Finoguenov}, A., {Rykoff}, E., {Clerc}, N., {et~al.} 2019{\natexlab{b}}, arXiv
  e-prints, arXiv:1912.03262

\bibitem[{{Fo{\"e}x} {et~al.}(2014){Fo{\"e}x}, {Motta}, {Jullo}, {Limousin}, \&
  {Verdugo}}]{Foex2014A&A...572A..19Fconc}
{Fo{\"e}x}, G., {Motta}, V., {Jullo}, E., {Limousin}, M., \& {Verdugo}, T.
  2014, \aap, 572, A19

\bibitem[{{Gao} {et~al.}(2005){Gao}, {Springel}, \& {White}}]{Gao2005}
{Gao}, L., {Springel}, V., \& {White}, S. D.~M. 2005, \mnras, 363, L66

\bibitem[{{Giocoli} {et~al.}(2012){Giocoli}, {Tormen}, \&
  {Sheth}}]{Giocoli2012concentration}
{Giocoli}, C., {Tormen}, G., \& {Sheth}, R.~K. 2012, \mnras, 422, 185

\bibitem[{{Heitmann} {et~al.}(2015){Heitmann}, {Frontiere}, {Sewell}, {Habib},
  {Pope}, {Finkel}, {Rizzi}, {Insley}, \& {Bhattacharya}}]{Heitmann2015}
{Heitmann}, K., {Frontiere}, N., {Sewell}, C., {et~al.} 2015, \apjs, 219, 34

\bibitem[{{Henson} {et~al.}(2017){Henson}, {Barnes}, {Kay}, {McCarthy}, \&
  {Schaye}}]{Henson2017MNRAS.465.3361H}
{Henson}, M.~A., {Barnes}, D.~J., {Kay}, S.~T., {McCarthy}, I.~G., \& {Schaye},
  J. 2017, \mnras, 465, 3361

\bibitem[{{Hitomi Collaboration} {et~al.}(2018){Hitomi Collaboration},
  {Aharonian}, {Akamatsu}, {Akimoto}, {Allen}, {Angelini}, {Audard}, {Awaki},
  {Axelsson}, {Bamba}, {Bautz}, {Blandford}, {Brenneman}, {Brown}, {Bulbul},
  {Cackett}, {Canning}, {Chernyakova}, {Chiao}, {Coppi}, {Costantini}, {de
  Plaa}, {de Vries}, {den Herder}, {Done}, {Dotani}, {Ebisawa}, {Eckart},
  {Enoto}, {Ezoe}, {Fabian}, {Ferrigno}, {Foster}, {Fujimoto}, {Fukazawa},
  {Furuzawa}, {Galeazzi}, {Gallo}, {Gandhi}, {Giustini}, {Goldwurm}, {Gu},
  {Guainazzi}, {Haba}, {Hagino}, {Hamaguchi}, {Harrus}, {Hatsukade}, {Hayashi},
  {Hayashi}, {Hayashi}, {Hayashida}, {Hiraga}, {Hornschemeier}, {Hoshino},
  {Hughes}, {Ichinohe}, {Iizuka}, {Inoue}, {Inoue}, {Inoue}, {Ishida},
  {Ishikawa}, {Ishisaki}, {Iwai}, {Kaastra}, {Kallman}, {Kamae}, {Kataoka},
  {Katsuda}, {Kawai}, {Kelley}, {Kilbourne}, {Kitaguchi}, {Kitamoto},
  {Kitayama}, {Kohmura}, {Kokubun}, {Koyama}, {Koyama}, {Kretschmar}, {Krimm},
  {Kubota}, {Kunieda}, {Laurent}, {Lee}, {Leutenegger}, {Limousin},
  {Loewenstein}, {Long}, {Lumb}, {Madejski}, {Maeda}, {Maier}, {Makishima},
  {Markevitch}, {Matsumoto}, {Matsushita}, {McCammon}, {McNamara}, {Mehdipour},
  {Miller}, {Miller}, {Mineshige}, {Mitsuda}, {Mitsuishi}, {Miyazawa},
  {Mizuno}, {Mori}, {Mori}, {Mukai}, {Murakami}, {Mushotzky}, {Nakagawa},
  {Nakajima}, {Nakamori}, {Nakashima}, {Nakazawa}, {Nobukawa}, {Nobukawa},
  {Noda}, {Odaka}, {Ohashi}, {Ohno}, {Okajima}, {Ota}, {Ozaki}, {Paerels},
  {Paltani}, {Petre}, {Pinto}, {Porter}, {Pottschmidt}, {Reynolds},
  {Safi-Harb}, {Saito}, {Sakai}, {Sasaki}, {Sato}, {Sato}, {Sato}, {Sawada},
  {Schartel}, {Serlemtsos}, {Seta}, {Shidatsu}, {Simionescu}, {Smith}, {Soong},
  {Stawarz}, {Sugawara}, {Sugita}, {Szymkowiak}, {Tajima}, {Takahashi},
  {Takahashi}, {Takeda}, {Takei}, {Tamagawa}, {Tamura}, {Tanaka}, {Tanaka},
  {Tanaka}, {Tanaka}, {Tashiro}, {Tawara}, {Terada}, {Terashima}, {Tombesi},
  {Tomida}, {Tsuboi}, {Tsujimoto}, {Tsunemi}, {Tsuru}, {Uchida}, {Uchiyama},
  {Uchiyama}, {Ueda}, {Ueda}, {Uno}, {Urry}, {Ursino}, {Wang}, {Watanabe},
  {Werner}, {Wilkins}, {Williams}, {Yamada}, {Yamaguchi}, {Yamaoka},
  {Yamasaki}, {Yamauchi}, {Yamauchi}, {Yaqoob}, {Yatsu}, {Yonetoku},
  {Zhuravleva}, \& {Zoghbi}}]{Hitomi2018Perseus}
{Hitomi Collaboration}, {Aharonian}, F., {Akamatsu}, H., {et~al.} 2018, \pasj,
  70, 9

\bibitem[{{Hollowood} {et~al.}(2019){Hollowood}, {Jeltema}, {Chen}, {Farahi},
  {Evrard}, {Everett}, {Rozo}, {Rykoff}, {Bernstein}, {Bermeo-Hernandez},
  {Eiger}, {Giles}, {Israel}, {Michel}, {Noorali}, {Romer}, {Rooney}, \&
  {Splettstoesser}}]{Hollowood2019ApJS..244...22H}
{Hollowood}, D.~L., {Jeltema}, T., {Chen}, X., {et~al.} 2019, \apjs, 244, 22

\bibitem[{Hwang \& Lee(2007)}]{Hwang_2007_spin_galaxy}
Hwang, H.~S. \& Lee, M.~G. 2007, The Astrophysical Journal, 662, 236

\bibitem[{{Ider Chitham} {et~al.}(2020){Ider Chitham}, {Comparat},
  {Finoguenov}, {Clerc}, {Kirkpatrick}, {Damsted}, {Kukkola}, {Capasso},
  {Nandra}, {Merloni}, {Bulbul}, {Rykoff}, {Schneider}, \&
  {Brownstein}}]{IderChitham2020MNRAS.499.4768I}
{Ider Chitham}, J., {Comparat}, J., {Finoguenov}, A., {et~al.} 2020, \mnras,
  499, 4768

\bibitem[{{Ishiyama} {et~al.}(2015){Ishiyama}, {Enoki}, {Kobayashi}, {Makiya},
  {Nagashima}, \& {Oogi}}]{Ishiyama2015}
{Ishiyama}, T., {Enoki}, M., {Kobayashi}, M.~A.~R., {et~al.} 2015, \pasj, 67,
  61

\bibitem[{{Jenkins} {et~al.}(2001){Jenkins}, {Frenk}, {White}, {Colberg},
  {Cole}, {Evrard}, {Couchman}, \& {Yoshida}}]{Jenkins2001}
{Jenkins}, A., {Frenk}, C.~S., {White}, S.~D.~M., {et~al.} 2001, \mnras, 321,
  372

\bibitem[{{K{\"a}fer} {et~al.}(2020){K{\"a}fer}, {Finoguenov}, {Eckert},
  {Clerc}, {Ramos-Ceja}, {Sanders}, \& {Ghirardini}}]{Kaefer2020A&Awvdet}
{K{\"a}fer}, F., {Finoguenov}, A., {Eckert}, D., {et~al.} 2020, \aap, 634, A8

\bibitem[{{K{\"a}fer} {et~al.}(2019){K{\"a}fer}, {Finoguenov}, {Eckert},
  {Sanders}, {Reiprich}, \& {Nandra}}]{Kafer2019}
{K{\"a}fer}, F., {Finoguenov}, A., {Eckert}, D., {et~al.} 2019, \aap, 628, A43

\bibitem[{Kilbinger(2015)}]{Kilbinger_2015}
Kilbinger, M. 2015, Reports on Progress in Physics, 78, 086901

\bibitem[{{Klypin} {et~al.}(2016){Klypin}, {Yepes}, {Gottl{\"o}ber}, {Prada},
  \& {He{\ss}}}]{Klypin2016}
{Klypin}, A., {Yepes}, G., {Gottl{\"o}ber}, S., {Prada}, F., \& {He{\ss}}, S.
  2016, \mnras, 457, 4340

\bibitem[{{Knebe} {et~al.}(2011){Knebe}, {Knollmann}, {Muldrew}, {Pearce},
  {Aragon-Calvo}, {Ascasibar}, {Behroozi}, {Ceverino}, {Colombi}, {Diemand},
  {Dolag}, {Falck}, {Fasel}, {Gardner}, {Gottl{\"o}ber}, {Hsu}, {Iannuzzi},
  {Klypin}, {Luki{\'c}}, {Maciejewski}, {McBride}, {Neyrinck}, {Planelles},
  {Potter}, {Quilis}, {Rasera}, {Read}, {Ricker}, {Roy}, {Springel}, {Stadel},
  {Stinson}, {Sutter}, {Turchaninov}, {Tweed}, {Yepes}, \& {Zemp}}]{Knebe2011}
{Knebe}, A., {Knollmann}, S.~R., {Muldrew}, S.~I., {et~al.} 2011, \mnras, 415,
  2293

\bibitem[{{Knebe} {et~al.}(2013){Knebe}, {Pearce}, {Lux}, {Ascasibar},
  {Behroozi}, {Casado}, {Moran}, {Diemand}, {Dolag}, {Dominguez-Tenreiro},
  {Elahi}, {Falck}, {Gottl{\"o}ber}, {Han}, {Klypin}, {Luki{\'c}},
  {Maciejewski}, {McBride}, {Merch{\'a}n}, {Muldrew}, {Neyrinck}, {Onions},
  {Planelles}, {Potter}, {Quilis}, {Rasera}, {Ricker}, {Roy}, {Ruiz},
  {Sgr{\'o}}, {Springel}, {Stadel}, {Sutter}, {Tweed}, \&
  {Zemp}}]{Knebe2013MNRAS.435.1618K}
{Knebe}, A., {Pearce}, F.~R., {Lux}, H., {et~al.} 2013, \mnras, 435, 1618

\bibitem[{{Kravtsov} {et~al.}(1997){Kravtsov}, {Klypin}, \&
  {Khokhlov}}]{Kravtsov1997}
{Kravtsov}, A.~V., {Klypin}, A.~A., \& {Khokhlov}, A.~M. 1997, \apjs, 111, 73

\bibitem[{{Lang} {et~al.}(2015){Lang}, {Holley-Bockelmann}, \&
  {Sinha}}]{Lang2015concvoroni}
{Lang}, M., {Holley-Bockelmann}, K., \& {Sinha}, M. 2015, \apj, 811, 152

\bibitem[{{Laureijs} {et~al.}(2011){Laureijs}, {Amiaux}, {Arduini},
  {Augu{\`e}res}, {Brinchmann}, {Cole}, {Cropper}, {Dabin}, {Duvet}, {Ealet},
  {Garilli}, {Gondoin}, {Guzzo}, {Hoar}, {Hoekstra}, {Holmes}, {Kitching},
  {Maciaszek}, {Mellier}, {Pasian}, {Percival}, {Rhodes}, {Saavedra Criado},
  {Sauvage}, {Scaramella}, {Valenziano}, {Warren}, {Bender}, {Castander},
  {Cimatti}, {Le F{\`e}vre}, {Kurki-Suonio}, {Levi}, {Lilje}, {Meylan},
  {Nichol}, {Pedersen}, {Popa}, {Rebolo Lopez}, {Rix}, {Rottgering},
  {Zeilinger}, {Grupp}, {Hudelot}, {Massey}, {Meneghetti}, {Miller}, {Paltani},
  {Paulin-Henriksson}, {Pires}, {Saxton}, {Schrabback}, {Seidel}, {Walsh},
  {Aghanim}, {Amendola}, {Bartlett}, {Baccigalupi}, {Beaulieu}, {Benabed},
  {Cuby}, {Elbaz}, {Fosalba}, {Gavazzi}, {Helmi}, {Hook}, {Irwin}, {Kneib},
  {Kunz}, {Mannucci}, {Moscardini}, {Tao}, {Teyssier}, {Weller}, {Zamorani},
  {Zapatero Osorio}, {Boulade}, {Foumond}, {Di Giorgio}, {Guttridge}, {James},
  {Kemp}, {Martignac}, {Spencer}, {Walton}, {Bl{\"u}mchen}, {Bonoli},
  {Bortoletto}, {Cerna}, {Corcione}, {Fabron}, {Jahnke}, {Ligori}, {Madrid},
  {Martin}, {Morgante}, {Pamplona}, {Prieto}, {Riva}, {Toledo}, {Trifoglio},
  {Zerbi}, {Abdalla}, {Douspis}, {Grenet}, {Borgani}, {Bouwens}, {Courbin},
  {Delouis}, {Dubath}, {Fontana}, {Frailis}, {Grazian}, {Koppenh{\"o}fer},
  {Mansutti}, {Melchior}, {Mignoli}, {Mohr}, {Neissner}, {Noddle}, {Poncet},
  {Scodeggio}, {Serrano}, {Shane}, {Starck}, {Surace}, {Taylor},
  {Verdoes-Kleijn}, {Vuerli}, {Williams}, {Zacchei}, {Altieri}, {Escudero
  Sanz}, {Kohley}, {Oosterbroek}, {Astier}, {Bacon}, {Bardelli}, {Baugh},
  {Bellagamba}, {Benoist}, {Bianchi}, {Biviano}, {Branchini}, {Carbone},
  {Cardone}, {Clements}, {Colombi}, {Conselice}, {Cresci}, {Deacon}, {Dunlop},
  {Fedeli}, {Fontanot}, {Franzetti}, {Giocoli}, {Garcia-Bellido}, {Gow},
  {Heavens}, {Hewett}, {Heymans}, {Holland}, {Huang}, {Ilbert}, {Joachimi},
  {Jennins}, {Kerins}, {Kiessling}, {Kirk}, {Kotak}, {Krause}, {Lahav}, {van
  Leeuwen}, {Lesgourgues}, {Lombardi}, {Magliocchetti}, {Maguire}, {Majerotto},
  {Maoli}, {Marulli}, {Maurogordato}, {McCracken}, {McLure}, {Melchiorri},
  {Merson}, {Moresco}, {Nonino}, {Norberg}, {Peacock}, {Pello}, {Penny},
  {Pettorino}, {Di Porto}, {Pozzetti}, {Quercellini}, {Radovich}, {Rassat},
  {Roche}, {Ronayette}, {Rossetti}, {Sartoris}, {Schneider}, {Semboloni},
  {Serjeant}, {Simpson}, {Skordis}, {Smadja}, {Smartt}, {Spano}, {Spiro},
  {Sullivan}, {Tilquin}, {Trotta}, {Verde}, {Wang}, {Williger}, {Zhao},
  {Zoubian}, \& {Zucca}}]{Laureijs2011}
{Laureijs}, R., {Amiaux}, J., {Arduini}, S., {et~al.} 2011, arXiv e-prints,
  arXiv:1110.3193

\bibitem[{{LSST Science Collaboration} {et~al.}(2009){LSST Science
  Collaboration}, {Abell}, {Allison}, {Anderson}, {Andrew}, {Angel}, {Armus},
  {Arnett}, {Asztalos}, {Axelrod}, \& et~al.}]{2009lsst}
{LSST Science Collaboration}, {Abell}, P.~A., {Allison}, J., {et~al.} 2009,
  arXiv e-prints, arXiv:0912.0201

\bibitem[{{Ludlow} {et~al.}(2016){Ludlow}, {Bose}, {Angulo}, {Wang},
  {Hellwing}, {Navarro}, {Cole}, \& {Frenk}}]{Ludlow2016conc_cold_warm}
{Ludlow}, A.~D., {Bose}, S., {Angulo}, R.~E., {et~al.} 2016, \mnras, 460, 1214

\bibitem[{{Ludlow} {et~al.}(2014){Ludlow}, {Navarro}, {Angulo},
  {Boylan-Kolchin}, {Springel}, {Frenk}, \&
  {White}}]{Ludlow2014MNRAS.441..378L}
{Ludlow}, A.~D., {Navarro}, J.~F., {Angulo}, R.~E., {et~al.} 2014, \mnras, 441,
  378

\bibitem[{{Ludlow} {et~al.}(2013){Ludlow}, {Navarro}, {Boylan-Kolchin}, {Bett},
  {Angulo}, {Li}, {White}, {Frenk}, \& {Springel}}]{Ludlow2013MNRAS.432.1103L}
{Ludlow}, A.~D., {Navarro}, J.~F., {Boylan-Kolchin}, M., {et~al.} 2013, \mnras,
  432, 1103

\bibitem[{{Ludlow} {et~al.}(2012){Ludlow}, {Navarro}, {Li}, {Angulo},
  {Boylan-Kolchin}, \& {Bett}}]{Ludlow2012MNRAS.427.1322L}
{Ludlow}, A.~D., {Navarro}, J.~F., {Li}, M., {et~al.} 2012, \mnras, 427, 1322

\bibitem[{{Macci{\`o}} {et~al.}(2008){Macci{\`o}}, {Dutton}, \& {van den
  Bosch}}]{Maccio2008}
{Macci{\`o}}, A.~V., {Dutton}, A.~A., \& {van den Bosch}, F.~C. 2008, \mnras,
  391, 1940

\bibitem[{{Manolopoulou} \& {Plionis}(2017)}]{Manolopouloucluster_rot}
{Manolopoulou}, M. \& {Plionis}, M. 2017, \mnras, 465, 2616

\bibitem[{{McClintock} {et~al.}(2019){McClintock}, {Rozo}, {Becker}, {DeRose},
  {Mao}, {McLaughlin}, {Tinker}, {Wechsler}, \&
  {Zhai}}]{McClintock2019ApJ...872...53M}
{McClintock}, T., {Rozo}, E., {Becker}, M.~R., {et~al.} 2019, \apj, 872, 53

\bibitem[{{Meneghetti} \& {Rasia}(2013)}]{Meneghetti2013conc}
{Meneghetti}, M. \& {Rasia}, E. 2013, arXiv e-prints, arXiv:1303.6158

\bibitem[{{Merloni} {et~al.}(2012){Merloni}, {Predehl}, {Becker},
  {B{\"o}hringer}, {Boller}, {Brunner}, {Brusa}, {Dennerl}, {Freyberg},
  {Friedrich}, {Georgakakis}, {Haberl}, {Hasinger}, {Meidinger}, {Mohr},
  {Nandra}, {Rau}, {Reiprich}, {Robrade}, {Salvato}, {Santangelo}, {Sasaki},
  {Schwope}, {Wilms}, \& {German eROSITA Consortium}}]{Merloni2012}
{Merloni}, A., {Predehl}, P., {Becker}, W., {et~al.} 2012, arXiv e-prints,
  arXiv:1209.3114

\bibitem[{{Murray} {et~al.}(2013){Murray}, {Power}, \& {Robotham}}]{Murray2013}
{Murray}, S.~G., {Power}, C., \& {Robotham}, A.~S.~G. 2013, Astronomy and
  Computing, 3, 23

\bibitem[{{Navarro} {et~al.}(1996){Navarro}, {Frenk}, \&
  {White}}]{Navarro_Frenk_White_1996}
{Navarro}, J.~F., {Frenk}, C.~S., \& {White}, S. D.~M. 1996, \apj, 462, 563

\bibitem[{{Neto} {et~al.}(2007){Neto}, {Gao}, {Bett}, {Cole}, {Navarro},
  {Frenk}, {White}, {Springel}, \& {Jenkins}}]{Neto2007}
{Neto}, A.~F., {Gao}, L., {Bett}, P., {et~al.} 2007, \mnras, 381, 1450

\bibitem[{{Nishimichi} {et~al.}(2019){Nishimichi}, {Takada}, {Takahashi},
  {Osato}, {Shirasaki}, {Oogi}, {Miyatake}, {Oguri}, {Murata}, {Kobayashi}, \&
  {Yoshida}}]{Nishimichi2019ApJ...884...29N}
{Nishimichi}, T., {Takada}, M., {Takahashi}, R., {et~al.} 2019, \apj, 884, 29

\bibitem[{{Ondaro-Mallea} {et~al.}(2021){Ondaro-Mallea}, {Angulo}, {Zennaro},
  {Contreras}, \& {Aric{\`o}}}]{Ondaro-Mallea2021arXiv210208958O}
{Ondaro-Mallea}, L., {Angulo}, R.~E., {Zennaro}, M., {Contreras}, S., \&
  {Aric{\`o}}, G. 2021, arXiv e-prints, arXiv:2102.08958

\bibitem[{{Peebles}(1969)}]{Peebles1969angular_momentum}
{Peebles}, P.~J.~E. 1969, \apj, 155, 393

\bibitem[{{Phriksee} {et~al.}(2020){Phriksee}, {Jullo}, {Limousin}, {Shan},
  {Finoguenov}, {Komonjinda}, {Wannawichian}, \&
  {Sawangwit}}]{Phriksee2020MNRAS.491.1643P}
{Phriksee}, A., {Jullo}, E., {Limousin}, M., {et~al.} 2020, \mnras, 491, 1643

\bibitem[{{Planck Collaboration} {et~al.}(2014){Planck Collaboration},
  {Abergel}, {Ade}, {Aghanim}, {Alves}, {Aniano}, {Armitage-Caplan}, {Arnaud},
  {Ashdown}, {Atrio-Barandela}, \& et~al.}]{Planck_2014}
{Planck Collaboration}, {Abergel}, A., {Ade}, P.~A.~R., {et~al.} 2014, \aap,
  571, A11

\bibitem[{{Poveda-Ruiz} {et~al.}(2016){Poveda-Ruiz}, {Forero-Romero}, \&
  {Mu{\~n}oz-Cuartas}}]{Poveda-Ruiz2016ApJ...832..169P}
{Poveda-Ruiz}, C.~N., {Forero-Romero}, J.~E., \& {Mu{\~n}oz-Cuartas}, J.~C.
  2016, \apj, 832, 169

\bibitem[{{Prada} {et~al.}(2012){Prada}, {Klypin}, {Cuesta}, {Betancort-Rijo},
  \& {Primack}}]{Prada2012}
{Prada}, F., {Klypin}, A.~A., {Cuesta}, A.~J., {Betancort-Rijo}, J.~E., \&
  {Primack}, J. 2012, \mnras, 423, 3018

\bibitem[{{Predehl} {et~al.}(2020){Predehl}, {Andritschke}, {Arefiev},
  {Babyshkin}, {Batanov}, {Becker}, {B{\"o}hringer}, {Bogomolov}, {Boller},
  {Borm}, {Bornemann}, {Br{\"a}uninger}, {Br{\"u}ggen}, {Brunner}, {Brusa},
  {Bulbul}, {Buntov}, {Burwitz}, {Burkert}, {Clerc}, {Churazov}, {Coutinho},
  {Dauser}, {Dennerl}, {Doroshenko}, {Eder}, {Emberger}, {Eraerds},
  {Finoguenov}, {Freyberg}, {Friedrich}, {Friedrich}, {F{\"u}rmetz},
  {Georgakakis}, {Gilfanov}, {Granato}, {Grossberger}, {Gueguen}, {Gureev},
  {Haberl}, {H{\"a}lker}, {Hartner}, {Hasinger}, {Huber}, {Ji}, {Kienlin},
  {Kink}, {Korotkov}, {Kreykenbohm}, {Lamer}, {Lomakin}, {Lapshov}, {Liu},
  {Maitra}, {Meidinger}, {Menz}, {Merloni}, {Mernik}, {Mican}, {Mohr},
  {M{\"u}ller}, {Nandra}, {Nazarov}, {Pacaud}, {Pavlinsky}, {Perinati},
  {Pfeffermann}, {Pietschner}, {Ramos-Ceja}, {Rau}, {Reiffers}, {Reiprich},
  {Robrade}, {Salvato}, {Sanders}, {Santangelo}, {Sasaki}, {Scheuerle},
  {Schmid}, {Schmitt}, {Schwope}, {Shirshakov}, {Steinmetz}, {Stewart},
  {Str{\"u}der}, {Sunyaev}, {Tenzer}, {Tiedemann}, {Tr{\"u}mper}, {Voron},
  {Weber}, {Wilms}, \& {Yaroshenko}}]{Predehl2020arXiv201003477P}
{Predehl}, P., {Andritschke}, R., {Arefiev}, V., {et~al.} 2020, arXiv e-prints,
  arXiv:2010.03477

\bibitem[{{Press} \& {Schechter}(1974)}]{PressSchechter1974}
{Press}, W.~H. \& {Schechter}, P. 1974, \apj, 187, 425

\bibitem[{{Ragagnin} {et~al.}(2019){Ragagnin}, {Dolag}, {Moscardini},
  {Biviano}, \& {D'Onofrio}}]{Ragagnin2019conc_magneticum}
{Ragagnin}, A., {Dolag}, K., {Moscardini}, L., {Biviano}, A., \& {D'Onofrio},
  M. 2019, \mnras, 486, 4001

\bibitem[{{Rephaeli}(1995)}]{Rephaeli1995SZ_review}
{Rephaeli}, Y. 1995, \araa, 33, 541

\bibitem[{{Riebe} {et~al.}(2013){Riebe}, {Partl}, {Enke}, {Forero-Romero},
  {Gottl{\"o}ber}, {Klypin}, {Lemson}, {Prada}, {Primack}, {Steinmetz}, \&
  {Turchaninov}}]{Riebe2013}
{Riebe}, K., {Partl}, A.~M., {Enke}, H., {et~al.} 2013, Astronomische
  Nachrichten, 334, 691

\bibitem[{{Rodriguez-Puebla} {et~al.}(2016){Rodriguez-Puebla}, {Behroozi},
  {Primack}, {Klypin}, {Lee}, \& {Hellinger}}]{Rodriguez-Puebla2016}
{Rodriguez-Puebla}, A., {Behroozi}, P., {Primack}, J., {et~al.} 2016, ArXiv
  e-prints [\eprint[arXiv]{1602.04813}]

\bibitem[{{Salvati} {et~al.}(2020){Salvati}, {Douspis}, \&
  {Aghanim}}]{Salvati2020}
{Salvati}, L., {Douspis}, M., \& {Aghanim}, N. 2020, arXiv e-prints,
  arXiv:2005.10204

\bibitem[{{Sereno} {et~al.}(2015){Sereno}, {Giocoli}, {Ettori}, \&
  {Moscardini}}]{Sereno2015conc_selection}
{Sereno}, M., {Giocoli}, C., {Ettori}, S., \& {Moscardini}, L. 2015, \mnras,
  449, 2024

\bibitem[{{Shan} {et~al.}(2017){Shan}, {Kneib}, {Li}, {Comparat}, {Erben},
  {Makler}, {Moraes}, {Van Waerbeke}, {Taylor}, {Charbonnier}, \&
  {Pereira}}]{Shan2017ApJ840104S}
{Shan}, H., {Kneib}, J.-P., {Li}, R., {et~al.} 2017, \apj, 840, 104

\bibitem[{{Sheth} \& {Tormen}(1999)}]{ShethTormen1999}
{Sheth}, R.~K. \& {Tormen}, G. 1999, \mnras, 308, 119

\bibitem[{{Sheth} \& {Tormen}(2002)}]{Sheth2002}
{Sheth}, R.~K. \& {Tormen}, G. 2002, \mnras, 329, 61

\bibitem[{{Skillman} {et~al.}(2014){Skillman}, {Warren}, {Turk}, {Wechsler},
  {Holz}, \& {Sutter}}]{Skillman2014}
{Skillman}, S.~W., {Warren}, M.~S., {Turk}, M.~J., {et~al.} 2014, ArXiv
  e-prints [\eprint[arXiv]{1407.2600}]

\bibitem[{{Song} {et~al.}(2018){Song}, {Hwang}, {Park}, {Smith}, \&
  {Einasto}}]{Song2018ApJ...869..124Sspin}
{Song}, H., {Hwang}, H.~S., {Park}, C., {Smith}, R., \& {Einasto}, M. 2018,
  \apj, 869, 124

\bibitem[{{Spergel} {et~al.}(2015){Spergel}, {Gehrels}, {Baltay}, {Bennett},
  {Breckinridge}, {Donahue}, {Dressler}, {Gaudi}, {Greene}, {Guyon}, {Hirata},
  {Kalirai}, {Kasdin}, {Macintosh}, {Moos}, {Perlmutter}, {Postman},
  {Rauscher}, {Rhodes}, {Wang}, {Weinberg}, {Benford}, {Hudson}, {Jeong},
  {Mellier}, {Traub}, {Yamada}, {Capak}, {Colbert}, {Masters}, {Penny},
  {Savransky}, {Stern}, {Zimmerman}, {Barry}, {Bartusek}, {Carpenter}, {Cheng},
  {Content}, {Dekens}, {Demers}, {Grady}, {Jackson}, {Kuan}, {Kruk}, {Melton},
  {Nemati}, {Parvin}, {Poberezhskiy}, {Peddie}, {Ruffa}, {Wallace}, {Whipple},
  {Wollack}, \& {Zhao}}]{Spergel2015_WFIRST}
{Spergel}, D., {Gehrels}, N., {Baltay}, C., {et~al.} 2015, arXiv e-prints,
  arXiv:1503.03757

\bibitem[{{Springel}(2005)}]{Springel2005}
{Springel}, V. 2005, \mnras, 364, 1105

\bibitem[{{Springel} {et~al.}(2005){Springel}, {White}, {Jenkins}, {Frenk},
  {Yoshida}, {Gao}, {Navarro}, {Thacker}, {Croton}, {Helly}, {Peacock}, {Cole},
  {Thomas}, {Couchman}, {Evrard}, {Colberg}, \& {Pearce}}]{Springel2005b}
{Springel}, V., {White}, S.~D.~M., {Jenkins}, A., {et~al.} 2005, \nat, 435, 629

\bibitem[{{Sunyaev} {et~al.}(2003){Sunyaev}, {Norman}, \&
  {Bryan}}]{Sunyaev2003AstL...29..783Sspin}
{Sunyaev}, R.~A., {Norman}, M.~L., \& {Bryan}, G.~L. 2003, Astronomy Letters,
  29, 783

\bibitem[{{Thomas} {et~al.}(2001){Thomas}, {Muanwong}, {Pearce}, {Couchman},
  {Edge}, {Jenkins}, \& {Onuora}}]{Thomas2001}
{Thomas}, P.~A., {Muanwong}, O., {Pearce}, F.~R., {et~al.} 2001, \mnras, 324,
  450

\bibitem[{{Tinker} {et~al.}(2008){Tinker}, {Kravtsov}, {Klypin}, {Abazajian},
  {Warren}, {Yepes}, {Gottl{\"o}ber}, \& {Holz}}]{Tinker2008}
{Tinker}, J., {Kravtsov}, A.~V., {Klypin}, A., {et~al.} 2008, \apj, 688, 709

\bibitem[{{Tovmassian}(2015)}]{Tovmassian2015rotat}
{Tovmassian}, H.~M. 2015, Astrophysics, 58, 328

\bibitem[{Umetsu(2020)}]{umetsu2020clustergalaxy}
Umetsu, K. 2020, Cluster-Galaxy Weak Lensing

\bibitem[{{van Uitert} {et~al.}(2016){van Uitert}, {Gilbank}, {Hoekstra},
  {Semboloni}, {Gladders}, \& {Yee}}]{vanUitert2016A&A...586A..43V}
{van Uitert}, E., {Gilbank}, D.~G., {Hoekstra}, H., {et~al.} 2016, \aap, 586,
  A43

\bibitem[{{Virtanen} {et~al.}(2020){Virtanen}, {Gommers}, {Oliphant},
  {Haberland}, {Reddy}, {Cournapeau}, {Burovski}, {Peterson}, {Weckesser},
  {Bright}, {van der Walt}, {Brett}, {Wilson}, {Jarrod Millman}, {Mayorov},
  {Nelson}, {Jones}, {Kern}, {Larson}, {Carey}, {Polat}, {Feng}, {Moore}, {Vand
  erPlas}, {Laxalde}, {Perktold}, {Cimrman}, {Henriksen}, {Quintero}, {Harris},
  {Archibald}, {Ribeiro}, {Pedregosa}, {van Mulbregt}, \&
  {Contributors}}]{2020SciPy-NMeth}
{Virtanen}, P., {Gommers}, R., {Oliphant}, T.~E., {et~al.} 2020, Nature
  Methods, 17, 261

\bibitem[{{Wang} {et~al.}(2019){Wang}, {Bose}, {Frenk}, {Gao}, {Jenkins},
  {Springel}, \& {White}}]{Wang_Springel_White2019}
{Wang}, J., {Bose}, S., {Frenk}, C.~S., {et~al.} 2019, arXiv e-prints,
  arXiv:1911.09720

\bibitem[{{Weinberg} {et~al.}(2013){Weinberg}, {Mortonson}, {Eisenstein},
  {Hirata}, {Riess}, \& {Rozo}}]{weinberg_2013_review}
{Weinberg}, D.~H., {Mortonson}, M.~J., {Eisenstein}, D.~J., {et~al.} 2013,
  \physrep, 530, 87

\bibitem[{{Zhao} {et~al.}(2009){Zhao}, {Jing}, {Mo}, \&
  {B{\"o}rner}}]{Zhao2009conc_mah}
{Zhao}, D.~H., {Jing}, Y.~P., {Mo}, H.~J., \& {B{\"o}rner}, G. 2009, \apj, 707,
  354

\bibitem[{{Zhao} {et~al.}(2003){Zhao}, {Mo}, {Jing}, \&
  {B{\"o}rner}}]{Zhao2003MNRAS.339...12Z}
{Zhao}, D.~H., {Mo}, H.~J., {Jing}, Y.~P., \& {B{\"o}rner}, G. 2003, \mnras,
  339, 12

\bibitem[{{Zubeldia} \& {Challinor}(2019)}]{Zubeldia2019MNRAS}
{Zubeldia}, {\'I}. \& {Challinor}, A. 2019, \mnras, 489, 401

\end{thebibliography}



\appendix
\section{Figures and tables}
\label{appendix:tables}

In this section we collect the  figures and tables relative to this work. They describe the mean relations between concentration, offset parameter, spin, and mass, as well as the full probability density functions of these quantities. Moreover, we show additional plots describing the halo $\sigma-X_{\rm off}-\lambda$ function. 

\begin{table*}
        \centering
        \caption{Best-fit parameters for concentration--$\sigma$ relation and its PDF P(c).
        }
        \label{tab:conc_sigma_pars}
        \scalebox{0.90}{
        \begin{tabular}{ cc |  cc cc cc cc cc cc} 
        \hline
     & & & $a_0$ & & $b_0$ & &  \\
    \hline
     & & & 0.754091 $\pm$ 0.000004 & & 0.574413 $\pm$ 0.000002 & &   \\
    \hline
    & $A$ & $\alpha$ & $\beta$ & $x_0$ & $e_0$ & $e_1$ & $e_2$\\
    \hline
    z=0 & 0.041 $\pm$ 0 & 1.397 $\pm$ 0.001 & 2.604 $\pm$ 0.001 & 7.225 $\pm$ 0.002 & 0.089 $\pm$ 0.001 & 0.776 $\pm$ 0.001 & -0.959 $\pm$ 0.001 \\
    z=0.52 & 0.044 $\pm$ 0 & 2.501 $\pm$ 0.001 & 1.283 $\pm$ 0.001 & 2.394 $\pm$ 0.001 & 0.579 $\pm$ 0.001 & -0.325 $\pm$ 0.001 & 0.0334 $\pm$ 0.001\\
    z=1.03 & 3.37e-3 $\pm$ 0 & 4.14 $\pm$ 0.12 & 0.927 $\pm$ 0.002 & 0.688 $\pm$ 0.025 & 0.188 $\pm$ 0.002 & 0.081 $\pm$ 0.001 & 0.0813 $\pm$ 0.001 \\
    z=1.43 & 2.54e-3 $\pm$ 0 & 4.45 $\pm$ 0.13 & 0.924 $\pm$ 0.002 & 0.623 $\pm$ 0.031 & 0.198 $\pm$ 0.003 & 0.095 $\pm$ 0.001 & 0.103 $\pm$ 0.001 \\
\hline
        \end{tabular}}
        \footnotesize{The models are described by Eqs. \ref{eq:conc_peak} and \ref{eq:modified_schechter_conc}. Uncertainties are of the percentage accuracy. In order to have compact information, when uncertainties are smaller than 4 order of magnitudes with respect to the parameter, a value of 0 is written. 
        }        
\end{table*}

\begin{table*}
        \centering
        \caption{Best-fit parameters for $\lambda$-$\sigma$ relation and its PDF P($\lambda$) at different redshifts.}
        \label{tab:lambda_sigma_pars}
        \begin{tabular}{ cc cc cc cc cc} 
        \hline
    & $a_0$ & & $b_0$  &\\
    \hline
    & 4.5357e-2 $\pm$ 2e-6 & & -5.4328e-3 $\pm$ 1e-7  &\\
    \hline
    & A & $\alpha$ & $\beta$ & $x_0$ \\
    \hline
    z=0 & 0.274 $\pm$ 0.009 & 3.002 $\pm$ 0.013 & 0.773 $\pm$ 0.001 & 4.33e-3 $\pm$ 0 \\
    z=0.52 & 1.01 $\pm$ 0.02 & 2.623 $\pm$ 0.004 & 0.911 $\pm$ 0.001 & 7.46e-3 $\pm$ 0 \\
    z=1.03 & 1.769 $\pm$ 0.009 & 2.409 $\pm$ 0.002 & 1.006 $\pm$ 0.001 & 9.32e-3 $\pm$ 0 \\
    z=1.43 & 2.089 $\pm$ 0.007 & 2.351 $\pm$ 0.001 & 1.031 $\pm$ 0.001 & 9.34e-3 $\pm$ 0 \\
\hline
        \end{tabular}\\
        \footnotesize{The models are described by Eqs. \ref{eq:lambda_sigma_rel} and \ref{eq:modified_schechter}. Uncertainties are the percentage accuracy. In order to have compact information, when uncertainties are smaller than 4 order of magnitudes with respect to the parameter, a value of 0 is written.}        
\end{table*}

\begin{table*}
        \centering
        \caption{Best-fit parameters for $X_{\rm off}$-$\sigma$ relation and its PDF P($X_{\rm off}$).
        }
        \label{tab:xoff_sigma_pars}
        \begin{tabular}{ cc |  cc cc cc cc}
        \hline
    &  $a_0$& &  $b_0$ &    \\
    \hline
    &  -1.30418 $\pm$ 0.00001 & & 0.15084 $\pm$ 0.00001 &  \\
     \hline
    & $\log_{10}A$ & $\alpha$ & $\beta$ & $\log_{10}x_0$  \\
    \hline
    z=0 & -3.09 $\pm$ 0.26 & 3.71 $\pm$ 0.12 & 0.64 $\pm$ 0.03 & -2.31 $\pm$ 0.09 \\
    z=0.52 & -2.72 $\pm$ 0.17 & 3.69 $\pm$ 0.03 & 0.69 $\pm$ 0.01 & -2.11 $\pm$ 0.01   \\
    z=1.03 & -2.18 $\pm$ 0.05 & 3.44 $\pm$ 0.02 & 0.77 $\pm$ 0.01 & -1.88 $\pm$ 0.01 \\
    z=1.43 & -1.79 $\pm$ 0.02 & 3.19 $\pm$ 0.01 & 0.84 $\pm$ 0.01 & -1.70 $\pm$ 0.01 \\
\hline
        \end{tabular}\\
        \footnotesize{The models are described by Eqs. \ref{eq:xoff_sigma_rel} and \ref{eq:modified_schechter_xoff}. Uncertainties on the mean relation are under the percentage level accuracy. 
        }        
\end{table*}

\begin{table}
        \centering
        \caption{Model parameters with priors and posterior constraints at redshift zero.} 
        \label{tab:parameters}
        \begin{tabular}{cc cc cc} 
        \hline
   Parameter & Prior & Posterior \\
   \hline
    $\log_{10}A$ & (-23,-20) & -22.004$^{+0.006}_{-0.006}$\\
    $a$ & (0.5,1.0) & 0.885$^{+0.004}_{-0.004}$\\
    $q$ & (1.5,2.5) & 2.284$^{+0.016}_{-0.016}$\\
    $\log_{10}\mu$ & (-3.5,-3.0) & -3.326$^{+0.001}_{-0.001}$\\
    $\alpha$ & (5.4,5.8) & 5.623$^{+0.002}_{-0.002}$\\
    $\beta$ & (-0.5,-0.3) & -0.391$^{+0.001}_{-0.001}$\\
    $\gamma$ & (2.8,3.2) & 3.024$^{+0.003}_{-0.003}$\\
    $\delta$ & (1.0,1.4) & 1.209$^{+0.001}_{-0.001}$\\
    $e$ & (-1.2,-0.8) & -1.105$^{+0.005}_{-0.005}$\\
    \hline
        \end{tabular}\\
        \footnotesize{The full distribution of the posteriors is shown in Fig. \ref{fig:corner_z_0}.}        
\end{table}

\begin{figure*}
    \centering
    \includegraphics[width=2.0\columnwidth]{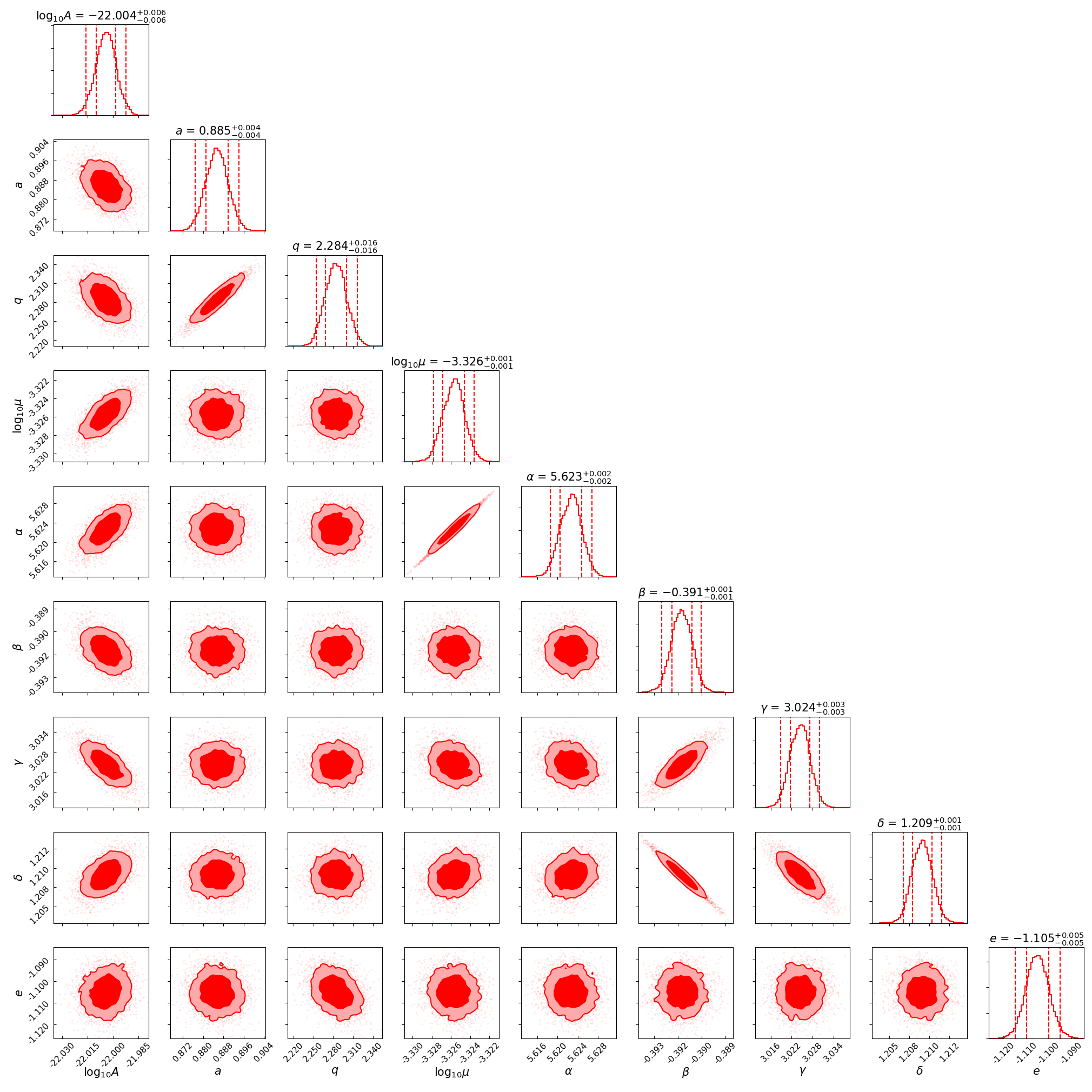}
        \caption{Marginalized posterior distributions of the best-fit parameters of the halo $\sigma-X_{\rm off}-\lambda$ function. The 0.68 and 0.95 confidence levels of the posteriors are shown as filled 2D contours. The 2.5th, 16th, 84th, and 97.5th percentiles of the one-dimensional 
        posterior distributions are indicated by the vertical lines on the diagonal plots. The model is given by Equation \ref{eq:model}. The parameters are also given in Table \ref{tab:parameters}.}
    \label{fig:corner_z_0}
\end{figure*}

\begin{table}
        \centering
        \caption{Model parameters with prior and posterior constraints for the redshift evolution of the halo $\sigma-X_{\rm off}-\lambda$ function.}
        \label{tab:parameters_zevo}
        \begin{tabular}{cc cc cc} 
        \hline
   Parameter & Prior & Posterior \\
   \hline
    $k_0$ & (-0.08,0.07) & -0.0441$\pm 0.0001$\\
    $k_1$ & (-0.25,0.05) & -0.161$\pm 0.001$\\
    $k_2$ & (-0.05,0.15) & 0.041$\pm 0.002$\\
    $k_3$ & (-0.18,0.02) & -0.1286$\pm 0.0002$\\
    $k_4$ & (-0.02,0.18) & 0.108$\pm 0.0002$\\
    $k_5$ & (-0.7,0.1) & -0.311$\pm 0.001$\\
    $k_6$ & (-0.1,0.2) & 0.0902$\pm 0.0004$\\   
    $k_7$ & (-0.2,0.1) & -0.0768$\pm 0.0004$\\
    $k_8$ & (-0.05,0.85) & 0.612$\pm 0.002$\\
    \hline
        \end{tabular}\\
        \footnotesize{In the posteriors, when the $1\sigma$ uncertainty on the one-dimensional distribution 
        is smaller than 3 order of magnitudes with respect to the parameter, we write null error. The redshift evolution of each parameter is shown in Fig. \ref{fig:zevo}.}        
\end{table}

\begin{figure*}
    \centering
    \includegraphics[width=2.0\columnwidth]{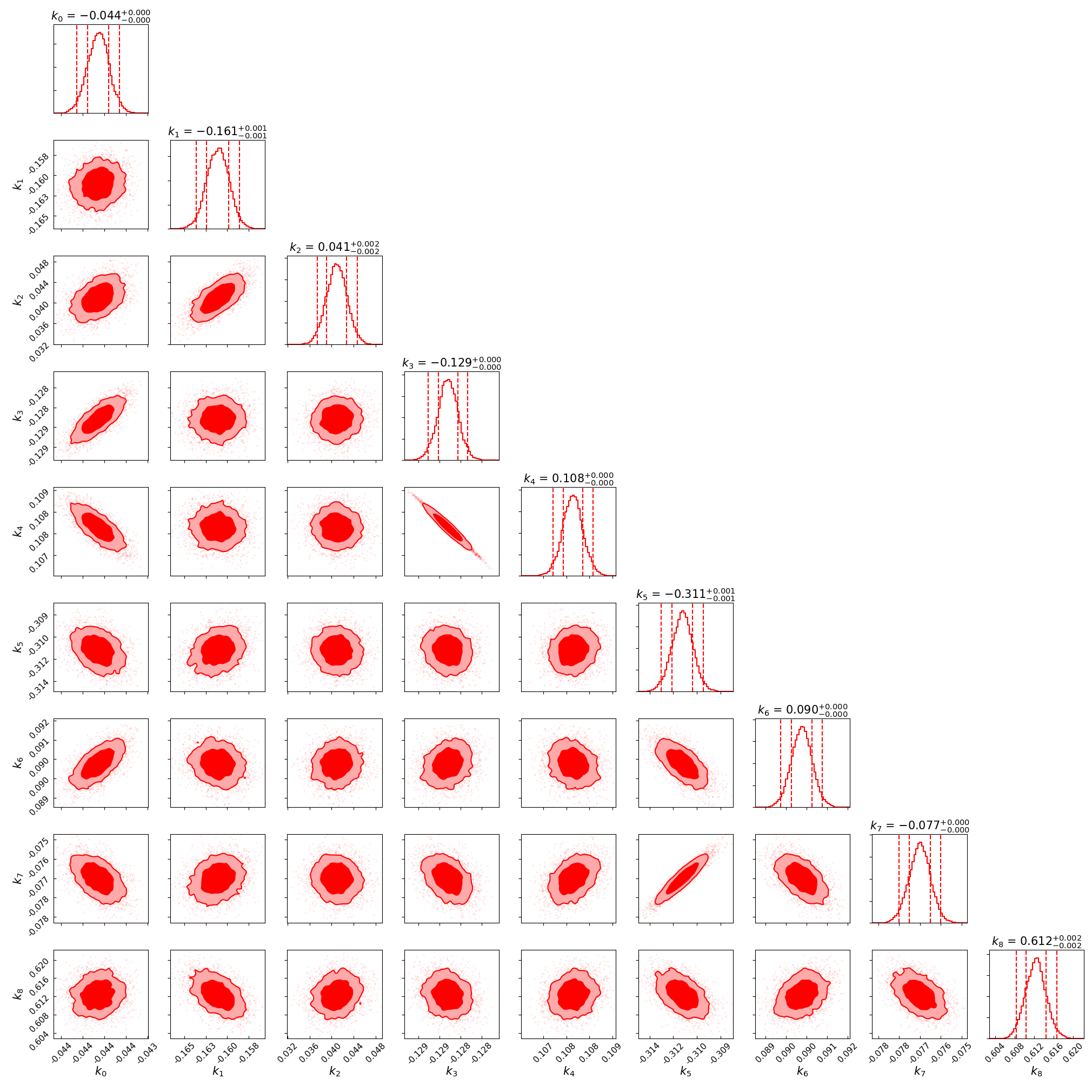}
        \caption{Marginalized posterior distributions of the best-fit parameters describing the redshift evolution of $h(\sigma,X_{\rm off},\lambda)$. The 0.68 and 0.95 confidence levels of the posteriors are shown as filled 2D contours. The model is given by Equation \ref{eq:zevolution}. The parameters are reported in Table \ref{tab:parameters_zevo}. 
    }
    \label{fig:corner_zevo}
\end{figure*}

\begin{table}
\centering
\caption{Full list of snapshots used, available in \textsc{HMD}, \textsc{BigMD}, \textsc{MDPL2}. }
\label{tab:snapshots:1}
\begin{tabular}{cc cc cc cc cc cc cc} 
\hline
a & z & T(Gyr) &  \textsc{HMDPL} & \textsc{BigMD} & \textsc{MDPL2}  \\   
\hline
1.0  & 0  & 13.82  & x & x & x \\
0.9567  & 0.04526  & 13.19  & x &  & x \\
0.956  & 0.04603  & 13.18  &  & x &  \\
0.8953  & 0.1169  & 12.27  &  & x &  \\
0.8951  & 0.1172  & 12.26  & x &  & x \\
0.8192  & 0.2207  & 11.09  & x &  & x \\
0.8173  & 0.2235  & 11.06  &  & x &  \\
0.7016  & 0.4253  & 9.198  & x &  & x \\
0.7003  & 0.428  & 9.177  &  & x &  \\
0.6583  & 0.5191  & 8.487  &  & x &  \\
0.6565  & 0.5232  & 8.458  & x &  & x \\


0.5876  & 0.7018  & 7.319  & x &  & x \\
0.5864  & 0.7053  & 7.299  &  & x &  \\
0.5623  & 0.7785  & 6.90  &  & x & \\
0.5622  & 0.7787  & 6.89  & x &  & x \\
0.5  & 1  & 5.88  &  & x &  \\
0.4922  & 1.032  & 5.753  & x &  & x \\
0.4123  & 1.425  & 4.482  & x &  & x \\
0.409  & 1.445  & 4.431  &  & x &  \\
\hline
\end{tabular}

\end{table}



\section{Offset in physical units}
\label{appendix:fig}
Here we present the results of the same analysis elaborated in Sections \ref{sec:relations} and \ref{sec:results} to the offset parameter in physical units $X_{\rm off,P}$, measured in kpc/$h$. This approach allows the comparison between the physics of dark matter simulations and observations.

\subsection{Offset--mass--redshift relation}

The relation between $X_{\rm off,P}$, mass and redshift is modeled by
\begin{equation}
    \log_{10}X_{\rm off,P}(\sigma,z) = \frac{b_0}{E(z)^{0.06}} \Big[1 + 2.39 \Big(\frac{\sigma}{a_{0}E(z)^{0.8}}\Big)^{c_0\sigma} \Big].
    \label{eq:xoff_sigma_rel_P}
\end{equation}
The distribution of $X_{\rm off,P}$ around its mean value is described by a modified Schechter function, but $X_{\rm off,P}$ is not normalized to the virial radius. Therefore, a mass dependence has to be included in the relation.
\begin{equation}
    P(X_{\rm off,P}) = A \Big(\frac{X_{\rm off,P}}{x_0\sigma^{e_0}} \Big)^{\alpha} \text{exp} \Big[-\Big(\frac{X_{\rm off,P}}{x_0\sigma^{e_0}} \Big)^{\beta} \Big].
    \label{eq:modified_schechter_xoff_P}    
\end{equation}

\begin{figure}
    \centering
    \includegraphics[width=\columnwidth]{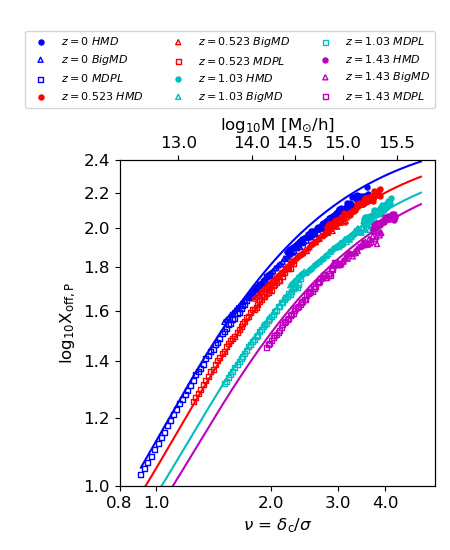}
        \caption{$X_{\rm off,P}$-$\sigma$ relation (Equation \ref{eq:xoff_sigma_rel_P}). Circular dots, triangles, and squares represent \textsc{HMD, BigMD}, and \textsc{MDPL2}, respectively. They are color-coded by redshift. Straight lines show the best-fit model. Parameters are given in Table \ref{tab:xoff_sigma_pars_P}.}
    \label{fig:xoff_sigma_relation_P}
\end{figure}

\begin{figure*}
    \centering
    \includegraphics[width=0.9\columnwidth]{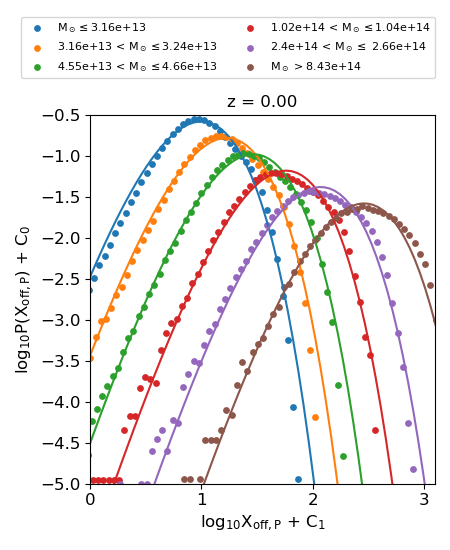}
    \includegraphics[width=0.9\columnwidth]{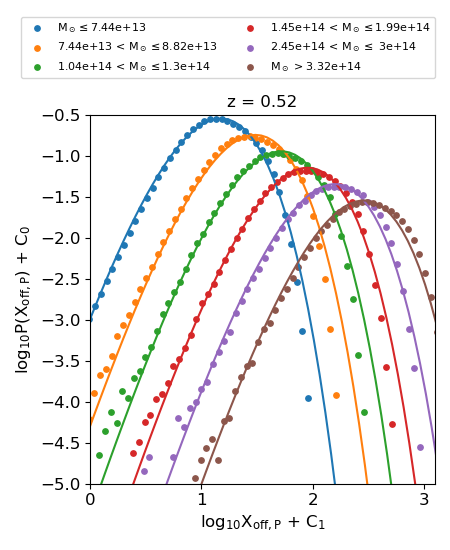}
    \includegraphics[width=0.9\columnwidth]{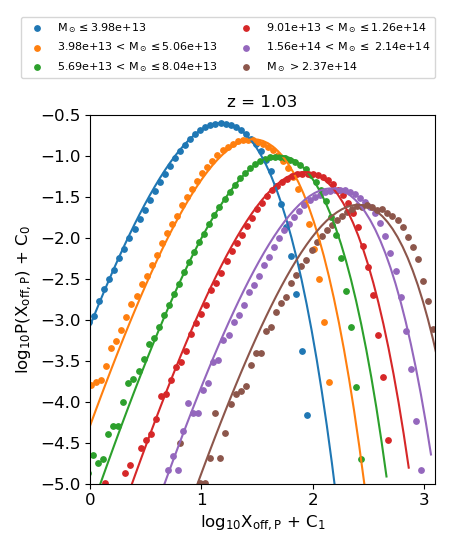}
    \includegraphics[width=0.9\columnwidth]{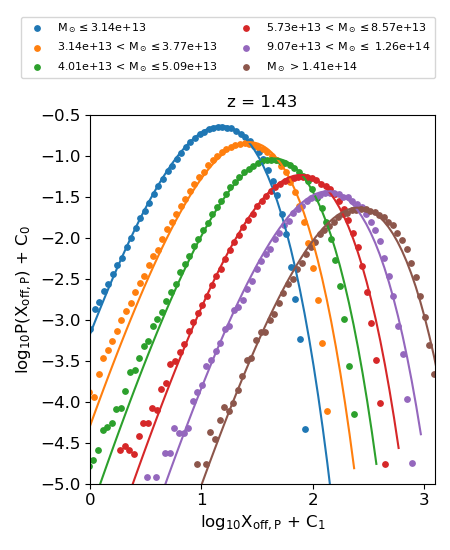}
        \caption{Probability density function of $X_{\rm off,P}$ (Equation \ref{eq:modified_schechter_xoff_P}). Each panel shows the distribution at a specific redshift. Each set is divided in mass slices, identified by color. Scatter points indicate the data, while straight lines represent the modified Schechter model \ref{eq:modified_schechter_xoff}. For clarity, each line and its fit are shifted by 0.2 dex along both axes. This means that the coefficient $C_0$  
        assumes values (+0.6,+0.4,+0.2,0.0,-0.2,-0.4), while $C_1$ is (-0.6,-0.4,-0.2,0.0,+0.2,+0.4). The red line is not shifted, and thus  is the one with the correct normalization.
        }
\label{fig:pdf_xoff_P}
\end{figure*}

The best-fit parameters are given in Table \ref{tab:xoff_sigma_pars_P}, and the results are shown in Figures \ref{fig:xoff_sigma_relation_P} and \ref{fig:pdf_xoff_P}.

\begin{table*}
        \centering
        \caption{Best-fit parameters for $X_{\rm off,P}$-$\sigma$ relation and P($X_{\rm off,P}$).
        }
        \label{tab:xoff_sigma_pars_P}
        \begin{tabular}{ cc |  cc cc cc cc cc}
        \hline
    & & $a_0$ &  $b_0$ & $c_0$ &   \\
    \hline
    & & 0.16523 $\pm$ 0.00004 & 0.74872 $\pm$ 0.00001 & -0.39607 $\pm$ 0.00003  & \\
     \hline
    & $A$ & $\alpha$ & $\beta$ & $x_0$ & $e_0$ \\
    \hline
    z=0 & -5.45 $\pm$ 0.01 & 10.56 $\pm$ 0.87 & 1.23 $\pm$ 0.09 & 2.34 $\pm$ 0.19 & -0.57 $\pm$ 0.05 \\
    z=0.52 & -5.45 $\pm$ 0.01 & 10.58 $\pm$ 0.87 & 1.23 $\pm$ 0.11 & 2.35 $\pm$ 0.19 & -0.43 $\pm$ 0.04  \\
    z=1.03 & -5.35 $\pm$ 0.01 & 10.54 $\pm$ 0.86 & 1.24 $\pm$ 0.11 & 2.36 $\pm$ 0.19 & -0.33 $\pm$ 0.03 \\
    z=1.43 & -5.32 $\pm$ 0.01 & 10.56 $\pm$ 0.86 & 1.32 $\pm$ 0.11 & 2.36 $\pm$ 0.19 & -0.46 $\pm$ 0.04 \\
\hline
        \end{tabular}\\
        \footnotesize{The models are described by Eqs. \ref{eq:xoff_sigma_rel_P} and \ref{eq:modified_schechter_xoff_P}. Uncertainties on the mean relation are under the percentage level accuracy.
        }        
\end{table*}

\subsection{Mass -- $X_{\rm off,P}$ -- $\lambda$ function}

In this section we collect figures and tables that describe h($\sigma$,$X_{\rm off,P},\lambda$). The analysis is similar to h($\sigma$,$X_{\rm off},\lambda$) explained in Sections \ref{sec:general_mf_framework} and \ref{sec:model}; the only difference is that the modified Schechter function that describes $X_{\rm off,P}$ needs a mass-dependent term. Therefore, we introduce an additional parameter and model the distribution according to 

\begin{align}
    &h(\sigma, X_{\rm off,P}, \lambda, z, A,a,q,\mu,\alpha,\beta,e_0,\gamma,\delta,e_1) = ...\nonumber \\
    &A\sqrt{\frac{2}{\pi}}\Big(\sqrt{a}\frac{\delta_c}{\sigma}\Big)^q \exp\Big[-\frac{a}{2}\frac{\delta_c^2}{\sigma^2}\Big]\Big(\frac{X_{\rm off,P}}{\mu \sigma^{e_0}} \Big)^{\alpha} ... \nonumber \\
    & \exp\Big[- \Big( \frac{X_{\rm off,P}}{\mu\sigma^{e_0}}\Big)^{0.05\alpha}\Big] \Big(\frac{\lambda}{0.7\mu}\Big)^{\gamma}exp\Big[- \Big( \frac{X_{\rm off,P}}{\mu\sigma^{e_1}}\Big)^{\beta} \Big(\frac{\lambda}{0.7\mu} \Big)^{\delta}\Big].
    \label{eq:model_P}
\end{align}

We recover the fiducial mass function at the $\sim 3.9\%$ level.

\begin{table}
        \centering
        \caption{h($\sigma$,$X_{\rm off,P},\lambda$) model parameters with priors and posterior costraints. } 
        \label{tab:parameters_P}
        \begin{tabular}{cc cc cc} 
        \hline
   Parameter & Prior & Posterior \\
   \hline
    $\log_{10}A$ & (-23,-20) & -22.004$^{+0.009}_{-0.009}$\\
    $a$ & (0.5,1.0) & 0.878$^{+0.004}_{-0.004}$\\
    $q$ & (1.5,2.5) & 2.257$^{+0.013}_{-0.013}$\\
    $\log_{10}\mu$ & (-3.5,-3.0) & -3.149$^{+0.002}_{-0.002}$\\
    $\alpha$ & (5.4,5.8) & 5.624$^{+0.002}_{-0.002}$\\
    $\beta$ & (-0.4,-0.3) & -0.365$^{+0.001}_{-0.001}$\\
    $e_0$ & (-2.0,-1.4) & -1.606$^{+0.002}_{-0.002}$\\   
    $\gamma$ & (2.8,3.2) & 3.095$^{+0.003}_{-0.003}$\\
    $\delta$ & (1.0,1.4) & 1.168$^{+0.002}_{-0.001}$\\
    $e_1$ & (-3.0,-2.5) & -2.270$^{+0.005}_{-0.005}$\\
    \hline
        \end{tabular}\\
        \footnotesize{The full distribution of the posteriors is shown in the triangular plot in Fig. \ref{fig:corner_z_0_P}.}        
\end{table}

\begin{figure*}
    \centering
    \includegraphics[width=0.8\columnwidth]{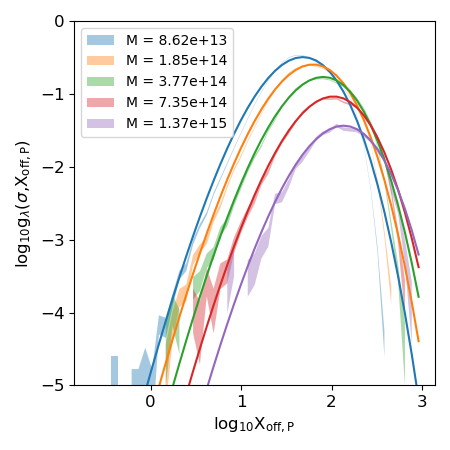}
    \includegraphics[width=0.8\columnwidth]{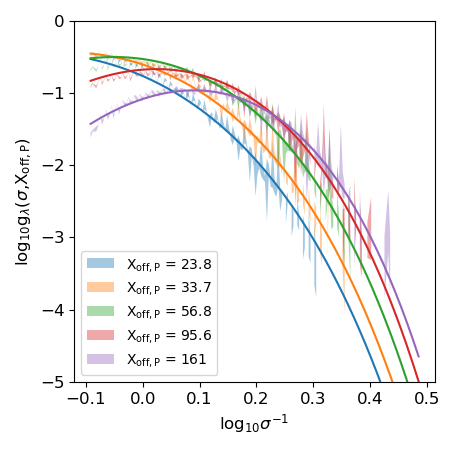}
    \includegraphics[width=0.8\columnwidth]{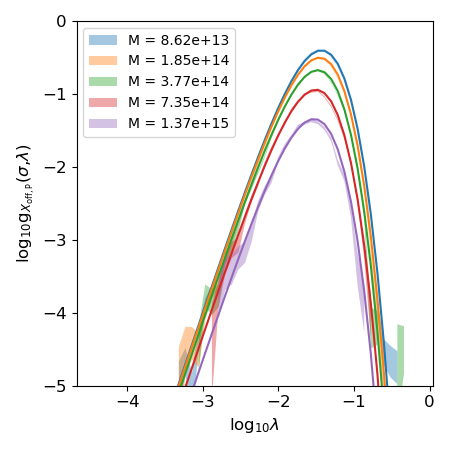}
    \includegraphics[width=0.8\columnwidth]{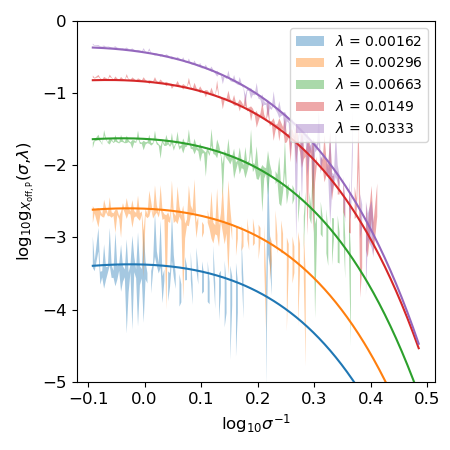}
    \includegraphics[width=0.8\columnwidth]{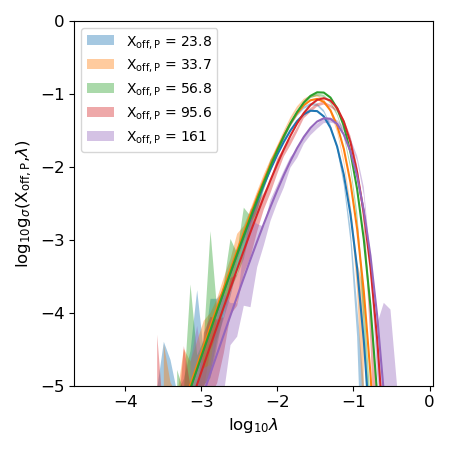}
    \includegraphics[width=0.8\columnwidth]{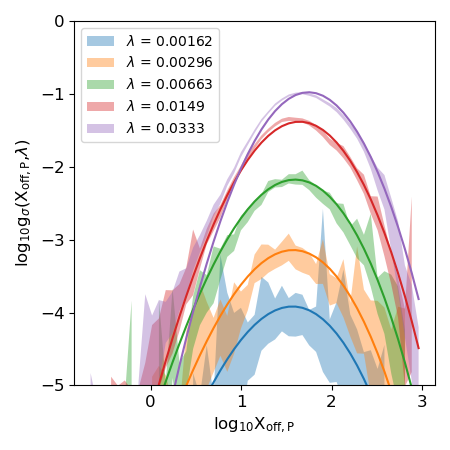}
        \caption{Single integration of the 3D model. In each panel straight lines indicate the best-fit model, while shaded areas represent the data with 1$\sigma$ uncertainties. Top left: $g_{\lambda}(\sigma,X_{\rm off,P})$ as a function of $X_{\rm off,P}$ in different mass slices. Top right: $g_{\lambda}(\sigma,X_{\rm off,P})$ as a function of $\sigma$ in different $X_{\rm off,P}$ slices. Middle left: $g_{X_{\rm off,P}}(\sigma,\lambda)$ as a function of $\lambda$ in different mass slices. Middle right: $g_{X_{\rm off,P}}(\sigma,\lambda)$ as a function of $\sigma$ in different $\lambda$ slices. Bottom left: $g_{\sigma}(X_{\rm off,P},\lambda)$ as a function of $\lambda$ in different $X_{\rm off,P}$ slices. Bottom right: $g_{\sigma}(X_{\rm off,P},\lambda)$ as a function of $X_{\rm off,P}$ in different $\lambda$ slices.}
        \label{fig:g_data_model_P}
\end{figure*}

\begin{figure*}
    \centering
    \includegraphics[width=\columnwidth]{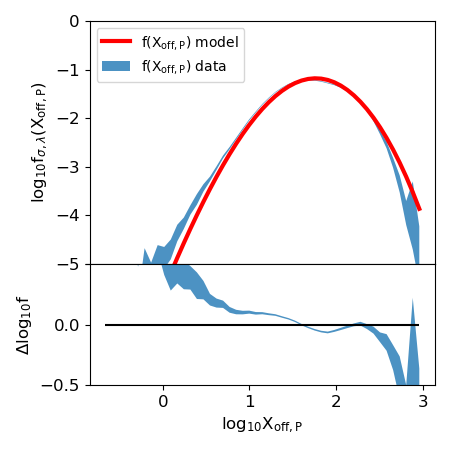}
    \includegraphics[width=\columnwidth]{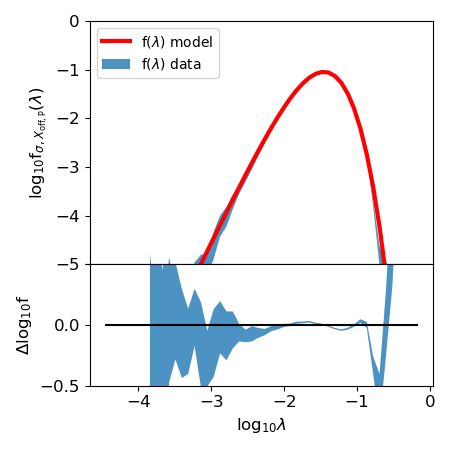}
        \caption{Comparison of $f(X_{\rm off,P})$ and $f(\lambda)$  between data and model. In the top panels the solid red lines indicate the integral on the best-fit model, while the shaded blue areas represent the integral on the 3D $h(\sigma,X_{\rm off,P},\lambda)$ data with 1$\sigma$ uncertainties. Each bottom panel shows the residual trend with $\sigma$ error; the straight black line represents the perfect match between data and model, with null residual. Top left: $f(X_{\rm off,P})$ as a function of $X_{\rm off,P}$. Bottom left: Residual between $f(X_{\rm off,P})$ data and model in logarithmic scale. Top right: $f(\lambda)$ as a function of $\lambda$. Bottom right: Residual between $f(\lambda)$ data and model in logarithmic scale.}
        \label{fig:f_data_model_P}
\end{figure*}

\begin{figure}
    \centering
    \includegraphics[width=1.0\columnwidth]{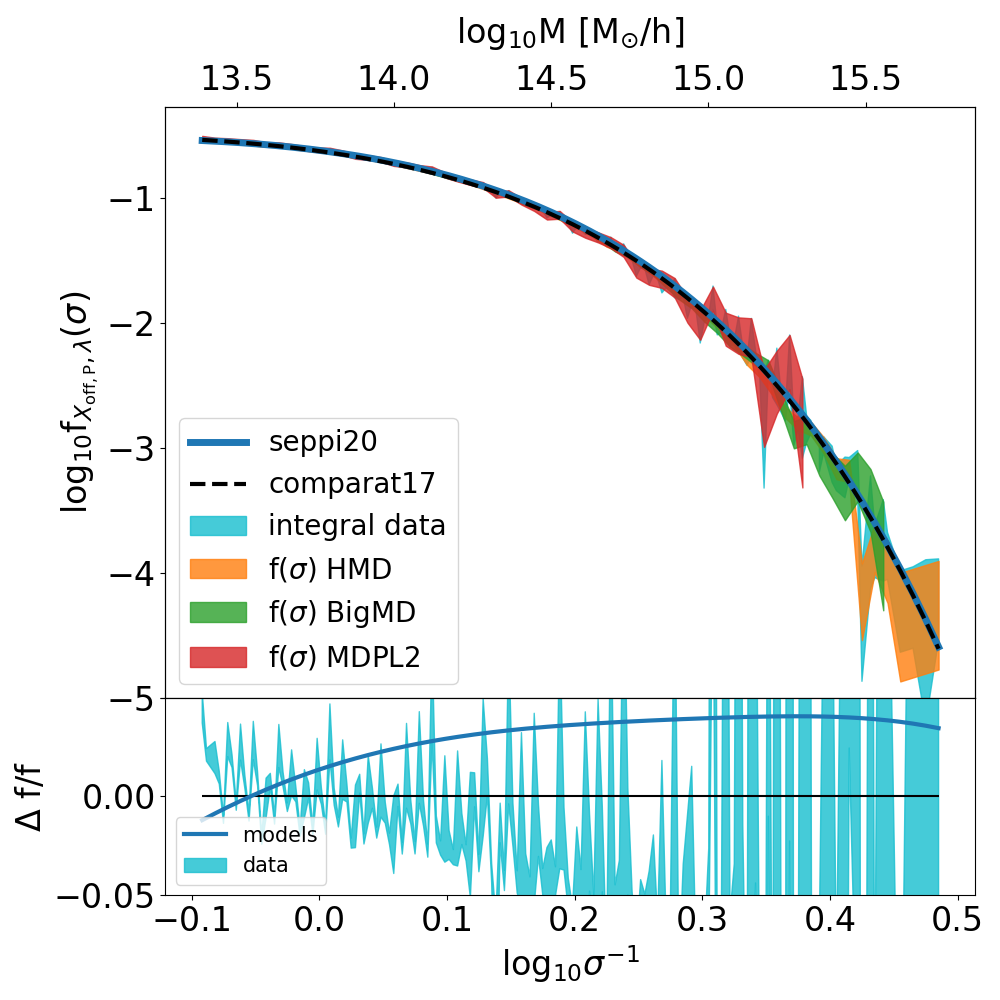}
        \caption{Comparison of multiplicty functions. Top: The three shaded regions show the 1$\sigma$ contours of $f(\sigma)$ data directly computed on different simulations (orange for \textsc{HMD}, green for \textsc{BigMD}, red for \textsc{MDPL2}); the light blue shaded region is the $1\sigma$ contour of the 2D integral computed on the concatenated sample containing all three simulations; the dashed pink line indicates the mass function from \citet{Comparat2017}, while the blue solid line is the $f(\sigma)$ we recover integrating our model along $X_{\rm off,P}$ and $\lambda$. Bottom: The blue thick line is the fractional difference between our $f(\sigma)$ and the \citet{Comparat2017} value. The light blue shaded area represent the $1\sigma$ contours of the residual between the integrated data and our  best-fit model; the black horizontal line indicates the perfect match with null residual.}
        \label{fig:integral_P}
\end{figure}

\begin{figure}
    \centering
    \includegraphics[width=\columnwidth]{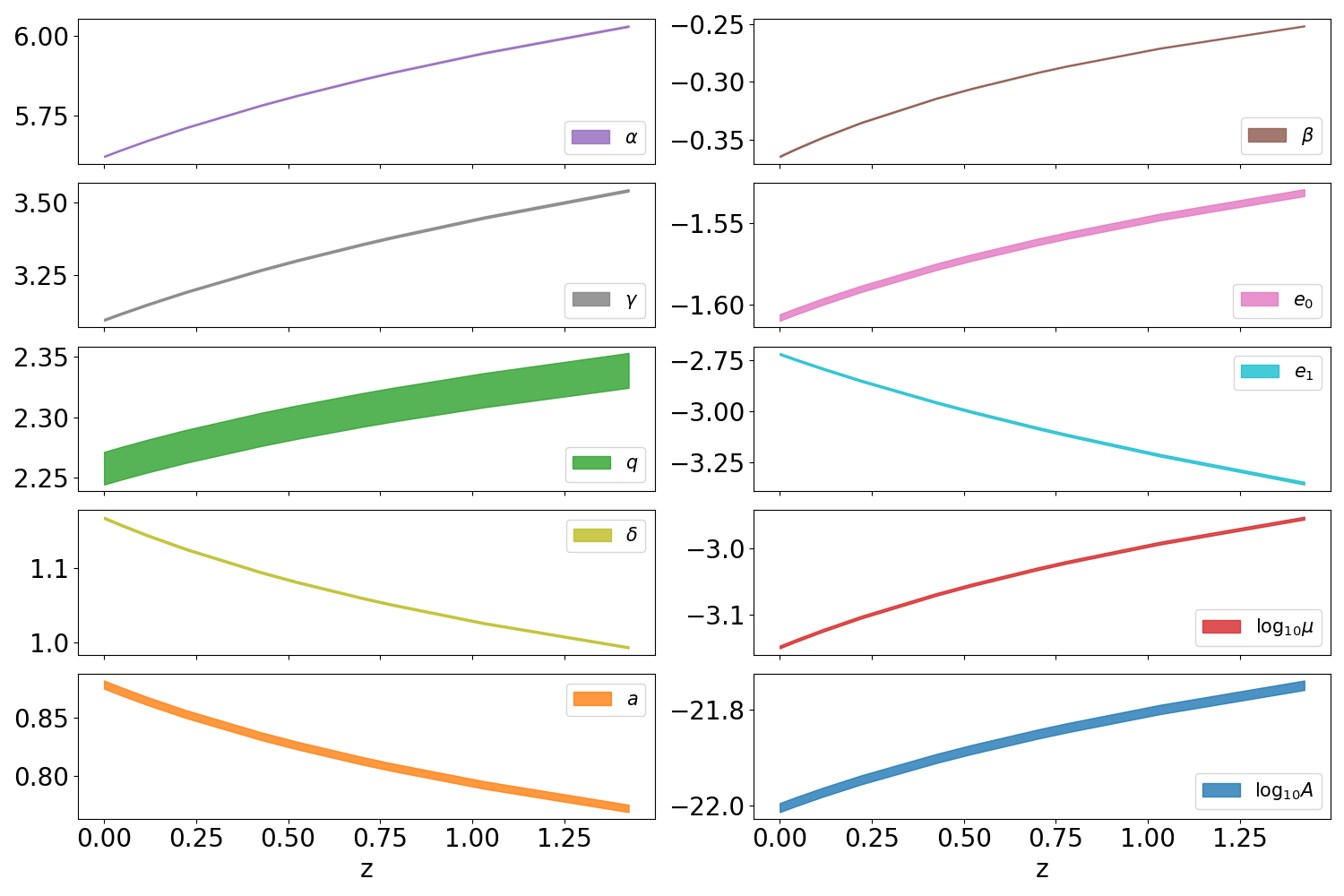}
        \caption{Redshift evolution of the best-fit parameters for h($\sigma$,$X_{\rm off,P},\lambda$). Each panel shows a single parameter. The values at z=0 are reported in Table \ref{tab:parameters_P}. The slopes of the redshift trends are given in Table \ref{tab:parameters_zevo_P}. }
        \label{fig:zevo_P}
\end{figure}

\begin{figure}
    \centering
    \includegraphics[width=\columnwidth]{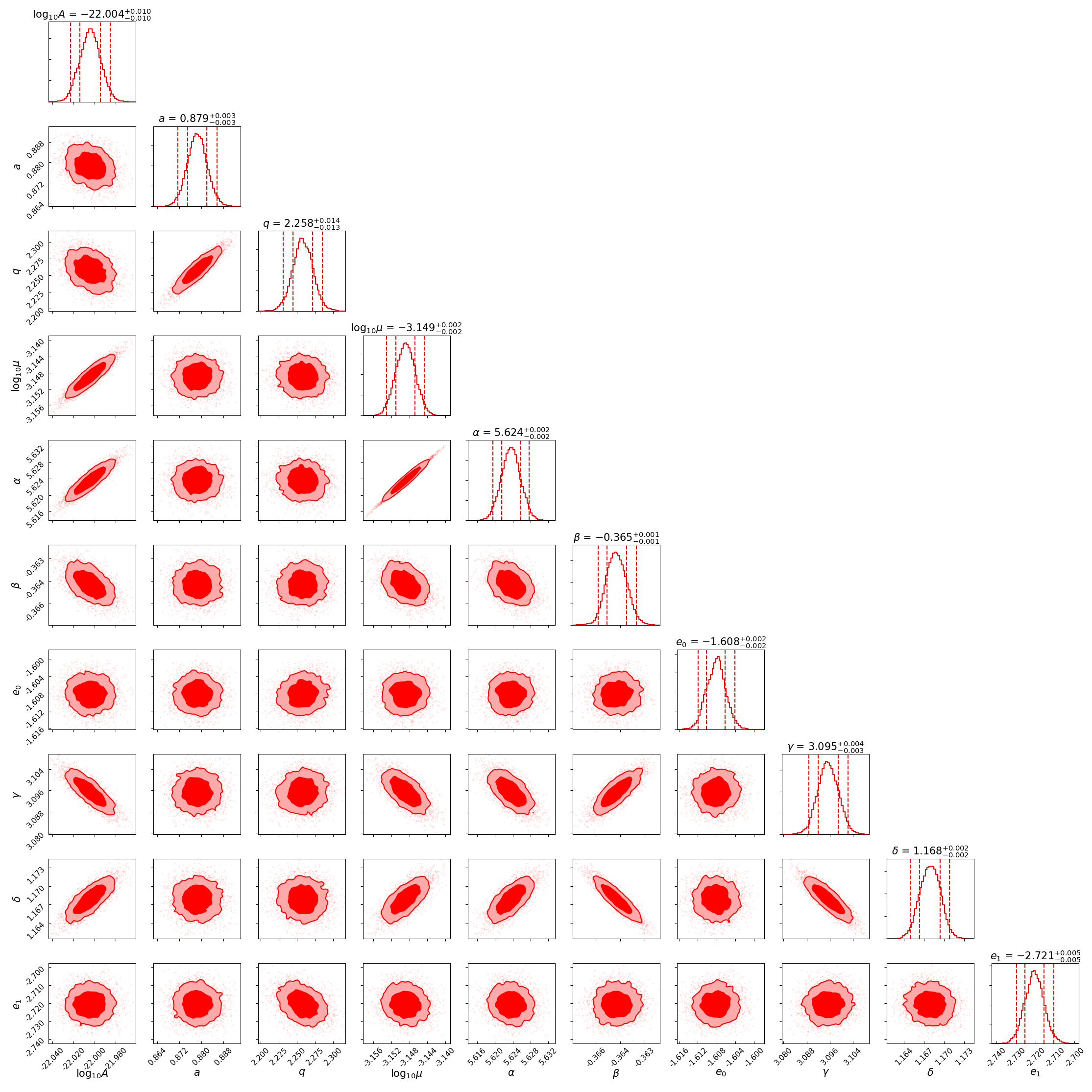}
        \caption{Marginalized posterior distributions of the h($\sigma$,$X_{\rm off,P},\lambda$) best-fit parameters at redshift 0. The 0.68 and 0.95 confidence levels of the posteriors are shown as filled 2D contours. The 2.5th, 16th, 84th, and 97.5th percentiles of the one-dimensional 
        posterior distributions are indicated by the vertical lines on the diagonal plots.}
    \label{fig:corner_z_0_P}
\end{figure}

\begin{table}
        \centering
        \caption{Model parameters with prior and posterior constraints for the redshift evolution of h($\sigma$,$X_{\rm off,P},\lambda$). }
        \label{tab:parameters_zevo_P}
        \begin{tabular}{cc cc cc} 
        \hline
   Parameter & Prior & Posterior \\
   \hline
    $k_0$ & (-0.08,0.07) & -0.0131$\pm 0.0001$\\
    $k_1$ & (-0.25,0.05) & -0.146$\pm0.001$\\
    $k_2$ & (-0.1,0.1) & 0.04$\pm 0.002$\\
    $k_3$ & (-0.15,0.05) & -0.0716$\pm 0.0003$\\
    $k_4$ & (-0.05,0.15) & 0.0789$\pm 0.0001$\\
    $k_5$ & (-0.7,0.1) & -0.4199$\pm 0.0005$\\
    $k_6$ & (-0.15,0.05) & -0.0554$\pm 0.001$\\    
    $k_7$ & (-0.05,0.25) & 0.1526$\pm 0.0002$\\
    $k_8$ & (-0.25,0.05) & -0.1834$\pm 0.0004$\\
    $k_9$ & (-0.05,0.35) & 0.235$\pm 0.001$\\
    \hline
        \end{tabular}\\
        \footnotesize{In the posteriors, when the $1\sigma$ uncertainty on the one-dimensional distribution is smaller than 3 order of magnitudes with respect to the parameter, we write null error. The redshift evolution of each parameter is shown in Fig. \ref{fig:zevo_P}.}        
\end{table}

\begin{figure}
    \centering
    \includegraphics[width=1\columnwidth]{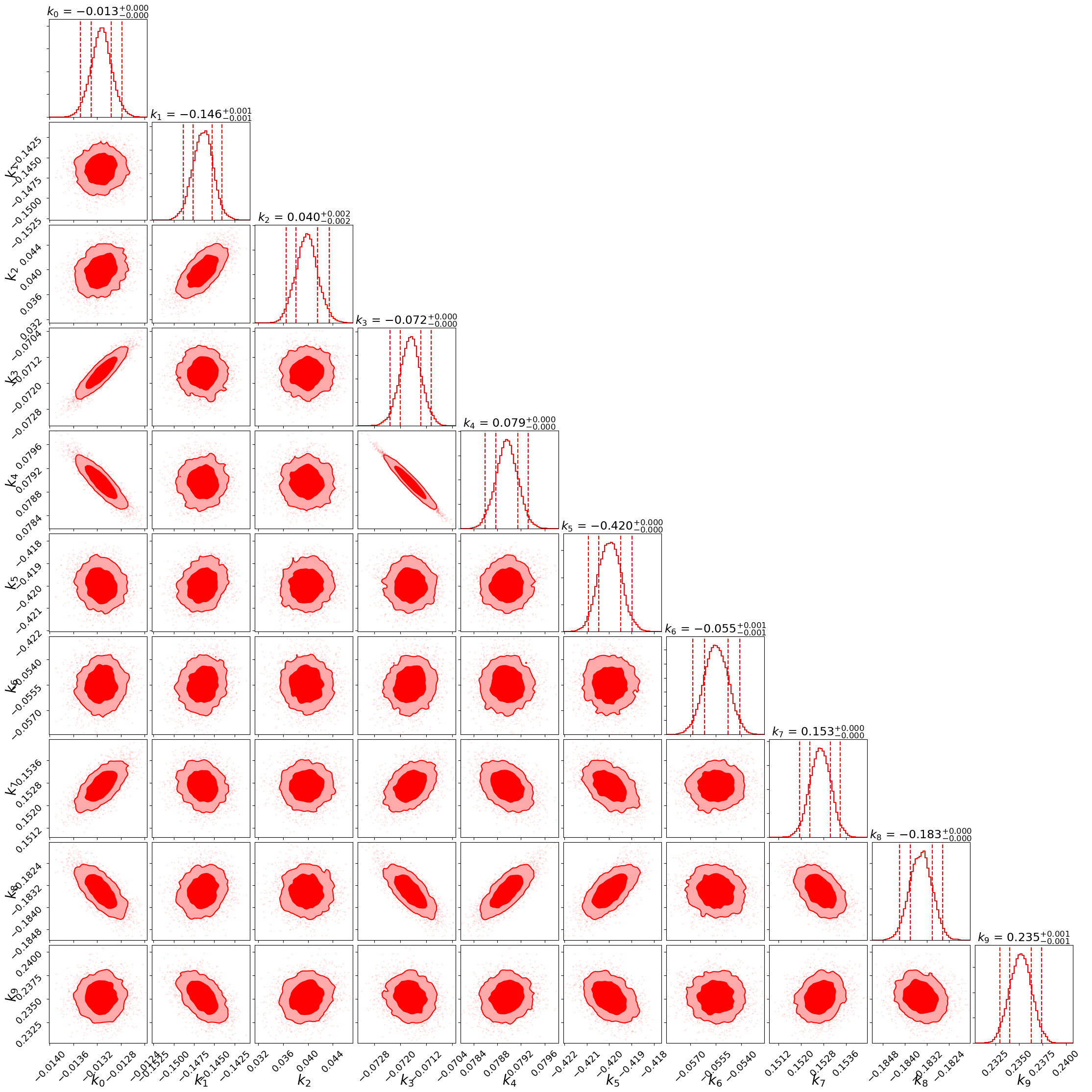}
        \caption{Marginalized posterior distributions of the best-fit parameters describing the redshift evolution of h($\sigma$,$X_{\rm off,P},\lambda$). The 0.68 and 0.95 confidence levels of the posteriors are shown as filled 2D contours. The 2.5th, 16th, 84th, and 97.5th percentiles of the one-dimensional posterior distributions are indicated by the vertical lines on the diagonal plots. }
    \label{fig:corner_zevo_P}
\end{figure}


\end{document}